\pdfoutput=1

\documentclass[11pt,twoside,a4paper,cmspaper,final,collab]{cms-tdr}
\def\svnVersion{269028}\def\cmsCernNo{2013-037}\def\cmsCernDate{\today}\def\cmsMessage{Published in Physical Review D as \href{http://dx.doi.org/10.1103/PhysRevD.90.092007}{\doi{10.1103/PhysRevD.90.092007.}}}

\begin{document}\cmsNoteHeader{SUS-14-002}

\hyphenation{had-ron-i-za-tion}
\hyphenation{cal-or-i-me-ter}
\hyphenation{de-vices}

\RCS$Revision: 266323 $
\RCS$HeadURL: svn+ssh://svn.cern.ch/reps/tdr2/papers/SUS-14-002/trunk/SUS-14-002.tex $
\RCS$Id: SUS-14-002.tex 266323 2014-11-04 01:22:20Z alverson $
\newlength\cmsFigWidth
\ifthenelse{\boolean{cms@external}}{\setlength\cmsFigWidth{0.49\textwidth}}{\setlength\cmsFigWidth{0.7\textwidth}}
\ifthenelse{\boolean{cms@external}}{\providecommand{\cmsLeft}{top\xspace}}{\providecommand{\cmsLeft}{left\xspace}}
\ifthenelse{\boolean{cms@external}}{\providecommand{\cmsRight}{bottom\xspace}}{\providecommand{\cmsRight}{right\xspace}}
\ifthenelse{\boolean{cms@external}}{\providecommand{\cmsLLeft}{Top\xspace}}{\providecommand{\cmsLLeft}{Left\xspace}}
\ifthenelse{\boolean{cms@external}}{\providecommand{\cmsRRight}{Bottom\xspace}}{\providecommand{\cmsRRight}{Right\xspace}}
\ifthenelse{\boolean{cms@external}}{\providecommand{\cmsMiddle}{middle\xspace}}{\providecommand{\cmsMiddle}{top right\xspace}}
\ifthenelse{\boolean{cms@external}}{\providecommand{\cmsLeftTop}{top\xspace}}{\providecommand{\cmsLeftTop}{top left\xspace}}
\ifthenelse{\boolean{cms@external}}{\providecommand{\CL}{C.L.\xspace}}{\providecommand{\CL}{CL\xspace}}
\ifthenelse{\boolean{cms@external}}{\providecommand{\CLend}{C.L.\xspace}}{\providecommand{\CLend}{CL.\xspace}}
\ifthenelse{\boolean{cms@external}}{\providecommand{\NA}{\ensuremath{\cdots}}}{\providecommand{\NA}{\text{---}}}
\newcommand{\red}[1]{{\color{red}#1}}
\newcommand{\lsp}{\ensuremath{\PXXSG}\xspace}
\newcommand{\metsig}{\ensuremath{{\mathcal{S}}_{\mathrm{MET}}}\xspace}
\newcommand{\met}{\MET}
\newcommand{\mdp}{\ensuremath{\Delta\phi_{\text{min}}}\xspace}
\newcommand{\dmH}{\ensuremath{\abs{\Delta m_{\bbbar}}}\xspace}
\newcommand{\amH}{\ensuremath{\langle m_{\bbbar} \rangle}\xspace}
\newcommand{\dRmax}{\ensuremath{\Delta R_{\text{max}}}\xspace}
\newcommand{\epem}{\ensuremath{\Pep\Pem}\xspace}
\newcommand{\mpmm}{\ensuremath{\Pgmp\Pgmm}\xspace}
\newcommand{\mgg}{\ensuremath{{m}_{\gamma\gamma}}\xspace}
\newcommand{\mbb}{\ensuremath{{m}_{\bbbar}}\xspace}
\newcommand{\mjj}{\ensuremath{{m}_{\mathrm{jj}}}\xspace}
\newcommand{\meg}{\ensuremath{{m}_{\Pe\gamma}}\xspace}
\newcommand{\mem}{\ensuremath{{m}_{\Pe\Pgm}}\xspace}
\newcommand{\mll}{\ensuremath{{m}_{\ell\ell}}\xspace}
\newcommand{\nossf}{\ensuremath{N_{\mathrm{OSSF}}}\xspace}
\newcommand{\mt}{\ensuremath{M_{\mathrm{T}}}\xspace}
\newcommand{\tauh}{\ensuremath{\tau_{\mathrm{h}}}\xspace}
\newcommand{\MTtwoj}{\ensuremath{M_{\mathrm{T2}}^{\mathrm{j\ell}}}\xspace}
\newcommand{\riso}{\ensuremath{R_{\text{iso}}}\xspace}
\newcommand{\risoch}{\ensuremath{R^{\text{ch}}_{\text{iso}}}\xspace}
\newcommand{\sthiggs}{\ensuremath{S_{\mathrm{T}}^{\Ph}}\xspace}
\newcommand{\mhiggsino}{\ensuremath{{m}_{\PSGczDo}}\xspace}
\newcommand{\isocone}{\ensuremath{R_{\text{cone}}}\xspace}
\newcommand{\mchionez}{\mhiggsino}
\newcommand{\mchitwoz}{\ensuremath{{m}_{\PSGczDt}}\xspace}
\newcommand{\mchionep}{\ensuremath{{m}_{\PSGcpm_1}}\xspace}
\newcommand{\PSGcmpone}{\ensuremath{\PSGc{^\mp_1}}\xspace} % chargino _1^\mp
\providecommand{\V}{\ensuremath{\cmsSymbolFace{V}}\xspace}
\cmsNoteHeader{SUS-14-002} % This is over-written in the CMS environment: useful as preprint no. for export versions
\title{Searches for electroweak neutralino and chargino production
in channels with Higgs, \texorpdfstring{\Z, and {\PW} bosons in $\Pp\Pp$ collisions at 8\TeV}{Z, and W bosons in pp collisions at 8 TeV}}

\date{\today}

\abstract{
Searches for supersymmetry (SUSY) are presented
based on the electroweak pair production of neutralinos and charginos,
leading to decay channels with Higgs, \cPZ, and {\PW} bosons and
undetected lightest SUSY particles (LSPs).
The data sample corresponds to an integrated luminosity
of about 19.5\fbinv of proton-proton collisions
at a center-of-mass energy of 8\TeV
collected in 2012 with the CMS detector at the LHC.
The main emphasis is neutralino pair production
in which each neutralino decays either to a Higgs boson (\Ph)
and an LSP or to a $\cPZ$ boson and an LSP,
leading to $\Ph\Ph$, $\Ph\cPZ$, and $\cPZ\cPZ$ states
with missing transverse energy ($\ETmiss$).
A second aspect is chargino-neutralino pair production,
leading to $\Ph\PW$ states with $\ETmiss$.
The decays of a Higgs boson to a bottom-quark pair,
to a photon pair,
and to final states with leptons are considered
in conjunction with hadronic and leptonic decay modes of the $\cPZ$ and $\PW$ bosons.
No evidence is found for supersymmetric particles,
and 95\% confidence level upper limits are evaluated for
the respective pair production cross sections
and for neutralino and chargino mass values.
}

\hypersetup{%
pdfauthor={CMS Collaboration},%
pdftitle={Searches for electroweak neutralino and chargino production
in channels with Higgs, Z, and W bosons in pp collisions at 8 TeV
},%
pdfsubject={CMS},%
pdfkeywords={CMS, physics, supersymmetry}}

\maketitle %maketitle comes after all the front information has been supplied

\section{Introduction}
\label{sec-introduction}

Supersymmetry (SUSY)~\cite{Ramond:1971gb,Golfand:1971iw,Neveu:1971rx,
Volkov:1972jx,Wess:1973kz,Wess:1974tw,Fayet:1974pd,Nilles:1983ge},
one of the most widely considered
extensions of the standard model (SM) of particle physics,
stabilizes the Higgs boson mass at the electroweak energy scale,
may predict unification of the strong, weak, and electromagnetic forces,
and might provide a dark matter candidate.
Supersymmetry postulates that each SM particle is paired with a SUSY partner
from which it differs in spin by one-half unit,
with otherwise identical quantum numbers.
For example,
squarks, gluinos, and winos are the SUSY partners
of quarks, gluons, and {\PW} bosons,
respectively.
Supersymmetric models contain extended Higgs sectors~\cite{Nilles:1983ge,Haber:1984rc},
with higgsinos the SUSY partners of Higgs bosons.
Neutralinos~\PSGcz (charginos \PSGcpm) arise from the mixture of
neutral (charged) higgsinos with the
SUSY partners of neutral (charged) electroweak vector bosons.

In this paper,
we consider R-parity-conserving models~\cite{Farrar:1978xj}.
In R-parity-conserving models,
SUSY particles are created in pairs.
Each member of the pair initiates a decay chain
that terminates with a stable lightest SUSY particle (LSP)
and SM particles.
If the LSP interacts only via the weak force,
as in the case of a dark matter candidate,
the LSP escapes detection,
potentially yielding large values of
missing momentum and energy.

Extensive searches for SUSY particles
have been performed at the CERN LHC,
but so far the searches have not uncovered evidence for their
existence~\cite{Aad:2013wta,Aad:2013ija,Aad:2014nua,Aad:2014qaa,Aad:2014vma,
Aad:2014iza,
Chatrchyan:2013wxa,Chatrchyan:2013xna,Chatrchyan:2013iqa,
Chatrchyan:2013fea,Chatrchyan:2014lfa,Khachatryan:2014doa}.
The recent discovery~\cite{Aad:2012tfa,Chatrchyan:2013lba,Chatrchyan:2012ufa}
of the Higgs boson,
with a mass of about 125\GeV,
opens new possibilities for SUSY searches.
In the SUSY context,
we refer to the 125\GeV boson as ``\Ph''~\cite{Branco:2011iw},
the lightest neutral CP-even state of an extended Higgs sector.
The \Ph boson is expected to have the properties of the SM Higgs boson
if all other Higgs bosons are much heavier~\cite{Martin:1997ns}.
Neutralinos and charginos are predicted to decay to
an \Ph or vector ($\V=\Z$, {\PW}) boson
over large regions of SUSY parameter
space~\cite{Matchev:1999ft,Asano:2010ut,Kats:2011qh,
Baer:2012ts,Byakti:2012qk,Howe:2012xe,Ruderman:2011vv}.
Pair production of neutralinos and/or charginos can thus
lead to $\Ph\Ph$, $\Ph\V$, and $\V\V^{(\prime)}$ states.
Requiring the presence of one or more \Ph bosons
provides a novel means to search for these channels.
Furthermore,
the observation of a Higgs boson in a SUSY-like process
would provide evidence that SUSY particles couple to the Higgs field,
a necessary condition for SUSY to stabilize the Higgs boson mass.
This evidence can not be provided by search channels without the Higgs boson.

In this paper,
searches are presented for electroweak pair production of
neutralinos and charginos
that decay to the $\Ph\Ph$, $\Ph\Z$, and $\Ph\PW$ states.
Related SUSY searches sensitive to the corresponding $\Z\Z$ state are
presented in Refs.~\cite{Chatrchyan:2014aea,Khachatryan:2014qwa}.
We assume the Higgs boson \Ph to have SM properties.
The data sample,
corresponding to an integrated luminosity of around 19.5\fbinv
of proton-proton collisions at $\sqrt{s}=8\TeV$,
was collected with the CMS detector at the LHC.
For most of the searches,
a large value of missing energy transverse to the direction
of the proton beam axis (\met) is required.

\begin{figure*}[thb]
\centering
\includegraphics[width=0.32\linewidth]{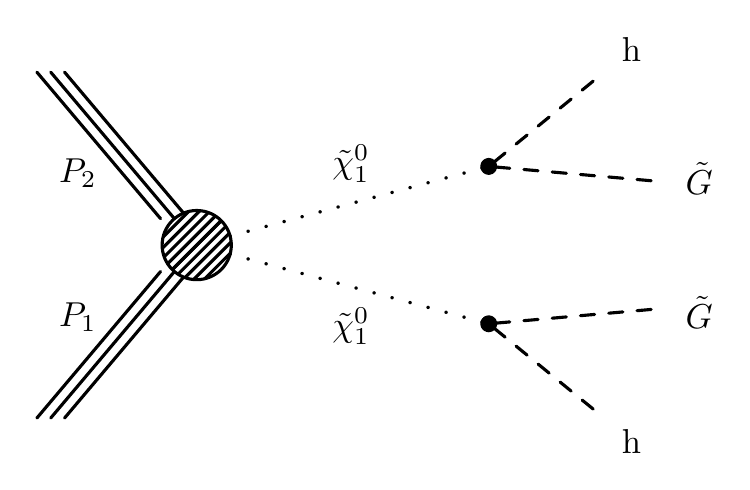}
\includegraphics[width=0.32\linewidth]{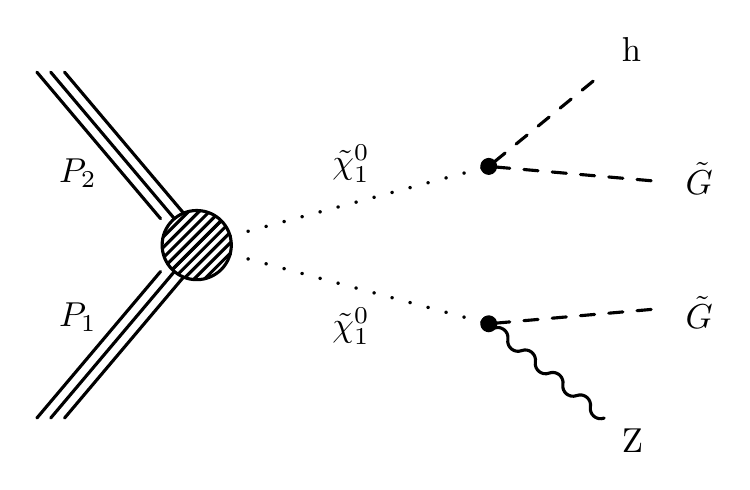}
\includegraphics[width=0.32\linewidth]{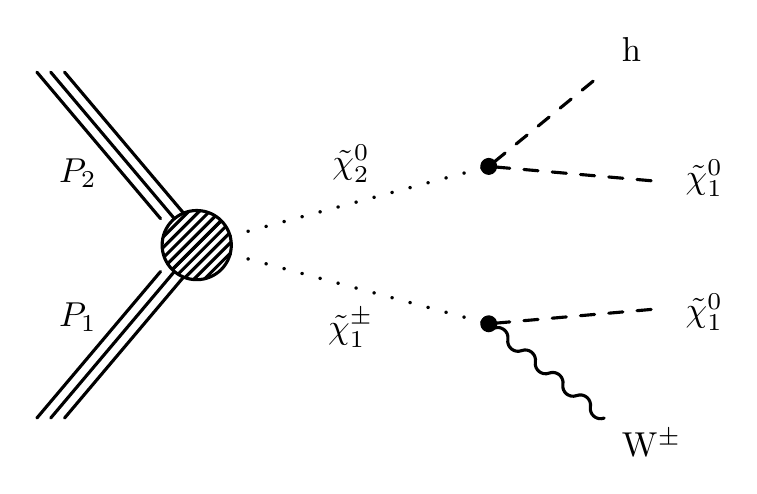}
\caption{
Event diagrams for the SUSY scenarios considered in this analysis.
(Left and center) $\Ph\Ph$ and $\Ph\Z$ production
in a GMSB model~\cite{Matchev:1999ft,Ruderman:2011vv},
where \Ph is the Higgs boson,
\PSGczDo is the lightest neutralino NLSP,
and \lsp is the nearly massless gravitino LSP.
The $\PSGczDo\PSGczDo$ state is created through
$\PSGczDo\PSGczDt$,
$\PSGczDo\PSGcpm_1$, $\PSGczDt\PSGcpm_1$,
and $\PSGcpm_1\PSGcmpone$ production
followed by the decay of the \PSGczDt and $\PSGcpm_1$ states to
the \PSGczDo and undetected SM particles,
with \PSGczDt and $\PSGcpm_1$ the second-lightest neutralino
and the lightest chargino,
respectively.
(Right) $\Ph\PW$ production through chargino-neutralino
$\PSGcpm_1\PSGczDt$
pair creation,
with \PSGczDo a massive neutralino~LSP.
}
\label{fig:event-diagrams}
\end{figure*}

The $\Ph\Ph$,
$\Ph\Z$,
and $\Z\Z$ topologies arise in a number of SUSY scenarios.
As a specific example,
we consider an R-parity-conserving gauge-mediated
SUSY-breaking (GMSB) model~\cite{Matchev:1999ft,Ruderman:2011vv}
in which the two lightest neutralinos \PSGczDo and \PSGczDt,
and the lightest chargino ${\PSGcpm}_1$,
are higgsinos.
In this model,
the \PSGczDo, \PSGczDt and ${\PSGcpm}_1$
are approximately mass degenerate,
with \PSGczDo the lightest of the three states.
The LSP is a gravitino \lsp~\cite{Giudice:1998bp},
the SUSY partner of a graviton.
The \PSGczDt and $\PSGcpm_1$ higgsinos decay to the
\PSGczDo state plus low-\pt SM particles,
where \pt represents momentum transverse to the beam axis.
The \PSGczDo higgsino,
which is the next-to-lightest SUSY particle (NLSP),
undergoes a two-body decay
to either an \Ph boson and \lsp or to a \Z boson and \lsp,
where \lsp is nearly massless, stable, and weakly interacting.
The pair production of any of the combinations
$\PSGczDo\PSGczDt$,
$\PSGczDo\PSGcpm_1$, $\PSGczDt\PSGcpm_1$,
or $\PSGcpm_1\PSGcmpone$ is allowed~\cite{Matchev:1999ft},
enhancing the effective cross section for
the $\PSGczDo\PSGczDo$ di-higgsino state and thus
for $\Ph\Ph$ and $\Ph\Z$ production
(Fig.~\ref{fig:event-diagrams} left and center).
The production of $\Z\Z$ combinations is also possible.
The final state includes
two LSP particles~\lsp,
leading to~\MET.
Note that $\PSGczDt\PSGczDt$
and direct $\PSGczDo\PSGczDo$ production are not allowed
in the pure higgsino limit,
as is considered here.

For the $\Ph\Ph$ combination,
we consider the
$\Ph(\to\bbbar)\Ph(\to\bbbar)$,
$\Ph(\to\gamma\gamma)\Ph(\to\bbbar)$,
and $\Ph(\to\gamma\gamma)\Ph(\to\Z\Z/\PW\PW/\tau\tau)$
decay channels,
with \bbbar a bottom quark-antiquark pair
and where the {$\Z\Z$}, {$\PW\PW$}, and $\tau\tau$ states decay to yield
at least one electron or muon.
For the $\Ph\Z$ combination,
we consider the
$\Ph(\to\gamma\gamma)\Z(\to\,$2~jets$)$,
$\Ph(\to\gamma\gamma)\Z(\to\Pe\Pe/\Pgm\Pgm/\Pgt\Pgt)$,
and $\Ph(\to\bbbar)\Z(\to\Pe\Pe/\Pgm\Pgm)$ channels,
where the $\tau\tau$ pair
yields at least one electron or muon.
We combine the results of the current study with those presented
for complementary Higgs and \Z boson decay modes in
Refs.~\cite{Chatrchyan:2014aea,Khachatryan:2014qwa}
to derive overall limits on electroweak GMSB
$\Ph\Ph$, $\Ph\Z$, and $\Z\Z$ production.

As a second specific example of a SUSY scenario with Higgs bosons,
we consider the R-parity-conserving
chargino-neutralino $\PSGcpm_1\PSGczDt$ electroweak pair production process
shown in Fig.~\ref{fig:event-diagrams} (right),
in which the $\PSGcpm_1$ chargino is wino-like and
the \PSGczDo neutralino is a massive, stable,
weakly interacting bino-like LSP,
where a bino is the SUSY partner of the {\PB} gauge boson.
This scenario represents the
SUSY process with the largest electroweak cross section~\cite{Beenakker:1999xh}.
It leads to the $\Ph\PW$ topology,
with \MET present because of the two LSP particles.
The decay channels considered are
$\Ph(\to\gamma\gamma)\PW(\to\,$2~jets$)$
and $\Ph(\to\gamma\gamma)\PW(\to\ell\nu)$,
with $\ell$ an electron, muon,
or leptonically decaying $\tau$ lepton.
We combine these results with those based on complementary decay modes
of this same scenario~\cite{Khachatryan:2014qwa}
to derive overall limits.

The principal backgrounds arise from the production of
a top quark-antiquark (\ttbar) pair,
a {\PW} boson, {\Z} boson, or photon in association with jets
({\PW}+jets, {\Z}+jets, and $\gamma$+jets),
and multiple jets through the strong interaction (QCD multijet).
Other backgrounds are due to events with a single top quark
and events with rare processes such as {$\ttbar\V$} or SM Higgs boson production.
The QCD multijet category as defined here excludes events in the other categories.
For events with a top quark or {\PW} boson,
significant \met can arise if a {\PW} boson decays leptonically,
producing a neutrino,
while for events with a \Z boson,
the decay of the \Z boson to two neutrinos can yield significant~\MET.
For $\gamma$+jets events,
{\Z}+jets events with
$\Z\to\ell^+\ell^-$ ($\ell=\Pe$, $\Pgm$),
and events with all-hadronic final states,
such as QCD multijet events,
significant \met can arise if the event contains
a charm or bottom quark that undergoes semileptonic decay,
but the principal source of \met is the
mismeasurement of jet~\pt (``spurious'' \met).

This paper is organized as follows.
In Sections~\ref{sec-detector}, \ref{sec-event-selection},
and~\ref{sec-event-simulation},
we discuss the detector and trigger,
the event reconstruction,
and the event simulation.
Section~\ref{sec-hh-4b} presents a search for
$\Ph\Ph$ SUSY events in which both Higgs bosons
decay to a \bbbar pair.
Section~\ref{sec-higgs-to-gg} presents searches
for $\Ph\Ph$, $\Ph\Z$, and $\Ph\PW$ SUSY events in
which one Higgs boson decays to a pair of photons.
A search for $\Ph\Z$ SUSY events with
a Higgs boson that decays to a \bbbar pair and a \Z boson that decays
to an \epem or \mpmm pair is presented in Section~\ref{sec-z-to-ll}.
In Section~\ref{sec-multilepton},
we briefly discuss the studies of
Refs.~\cite{Chatrchyan:2014aea,Khachatryan:2014qwa}
as they pertain to the SUSY scenarios considered here.
The interpretation of the results
is presented in Section~\ref{sec-interpretation} and
a summary in Section~\ref{sec-summary}.

\section{Detector and trigger}
\label{sec-detector}

A detailed description of the CMS detector is
given elsewhere~\cite{Chatrchyan:2008aa}.
A superconducting solenoid of 6\unit{m} internal diameter
provides an axial magnetic field of 3.8\unit{T}.
Within the field volume are a silicon pixel and strip tracker,
a crystal electromagnetic calorimeter,
and a brass-and-scintillator hadron calorimeter.
Muon detectors based on gas ionization chambers
are embedded in a steel flux-return yoke located outside the solenoid.
The CMS coordinate system is defined with the origin at the center
of the detector and with the $z$ axis along the direction of the
counterclockwise beam.
The transverse plane is perpendicular to the beam axis,
with $\phi$ the azimuthal angle (measured in radians),
$\theta$ the polar angle,
and $\eta=-\ln[\tan(\theta/2)]$ the pseudorapidity.
The tracking system covers the region $\abs{\eta}<2.5$,
the muon detector $\abs{\eta}<2.4$,
and the calorimeters $\abs{\eta}<3.0$.
Steel-and-quartz-fiber forward calorimeters cover $3<\abs{\eta}<5$.
The detector is nearly hermetic,
permitting accurate
measurements of energy balance in the transverse plane.

The trigger is based on the identification of events
with one or more jets, bottom-quark jets ($\cPqb$ jets),
photons, or charged leptons.
The main trigger used for the $\Ph\Ph\to\bbbar\bbbar$ analysis
(Section~\ref{sec-hh-4b})
requires the presence of at least two jets with $\pt>30\GeV$,
including at least one tagged $\cPqb$ jet,
and $\met>80\GeV$.
For the diphoton studies (Section~\ref{sec-higgs-to-gg}),
there must be at least one photon with $\pt>36\GeV$ and another with $\pt>22\GeV$.
The study utilizing $\Z\to\ell^+\ell^-$ events
(Section~\ref{sec-z-to-ll})
requires at least one electron or muon with $\pt>17\GeV$ and
another with $\pt>8\GeV$.
Corrections are applied to the selection efficiencies to
account for trigger inefficiencies.

\section{Event reconstruction}
\label{sec-event-selection}

The particle-flow (PF) method~\cite{cms-pas-pft-09-001,cms-pas-pft-10-001}
is used to reconstruct and identify charged and neutral hadrons,
electrons (with associated bremsstrahlung photons), muons, and photons,
using an optimized combination of information from CMS subdetectors.
The reconstruction of photons for the
$\Ph\to\gamma\gamma$-based searches
is discussed in Section~\ref{sec-higgs-to-gg}.
Hadronically decaying $\tau$ leptons (\tauh) are reconstructed using PF objects
(we use the ``hadron-plus-strips'' $\tau$-lepton reconstruction
algorithm~\cite{Chatrchyan:2012zz}
with loose identification requirements).
The event primary vertex,
taken to be the reconstructed vertex
with the largest sum of charged-track $\pt^2$ values,
is required to contain at least four charged tracks
and to lie within 24\cm of the origin in the direction along the beam axis
and 2\cm in the perpendicular direction.
Charged hadrons from extraneous {\Pp\Pp} interactions within
the same or a nearby bunch crossing (``pileup'') are removed~\cite{CMS-PAS-JME-13-005}.
The PF objects serve as input for jet reconstruction,
based on the anti-\kt algorithm~\cite{bib-antikt,Cacciari:2011ma},
with a distance parameter of~0.5.
Jets are required to satisfy basic quality criteria
(jet ID~\cite{cms-pas-jme-10-003}),
which eliminate,
for example,
spurious events caused by calorimeter noise.
Contributions to an individual jet's \pt from pileup
interactions are subtracted~\cite{Cacciari:2007fd}.
Finally,
jet energy corrections are applied as a function of \pt and~$\eta$
to account for residual effects of
non-uniform detector response~\cite{Chatrchyan:2011ds}.

The missing transverse energy \MET is defined as the modulus
of the vector sum of the transverse momenta of all PF objects.
The \MET vector is the negative of that same vector sum.
We also make use of the
{\met} significance variable~{\metsig}~\cite{Chatrchyan:2011tn},
which represents a $\chi^2$ difference between the observed
result for \met and the ${\met}=0$ hypothesis.
The \metsig variable provides an event-by-event assessment of the
consistency of the observed \met with zero,
given the measured content of the event and the known measurement resolutions.
Because it accounts for finite jet resolution on an event-by-event basis,
{\metsig} provides better discrimination between signal and background
events than does \met,
for background events with spurious \met.

The identification of $\cPqb$ jets is performed
using the combined secondary vertex (CSV)
algorithm~\cite{Chatrchyan:2012jua,CMS-PAS-BTV-13-001},
which computes a discriminating variable for each jet based on
displaced secondary vertices,
tracks with large impact parameters,
and kinematic variables,
such as jet mass.
Three operating points are defined,
denoted ``loose,'' ``medium,'' and ``tight.''
These three working points yield average
signal efficiencies for $\cPqb$ jets
(misidentification probabilities for light-parton jets)
of approximately 83\% (10\%), 70\% (1.5\%), and 55\% (0.1\%), respectively,
for jets with $\pt>60\GeV$~\cite{CMS-PAS-BTV-13-001}.

We also make use of isolated electrons and muons,
either vetoing events with such leptons
in order to reduce background from SM \ttbar and electroweak boson production
(Sections~\ref{sec-hh-4b}, \ref{sec-hhggbb}, and~\ref{sec-gg-2j}),
or selecting these events because
they correspond to the targeted signal process
(Sections~\ref{sec-hgglepton} and~\ref{sec-z-to-ll}).
Isolated electron and muon identification is based on the variable \riso,
which is the scalar sum of the \pt values of charged hadrons,
neutral hadrons, and photons within a cone of radius
$\isocone\equiv\sqrt{\smash[b]{(\Delta\phi)^2+(\Delta\eta)^2}}$
around the lepton direction,
corrected for the contributions of pileup interactions,
divided by the lepton \pt value itself.
For the analyses presented here,
$\isocone=0.3$\,(0.4) for electrons (muons),
unless stated otherwise.

\section{Event simulation}
\label{sec-event-simulation}

Monte Carlo (MC) simulations of signal and background
processes are used to optimize selection criteria,
validate analysis performance,
determine signal efficiencies,
and evaluate some backgrounds and systematic uncertainties.

Standard model background events are simulated with
the {\MADGRAPH} {5.1.3.30}~\cite{Alwall:2014hca},
\POWHEG 301~\cite{bib-powheg,Alioli:2009je,Luisoni:2013cuh},
and \PYTHIA 6.4.26~\cite{bib-pythia} generators.
The \ttbar events (generated with {\MADGRAPH}) incorporate up to three 
additional partons,
including {\cPqb} quarks,
at the matrix element level.
The \ttbar+\bbbar events account for contributions from gluon splitting. 
The SM processes are normalized to cross section calculations
valid to next-to-leading (NLO) or
next-to-next-to-leading order~\cite{Frixione:2002ik,
Frixione:2003ei,Czakon:2013goa,Campbell:2012dh,
Garzelli:2012bn,Campbell:2011bn,Gavin:2012sy},
depending on availability,
and otherwise to leading order.
For the simulation of SM events,
the \GEANTfour~\cite{Agostinelli:2002hh} package is
used to model the detector and detector response.

Signal events are simulated with the \MADGRAPH 5.1.5.4 generator,
with a Higgs boson mass of 126\GeV~\cite{Chatrchyan:2013mxa}.
Up to two partons from initial-state radiation (ISR) are allowed.
To reduce computational requirements,
the detector and detector response for signal events
are modeled with the CMS fast simulation
program~\cite{Abdullin:2011zz},
with the exception of the signal events for the
$\Ph\Ph\to\bbbar\bbbar$ study
(Section~\ref{sec-hh-4b}),
for which \GEANTfour modeling is used.
For the quantities based on the fast simulation,
the differences with respect to the GEANT-based results are
found to be small ($\lesssim$5\%).
Corrections are applied,
as appropriate,
to account for the differences.
The signal event rates are normalized to the
NLO plus next-to-leading-logarithmic (NLO+NLL)
cross sections~\cite{Beenakker:1999xh,Fuks:2012qx,Fuks:2013vua}
for the GMSB $\Ph\Ph$, $\Ph\Z$, and $\Z\Z$ channels,
and to the
NLO cross sections~\cite{Beenakker:1999xh,Kramer:2012bx}
for the electroweak $\Ph\PW$ channel.
For the GMSB scenarios
[Fig.~\ref{fig:event-diagrams} (left) and (center)],
the \PSGczDo, \PSGczDt, and $\PSGcpm_1$ particles are taken to be
mass-degenerate pure higgsino states,
such that any SM particles arising from the decays of
the \PSGczDt and $\PSGcpm_1$ states to the \PSGczDo state are
too soft to be detected.
Signal MC samples are generated for
a range of higgsino mass values $m_{\PSGczDo}$,
taking the LSP (gravitino~\lsp) mass to be 1\GeV (\ie, effectively zero).
The decays of the \PSGczDo higgsinos are described
with a pure phase-space matrix element.
For the electroweak $\Ph\PW$ scenario
[Fig.~\ref{fig:event-diagrams} (right)],
we make the simplifying assumption
$m_{\PSGczDt}=m_{\PSGcpm_1}$~\cite{Khachatryan:2014qwa}
and generate event samples for a range of
\PSGczDt and LSP (\PSGczDo) mass values,
with the decays of the $\PSGcpm_1$ chargino and \PSGczDt neutralino described
using the \textsc{bridge} v2.24 program~\cite{Meade:2007js}.
Note that we often consider small LSP masses in this study,
viz.,
${m}_{\lsp}=1\GeV$ for the GMSB scenario,
and,
in some cases,
${m}_{\PSGczDo}=1\GeV$
for the electroweak $\Ph\PW$ scenario
[see Figs.~\ref{fig:higgs-mt}, \ref{fig:higgs-ggl-met},
\ref{fig:hw-limits-01} (bottom), and~\ref{fig:hw-limits-02}, below].
These scenarios are not excluded by limits~\cite{ALEPH:2005ab}
on \Z boson decays to undetected particles for the cases considered here,
in which the LSP is either a gravitino or a bino-like neutralino~\cite{Carena:2012np}.

All MC samples incorporate the CTEQ6L1 or
CTEQ6M~\cite{Pumplin:2002vw,Nadolsky:2008zw}
parton distribution functions,
with \PYTHIA used for parton showering and hadronization.
The MC events are corrected to account for pileup interactions,
such that they describe the distribution of reconstructed
vertices observed in data.
The simulations are further adjusted so that the {\cPqb}-jet tagging and
misidentification efficiencies match those
determined from control samples in the data.
The $\cPqb$-jet tagging efficiency correction factor depends
slightly on the jet \pt and $\eta$ values
and has a typical value of 0.99, 0.95, and 0.93
for the loose, medium, and tight CSV operating points~\cite{Chatrchyan:2012jua}.
Additional corrections are applied so that the jet energy resolution
in signal samples corresponds to the observed results.
A further correction,
implemented as described in Appendix~B of Ref.~\cite{Chatrchyan:2013xna},
accounts for mismodeling of ISR in signal events.

\section{Search in the \texorpdfstring{$\Ph\Ph\to\bbbar\bbbar$}{hh to b bbar b bbar} channel}
\label{sec-hh-4b}

With a branching fraction of about 0.56~\cite{Heinemeyer:2013tqa},
$\Ph\to\bbbar$ decays
represent the most likely decay mode of the Higgs boson.
The $\Ph(\to\bbbar)\Ph(\to\bbbar)$
final state thus provides a sensitive search channel for
SUSY $\Ph\Ph$ production.
For this channel,
the principal visible objects are the four $\cPqb$ jets.
Additional jets may arise from
ISR, final-state radiation, or pileup interactions.
For this search,
jets (including $\cPqb$ jets) must satisfy $\pt>20\GeV$ and $\abs{\eta}<2.4$.
In addition,
we require the following:
\begin{itemize}
\item exactly four or five jets,
   where $\pt>50\GeV$ for the two highest \pt jets;
\item \met significance $\metsig > 30$;
\item no identified, isolated electron or muon candidate with $\pt>10\GeV$;
  electron candidates are restricted to $\abs{\eta}<2.5$ and
  muon candidates to $\abs{\eta}<2.4$;
  the isolation requirements are $\riso<0.15$ for electrons and
  $\riso<0.20$ for muons;
\item no \tauh candidate
  with $\pt>20\GeV$ and $\abs{\eta}<2.4$;
\item no isolated charged particle with $\pt>10\GeV$ and $\abs{\eta}<2.4$,
  where the isolation condition is based on the scalar sum \risoch of 
  charged-particle \pt values in a cone of radius $\isocone=0.3$ around
  the charged-particle direction,
  excluding the charged particle itself,
  divided by the charged-particle \pt value;
  we require $\risoch<0.10$;
\item $\mdp>0.5$ for events with $30<\metsig<50$
  and $\mdp>0.3$ for $\metsig>$50,
  where \mdp is the smallest difference in $\phi$ between the \met vector
  and any jet in the event;
  for the \mdp calculation
  we use less restrictive criteria for jets compared with
  the standard criteria:
  $\abs{\eta}<5.0$,
  no rejection of jets from pileup interactions,
  and no jet ID requirements,
  with all other conditions unchanged.
\end{itemize}
The isolated charged-particle requirement rejects events with
a \tauh decay to a single charged track
as well as events with an isolated electron or muon
in cases where the lepton is not identified.
The \mdp restriction eliminates QCD multijet and all-hadronic \ttbar events,
whose contribution is expected to be large at small values of~\metsig.
The use of less restrictive jet requirements for the
\mdp calculation yields more efficient rejection of these backgrounds.

Three mutually exclusive samples of events with tagged $\cPqb$ jets
are defined:
\begin{itemize}
\item 2$\cPqb$ sample:
  Events in this sample must contain exactly two tight $\cPqb$ jets
  and no medium $\cPqb$ jets;
\item 3$\cPqb$ sample:
  Events in this sample must contain two jets that are tight $\cPqb$ jets,
  a third jet that is either a tight or a medium $\cPqb$ jet,
  and no other tight, medium, or loose $\cPqb$ jet;
\item 4$\cPqb$ sample:
  Events in this sample must contain two jets that are tight $\cPqb$ jets,
  a third jet that is either a tight or medium $\cPqb$ jet,
  and a fourth jet that is either a tight, medium, or loose $\cPqb$ jet.
\end{itemize}
The sample most sensitive to signal events is the 4$\cPqb$ sample.
The 3$\cPqb$  sample is included to improve the signal efficiency.
The 2$\cPqb$  sample is depleted in signal events and
is used to help evaluate the background,
as described below.
The dominant background arises from \ttbar events
in which one top quark decays hadronically
while the other decays to a state with a lepton $\ell$ through
$\cPqt\to\cPqb\ell\nu$,
where the lepton is not identified and the neutrino provides a
source of genuine~\met.

\begin{figure*}[tbh]
\centering
\includegraphics[width=0.49\linewidth]{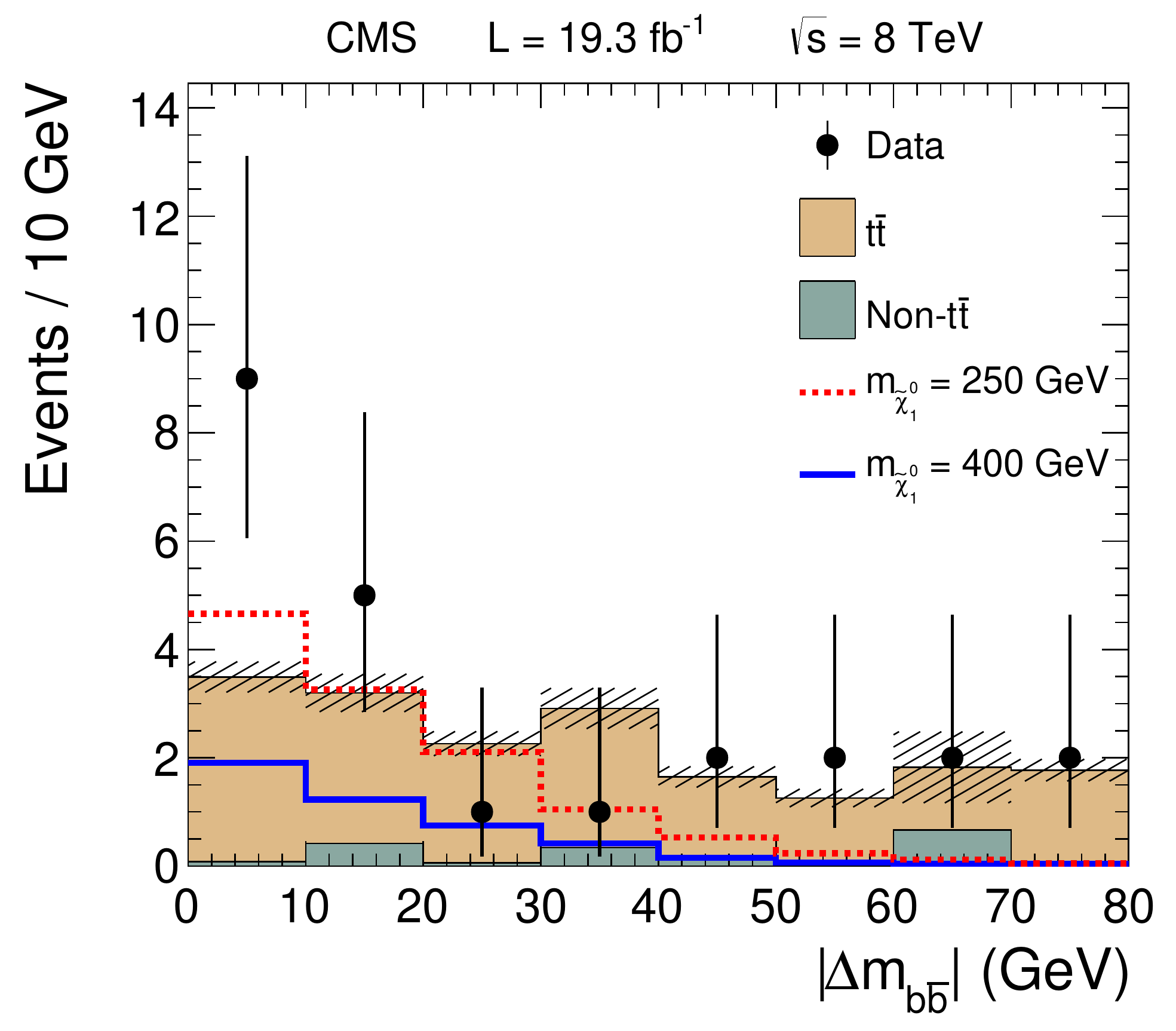}
\includegraphics[width=0.49\linewidth]{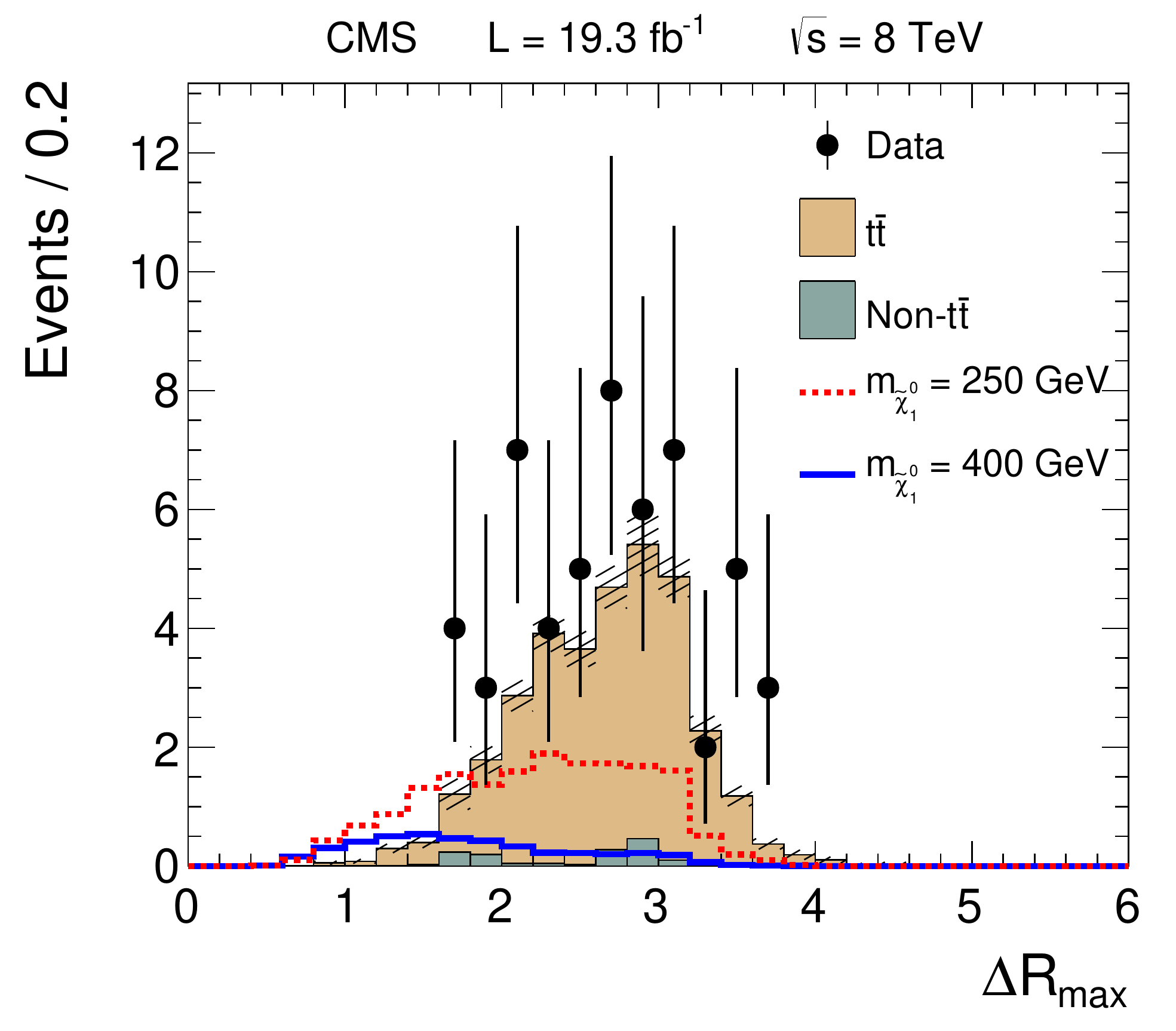} \\
\includegraphics[width=0.49\linewidth]{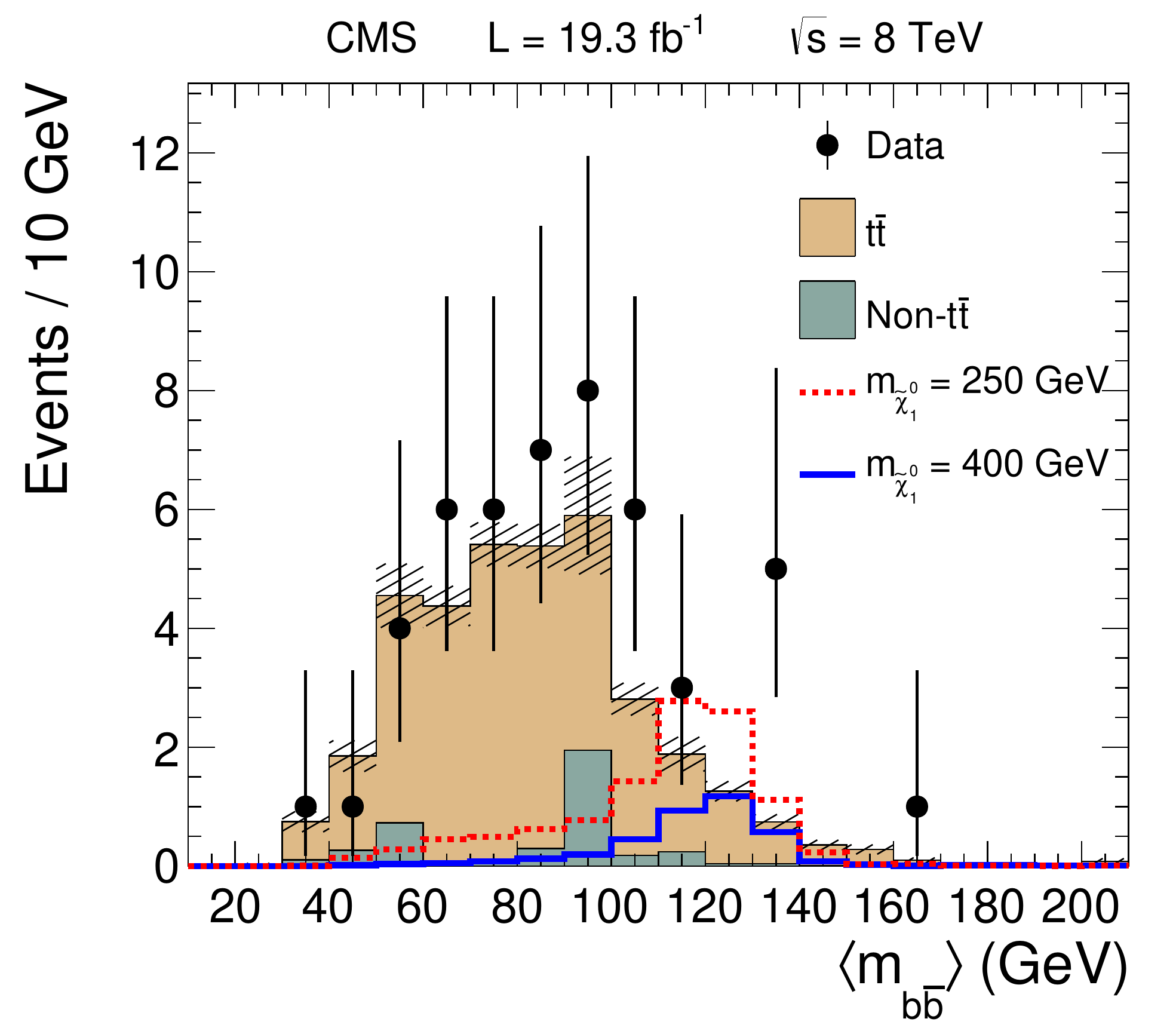}
\caption{
Distributions of events in the 4$\cPqb$  sample
of the $\Ph\Ph\to\bbbar\bbbar$ analysis,
after all signal region requirements are applied
except for that on the displayed variable,
in comparison with simulations
of background and signal events:
(\cmsLeftTop)~\dmH,
(\cmsMiddle)~\dRmax,
and (\cmsRight)~\amH.
For the signal events,
results are shown for higgsino (\PSGczDo) mass values of 250 and 400\GeV,
with an LSP (gravitino) mass of 1\GeV.
The background distributions are stacked while the signal
distributions are not.
The hatched bands indicate the statistical uncertainty
of the total SM simulated prediction.
}
\label{fig:hh-bbbb-nminusone}
\end{figure*}

\begin{figure*}[tbp]
  \centering
   \includegraphics[width=0.46\linewidth]{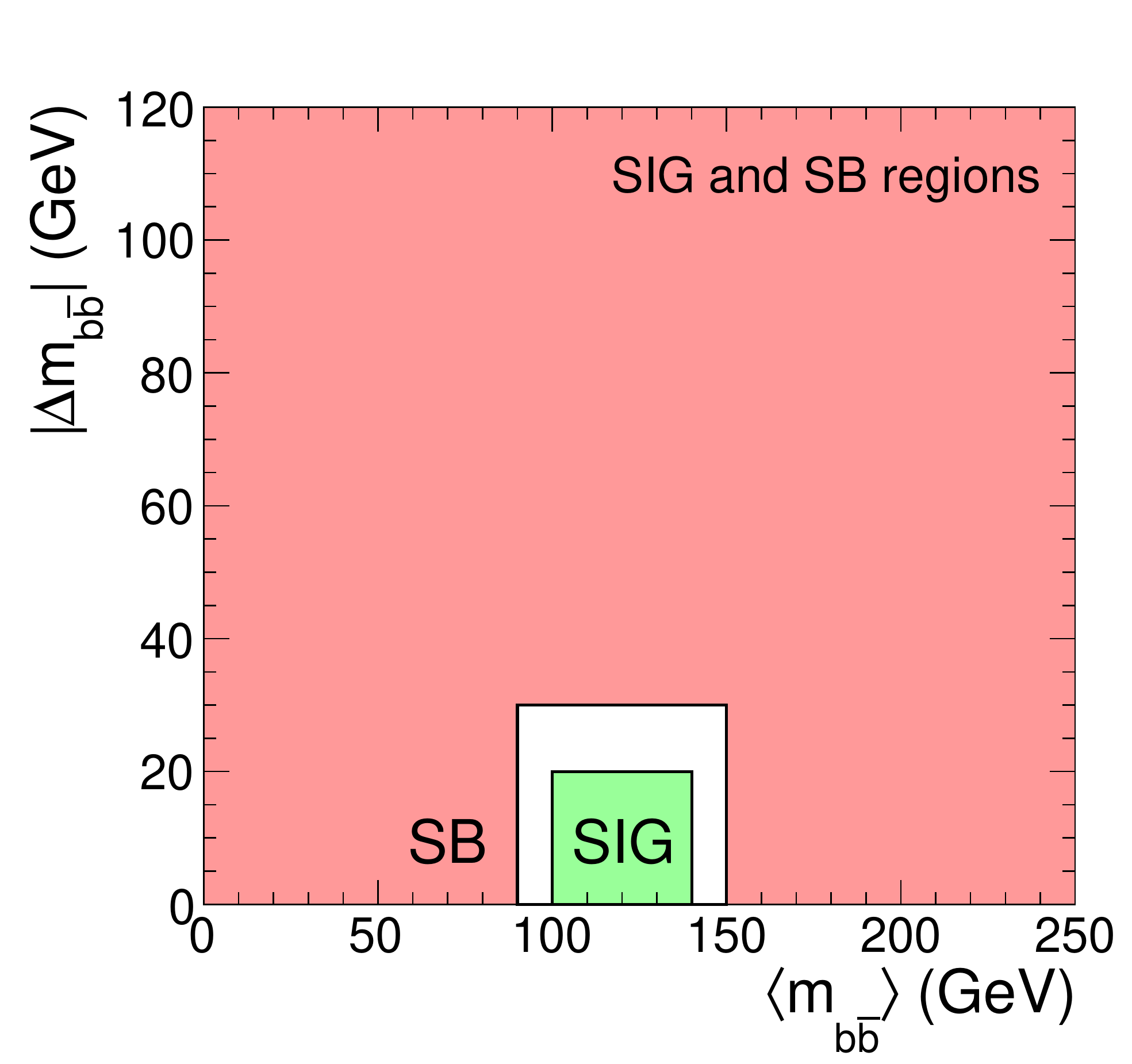}
   \includegraphics[width=0.46\linewidth]{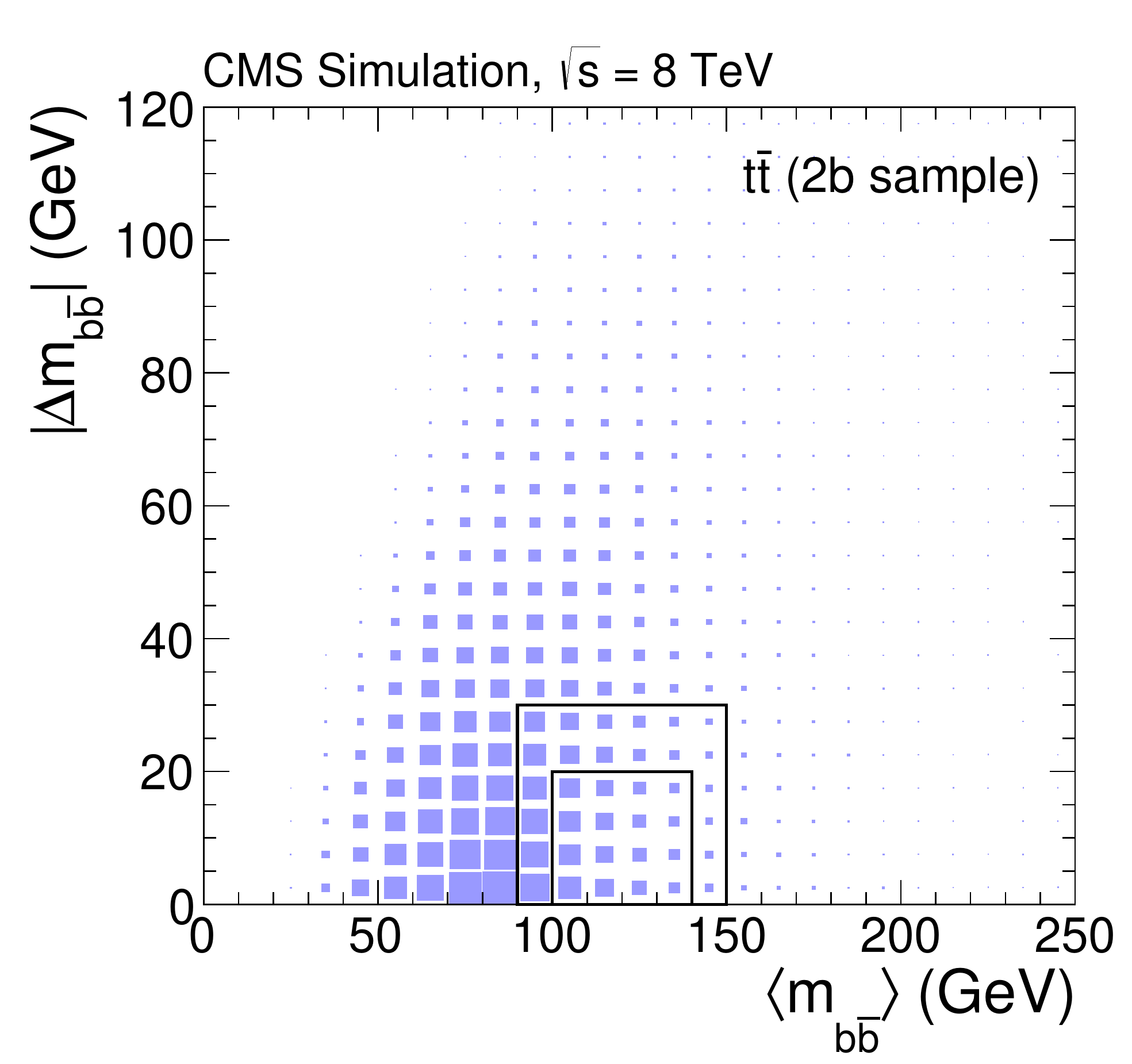} \\
   \includegraphics[width=0.46\linewidth]{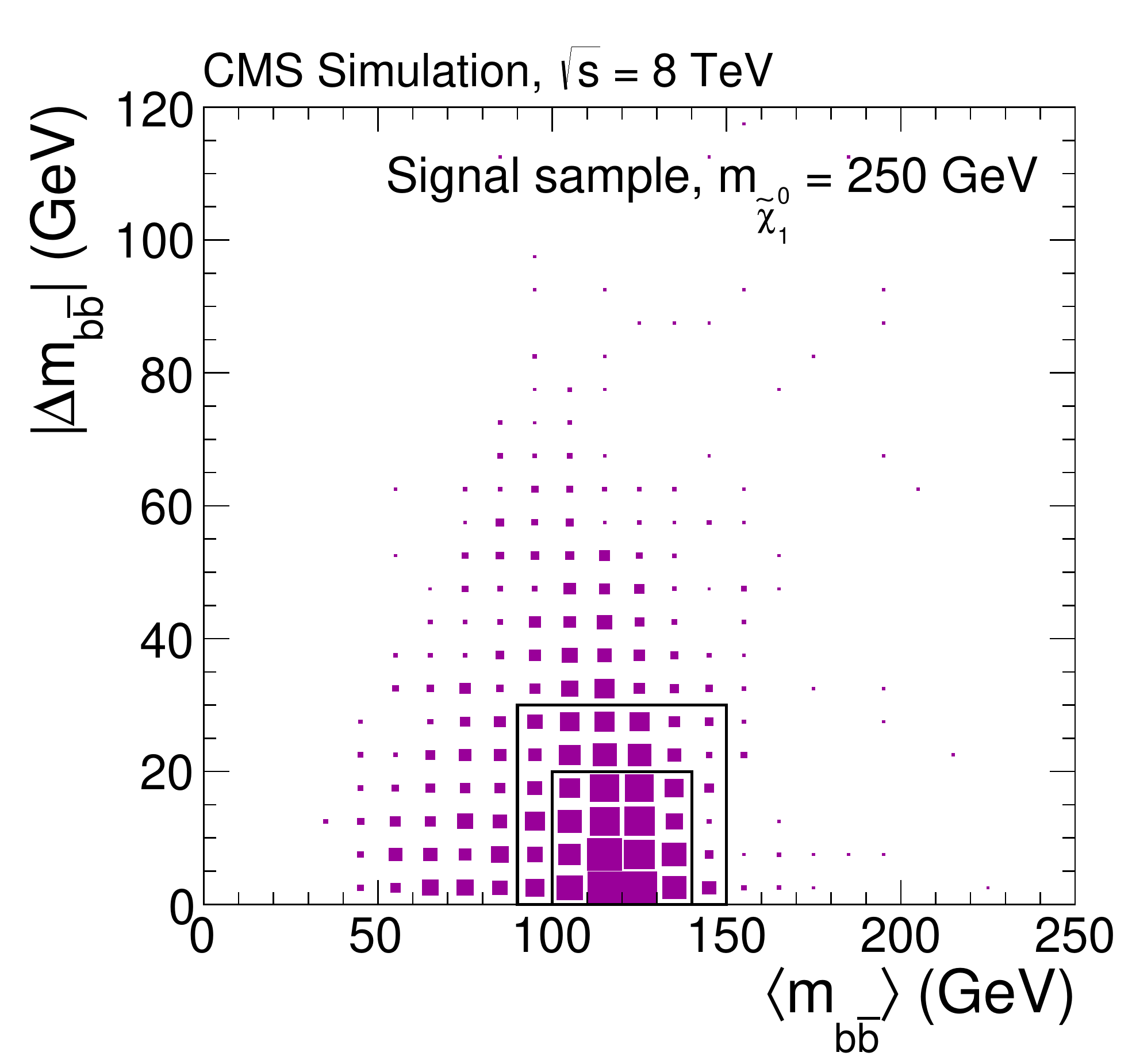}
   \includegraphics[width=0.46\linewidth]{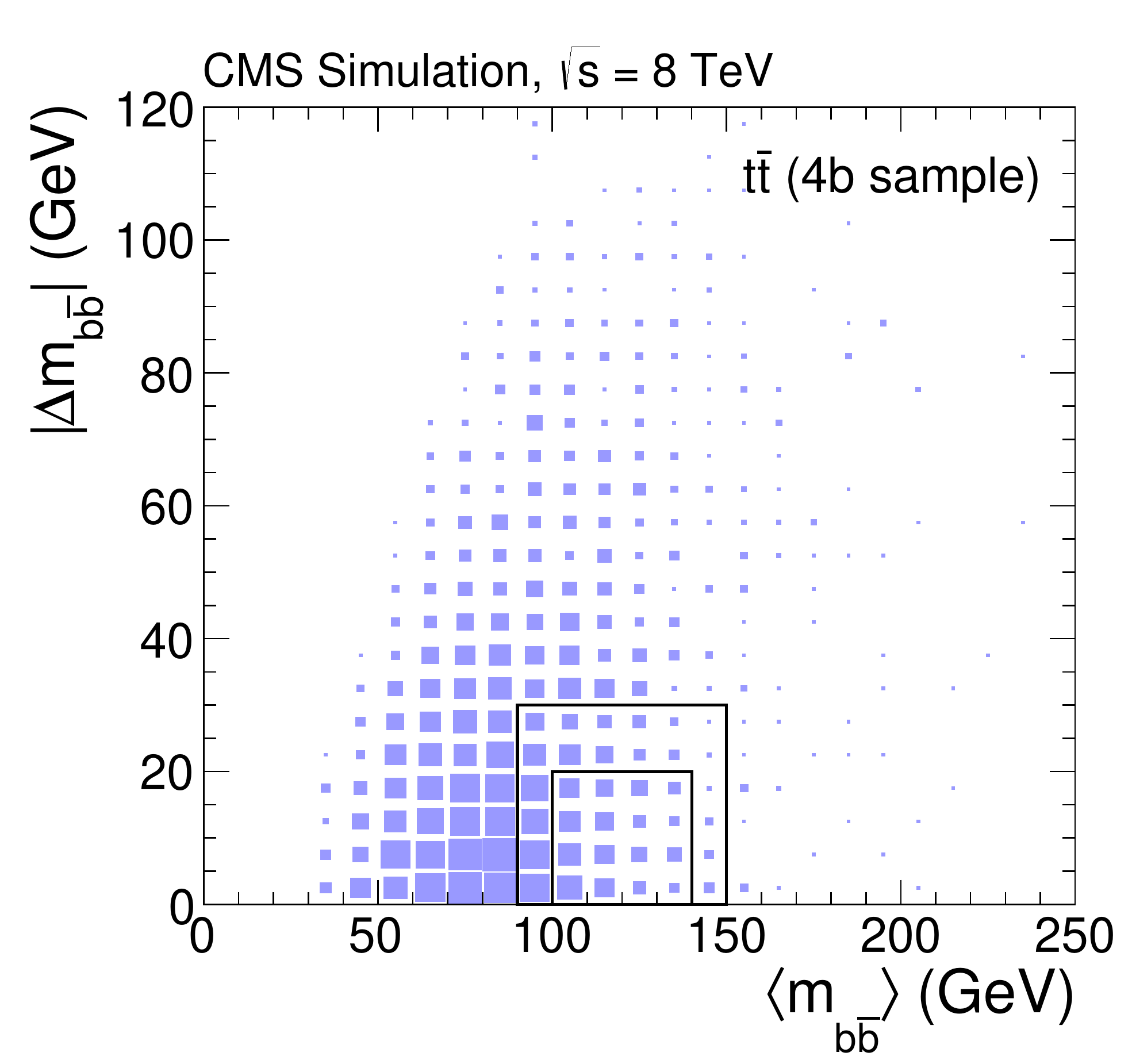}
  \caption{
(Top left) Illustration of the signal (SIG) and sideband (SB)
regions in the \dmH versus \amH plane
of the $\Ph\Ph\to\bbbar\bbbar$ analysis.
(Top right and bottom right)
Distributions of simulated \ttbar events in the 2$\cPqb$  and 4$\cPqb$  samples.
(Bottom left)
Distribution of simulated signal events in the 4$\cPqb$  sample
for a higgsino (\PSGczDo) mass of 250\GeV
and an LSP (gravitino) mass of 1\GeV.
The plots employ an arbitrary integrated luminosity.
The size of a box is proportional to the relative number of events.
}
  \label{fig:abcd-2d}
\end{figure*}

To reconstruct the two Higgs boson candidates in an event,
we choose the four most \cPqb-like jets
based on the value of the CSV discriminating variable.
These four jets can be grouped in three unique ways to form
a pair of Higgs boson candidates.
Of the three possibilities, we choose the one with the smallest
difference $\dmH \equiv \abs{m_{\bbbar,1} - m_{\bbbar,2}}$
between the two candidate masses,
where \mbb is the invariant mass of two tagged $\cPqb$ jets.
We calculate the distance
$\Delta R \equiv \sqrt{\smash[b]{ (\Delta\phi)^2 + (\Delta\eta)^2}}$
between the two jets
for each $\Ph\to\bbbar$ candidate.
We call the larger of these two values \dRmax.
In signal events,
the two $\cPqb$ jets from the decay of a Higgs boson
generally have similar directions since the Higgs boson
is not normally produced at rest.
Thus the two $\Delta R$ values tend to be small,
making \dRmax small.
In contrast, for the dominant background,
from the class of \ttbar events described above,
three jets tend to lie in the same hemisphere,
while the fourth jet lies in the opposite hemisphere,
making \dRmax relatively large.

A signal region (SIG) is defined using the variables \dmH, \dRmax,
and the average of the two Higgs boson candidate mass values
$\amH \equiv (m_{\bbbar,1} + m_{\bbbar,2})/2$.
We require
\begin{itemize}
 \item $\dmH < 20 \GeV$ ;
 \item $\dRmax < 2.2$ ;
 \item $100 < \amH < 140\GeV$.
\end{itemize}
These requirements are determined through an optimization procedure
that takes into consideration both the higgsino discovery potential
and the ability to set stringent limits in the case of non-observation.
Distributions of these variables
for events in the 4$\cPqb$  event sample
are shown in Fig.~\ref{fig:hh-bbbb-nminusone}.

A sideband region (SB) is defined by applying the SIG-region
criteria except using the area outside
the following rectangle in the {\dmH}-{\amH} plane:
\begin{itemize}
  \item $\dmH < 30$\GeV;
  \item  $90<\amH<150$\GeV.
\end{itemize}
Schematic representations of the SIG and SB
regions are shown in Fig.~\ref{fig:abcd-2d} (upper left).

To illustrate the basic principle of
the background determination method,
consider the 4$\cPqb$  and 2$\cPqb$  samples.
We can define four observables,
denoted A, B, C, and D:
\begin{itemize}
\item A: number of background events in the 4$\cPqb$-SIG region;
\item B: number of background events in the 4$\cPqb$-SB region;
\item C: number of background events in the 2$\cPqb$-SIG region;
\item D: number of background events in the 2$\cPqb$-SB region.
\end{itemize}
We assume that the ratio of the number of background events in the SIG
region to that in the SB region,
denoted as the SIG/SB ratio,
is the same for the 2$\cPqb$  and 4$\cPqb$  samples.
This assumption is supported by (for example) the similarity between the
2$\cPqb$  and 4$\cPqb$  results shown
in the top-right and bottom-right plots of Fig.~\ref{fig:abcd-2d}.
We further assume that the 2$\cPqb$-SIG and all SB regions
are dominated by background.
The prediction for the number of background events in the 4$\cPqb$-SIG region
is then given by the algebraic expression A$\,=\,$B$\,$C/D.
The same result applies replacing the 4$\cPqb$  sample by the 3$\cPqb$  sample
in the above discussion.

\begin{table*}[tbhp]
\centering
\topcaption{
Observed numbers of events and corresponding SM background estimates
in bins of {\met} significance~\metsig
for the $\Ph\Ph\to\bbbar\bbbar$ analysis.
For the SM background estimate,
the first uncertainty is statistical and the second systematic.
Numerical results for example signal scenarios are given in
Tables~\ref{tab:sigeff250} and~\ref{tab:sigeff400}
of the Appendix.
}
\label{tab:results}
\begin{scotch}{cccccc}
 \metsig bin & \metsig  & SM background & Data      & SM background & Data\\
             &  range   &  (3$\cPqb$-SIG)    & (3$\cPqb$-SIG) & (4$\cPqb$-SIG)     & (4$\cPqb$-SIG) \\
\hline
& & & & & \\[-3mm]
1 & 30--50   & $6.7^{+1.4+1.0}_{-1.1-0.7}$   & 4  & $2.9^{+0.8+0.5}_{-0.6-0.4}$    & 4  \\[1mm]
2 & 50--100  & $11.6^{+1.9+0.9}_{-1.6-0.7}$  & 15 & $4.9^{+1.1+1.4}_{-0.9-0.9}$    & 7  \\[1mm]
3 & 100--150 & $2.44^{+0.84+0.56}_{-0.64-0.35}$   & 1  & $0.59^{+0.39+0.09}_{-0.26-0.09}$ & 3  \\[1mm]
4 & $>$150    & $1.50^{+0.82+0.64}_{-0.54-0.32}$   & 0  & $0.40^{+0.39+0.26}_{-0.22-0.10}$ & 0  \\[1mm]
\end{scotch}
\end{table*}

\begin{figure}[tbh]
  \centering
   \includegraphics[width=\cmsFigWidth]{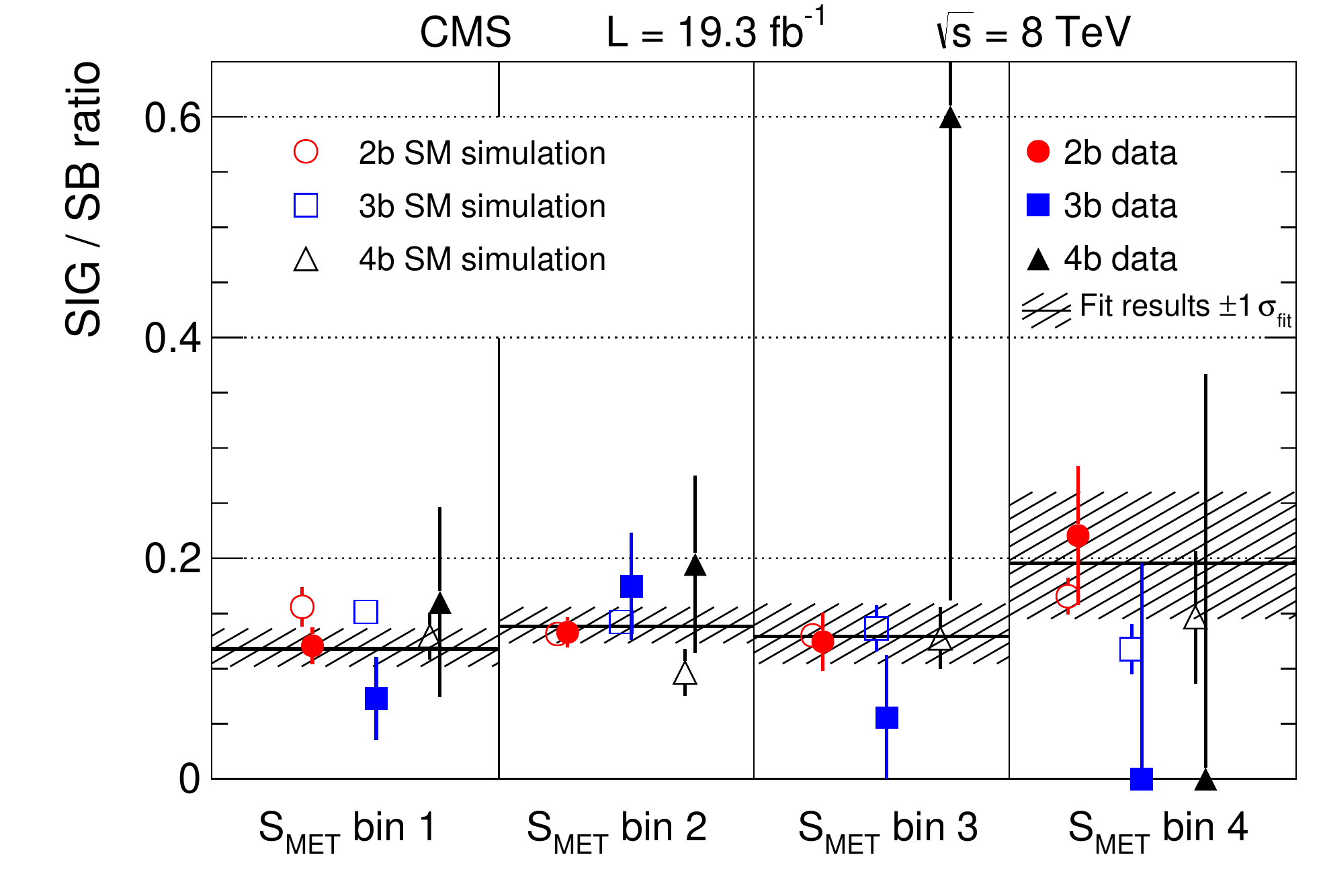}
\caption{
Ratio of the number of events in the signal (SIG) region
to that in the sideband (SB) region
as a function of \metsig bin
(see Table~\ref{tab:results}),
for the 2$\cPqb$, 3$\cPqb$, and 4$\cPqb$  event samples
of the $\Ph\Ph\to\bbbar\bbbar$ analysis.
The simulated results account for the various expected SM processes.
The results of a likelihood fit to data,
in which the SIG/SB ratio is determined separately for each bin,
are also shown.
}
\label{fig:ratio-sigsb-datamc}
\end{figure}

\begin{figure}[tbh]
\centering
\includegraphics[width=\cmsFigWidth]{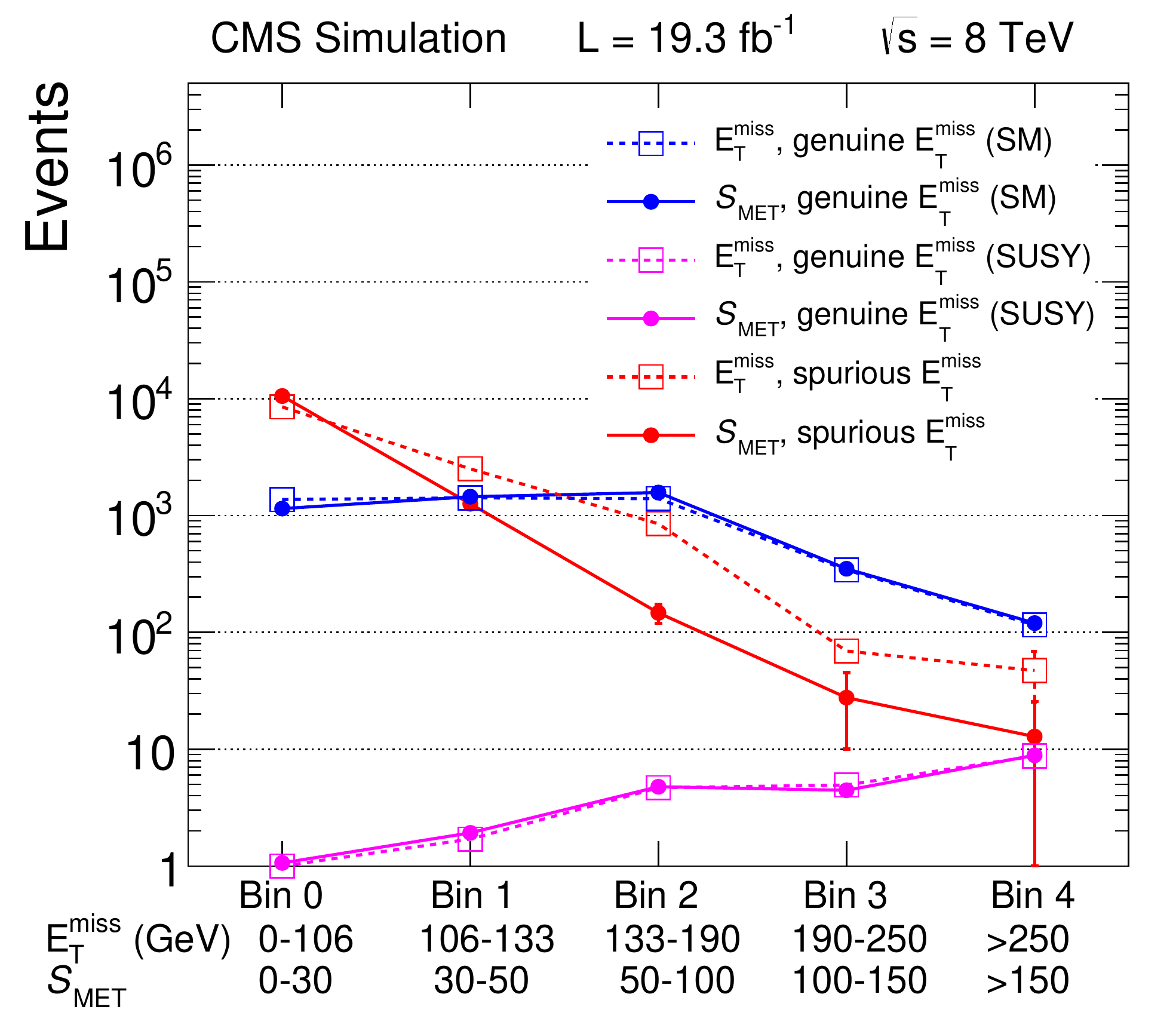}
\caption{
Distribution of simulated \ttbar [``genuine \met (SM)''],
signal [``genuine \met (SUSY)''],
and QCD multijet (``spurious \met'') events using loosened selection criteria (see text)
in bins of \metsig and \met.
The uncertainties are statistical.
The bin edges for \met have been adjusted so that the number of \ttbar events
in each bin is about the same as for the corresponding \metsig bin.
The signal events correspond to the higgsino pair production scenario
of Fig.~\ref{fig:event-diagrams} (left)
with a higgsino (\PSGczDo) mass of 250\GeV
and an LSP (gravitino) mass of 1\GeV.
}
\label{fig:metsig-vs-sig}
\end{figure}

\begin{figure}[tbhp]
\centering
\includegraphics[width=0.48\textwidth]{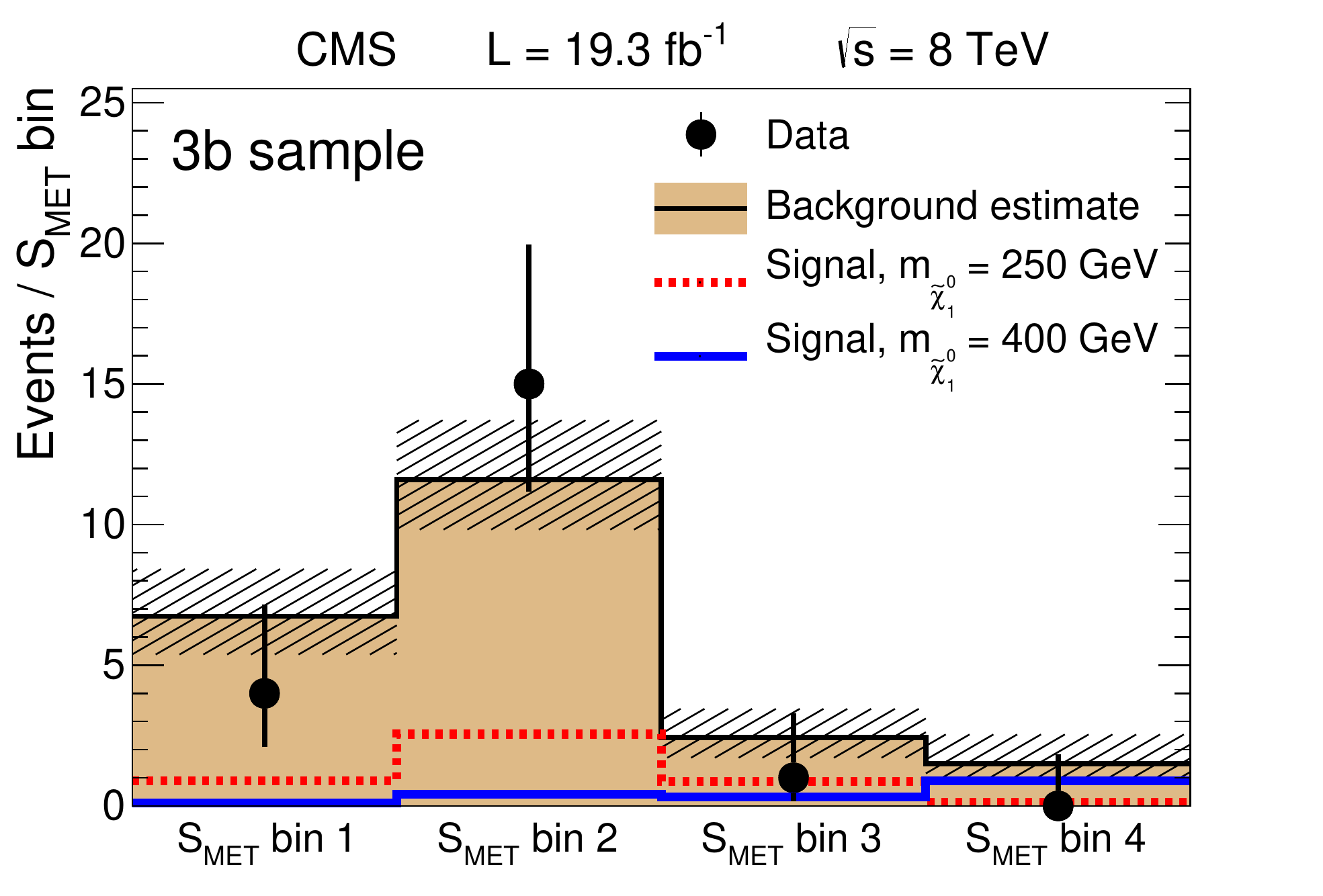}
\includegraphics[width=0.48\textwidth]{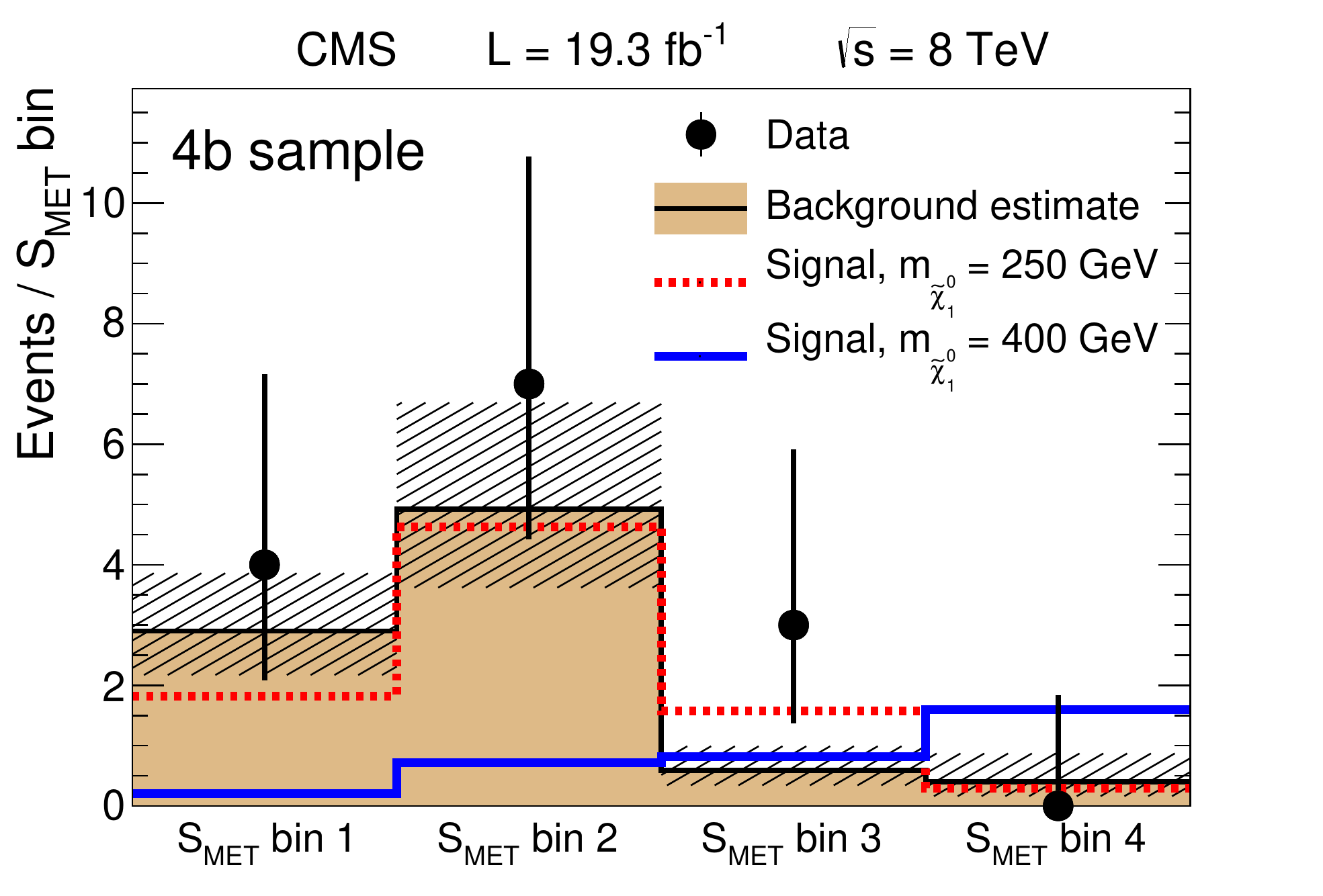}
\caption{
Observed numbers of events as a function of \met significance (\metsig) bin
for the $\Ph\Ph\to\bbbar\bbbar$ analysis,
in comparison with the SM background estimate from the likelihood fit,
for the (\cmsLeft) 3$\cPqb$-SIG and (\cmsRight) 4$\cPqb$-SIG regions.
The hatched bands show the total uncertainty of the background prediction,
with statistical and systematic terms combined.
The expected (unstacked) results for signal events,
with higgsino (\PSGczDo) mass values of 250 and 400\GeV
and an LSP (gravitino) mass of 1\GeV,
are also shown.
}
\label{fig:result-plots}
\end{figure}

In practice,
we examine the data in four bins of \metsig,
which are indicated in Table~\ref{tab:results}.
The background yields in the four \metsig bins
of the 2$\cPqb$-SIG, 3$\cPqb$-SIG, and 4$\cPqb$-SIG regions are
determined simultaneously in a likelihood fit,
with the SIG/SB ratios for the background in all three \cPqb-jet samples
constrained to a common value (determined in the fit)
for each \metsig bin separately.
Figure~\ref{fig:ratio-sigsb-datamc} shows the predictions
of the SM simulation for the SIG/SB ratios,
in the four bins of \metsig,
for the three \cPqb-jet samples
(for purposes of comparison, the data are also shown).
It is seen that for each individual bin of \metsig,
the SIG/SB ratio of SM events is
predicted to be about the same for all three \cPqb-jet samples,
\ie, within \metsig bin~1,
the 2$\cPqb$, 3$\cPqb$, and 4$\cPqb$  results are all about the same,
within \metsig bin~2 they are all about the same, etc.,
supporting the key assumption of the method.
Figure~\ref{fig:ratio-sigsb-datamc} includes the
results determined from the likelihood fit
for the SIG/SB ratio in each bin,
assuming the SUSY signal yield to be zero.
Note that in setting limits
(Section~\ref{sec-interpretation}),
the contributions of signal events to both the signal
and sideband regions are taken into account,
and thus,
e.g., 
the level of signal contribution
to the SB regions does not affect the results.

The four bins of \metsig correspond roughly to \met ranges of
106--133\GeV, 133--190\GeV, 190--250\GeV, and $>$250\GeV, respectively,
as determined from a sample of events selected with loosened criteria.
For this result,
the edges of the \met ranges are adjusted so that the number of
selected \ttbar MC events is about the same
within the respective \met and \metsig bins.
The loosened selection criteria,
specifically no requirement on \mdp and a requirement of least two tight $\cPqb$ jets
with no other \cPqb-jet restrictions,
permit more QCD multijet events to enter the sample,
allowing the relative merits of the \met and \metsig variables to be tested.
The results are illustrated in Fig.~\ref{fig:metsig-vs-sig}.
The \metsig variable is seen to provide better rejection of
background events with spurious \met than does \met,
as mentioned in Section~\ref{sec-event-selection}.

To evaluate the systematic uncertainty of the background estimate,
we consider two terms,
determined from simulation,
which are treated as separate nuisance parameters in the likelihood fit.
The first term is determined for each bin of \metsig in the 4$\cPqb$  (3$\cPqb$) sample.
It is given by the difference from unity of the double ratio $R$,
where $R$ is the SIG/SB ratio of 4$\cPqb$  (3$\cPqb$) events
divided by the SIG/SB ratio of 2$\cPqb$  events
(``non-closure result''),
or else by the statistical uncertainty of $R$,
whichever is larger.
The size of this uncertainty varies between 14 and 40\%,
with a typical value of 25\%.
The second term accounts for potential differences between
the SIG/SB ratio of \ttbar and QCD multijet events
as well as for the possibility that the fraction of \ttbar
and QCD multijet events differs between the 2$\cPqb$, 3$\cPqb$, and 4$\cPqb$  samples.
Based on studies with a QCD multijet data control sample,
the fraction of background events due to QCD multijet
events is conservatively estimated to be less than 20\%.
We reevaluate the background assuming that the fraction
of QCD multijets varies by the full 20\% between the 2$\cPqb$  and 4$\cPqb$  samples
and find the non-closure to be~7\%,
which we define as the associated uncertainty.

The observed numbers of events in the
3$\cPqb$-SIG and 4$\cPqb$-SIG regions are shown in
Fig.~\ref{fig:result-plots} as a function of \metsig,
in comparison with
the SM background predictions from the likelihood fit
and the predictions of two signal scenarios.
Numerical values are given in Table~\ref{tab:results}.

\section{\texorpdfstring{Search in the $\Ph\Ph$, $\Ph\Z$, and $\Ph\PW$
channels  with one $\Ph\to\gamma\gamma$ decay}{Search in the hh, hZ, and hW channels with one h to gamma gamma decay}}
\label{sec-higgs-to-gg}

We next describe searches for
$\Ph\Ph$,
$\Ph\Z$,
and $\Ph\PW$ states
in channels with one Higgs boson that decays to photons.
While the $\Ph\to\gamma\gamma$ branching fraction
is small~~\cite{Heinemeyer:2013tqa},
the expected diphoton invariant-mass signal peak is narrow,
allowing the SM background to be reduced.
For $\Ph\Ph$ production,
we search in channels in which the
second Higgs boson decays to
$\bbbar$, $\PW\PW$, $\Z\Z$, or $\tau\tau$,
where, in the case of these last three modes,
at least one electron or muon is required to be present in the final state.
For the $\Ph\Z$ and $\Ph\PW$ combinations,
we search in the channels in which the {\Z} or {\PW}
boson decays either to two light-flavor jets
or leptonically,
where the leptonic decays yield at least one electron or muon.

Photon candidates are reconstructed from ``superclusters'' of energy
deposited in the electromagnetic
calorimeter~\cite{CMS-PAS-EGM-10-005,Chatrchyan:2013dga},
with energies determined using a multivariate
regression technique~\cite{Chatrchyan:2013lba,Chatrchyan:2013dga}.
To reduce contamination from electrons misidentified as photons,
photon candidates are rejected if they register hit patterns
in the pixel detector that are consistent with a track.
The photon candidates
are required to satisfy loose identification criteria
based primarily on their shower shape and isolation~\cite{bib-cms-dp-2013-010}.
Signal events tend to produce decay products in the central region of the detector,
because of the large masses of the produced SUSY particles.
Therefore,
photon candidates are restricted to $\abs{\eta}<1.44$.

Events must contain at least one photon candidate
with $\pt>40\GeV$ and another with $\pt>25\GeV$.
The $\Ph\to\gamma\gamma$ boson candidate is formed from
the two highest \pt photons in the event.
The resulting diphoton invariant mass \mgg is required to
appear in the Higgs boson mass region defined by $120<\mgg<131\GeV$.

For the searches described in this section,
jets must have $\pt>30\GeV$ and $\abs{\eta}<2.4$.
Tagged $\cPqb$ jets are defined using the CSV-medium criteria.

\subsection{\texorpdfstring{$\Ph\Ph\to\gamma\gamma\bbbar$}{hh to gamma gamma b bbar}}
\label{sec-hhggbb}

\begin{figure}[tbhp]
  \centering
   \includegraphics[width=\cmsFigWidth]{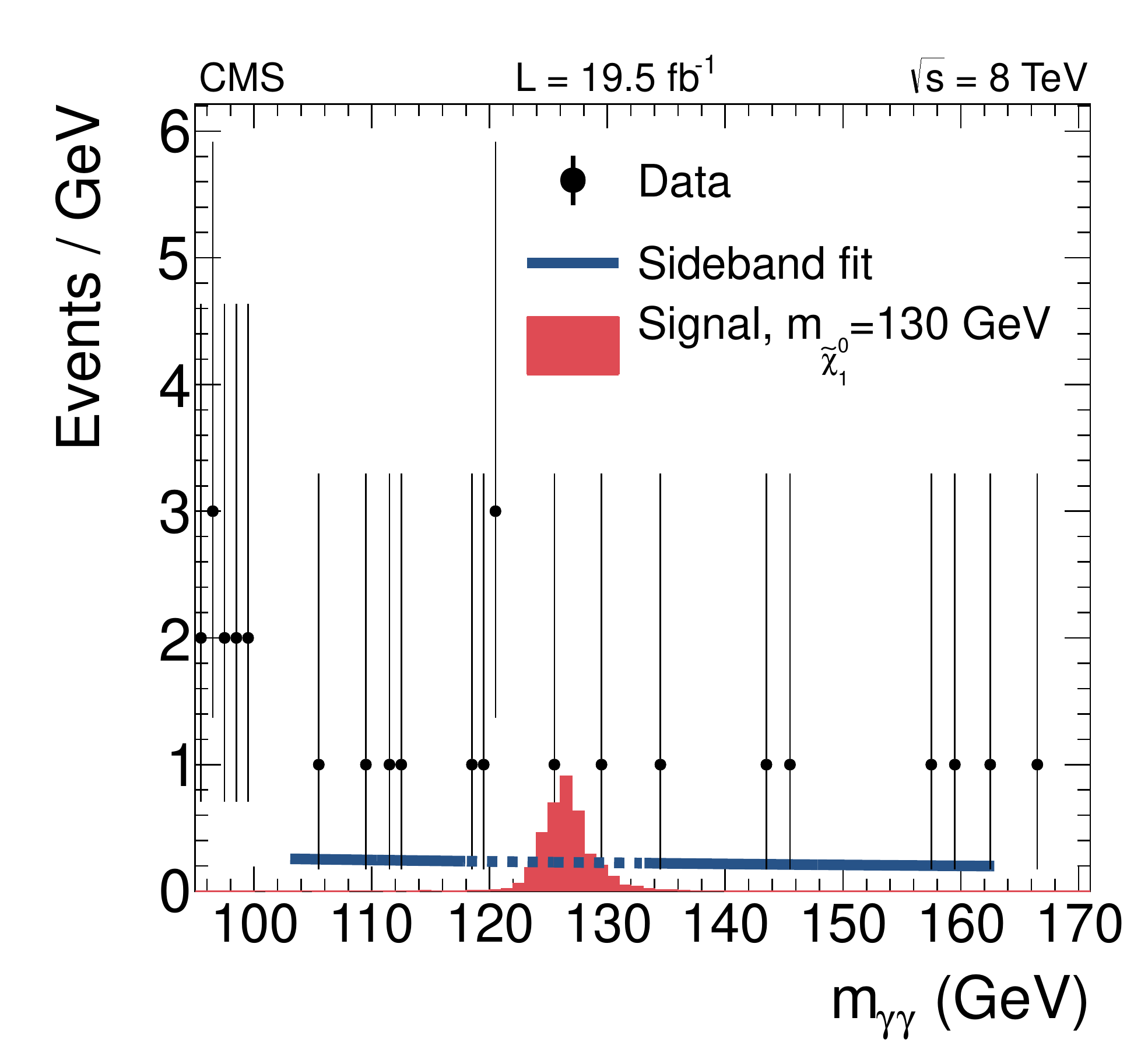}
  \caption{
Distribution of diphoton invariant mass \mgg
after all selection criteria are applied except for that on~\mgg,
for the $\Ph(\to\gamma\gamma)\Ph(\to\bbbar)$ search.
The result of a fit to a power-law function using data
in the sideband regions (see text)
is indicated by the solid line.
The dotted line shows an interpolation of the fitted function
into the Higgs boson mass region excluded from the fit.
The expected results for signal events,
with a higgsino (\PSGczDo) mass value of 130\GeV
and an LSP (gravitino) mass of 1\GeV,
are also shown.
}
  \label{fig:hgg-mgg}
\end{figure}

For the search in the
$\Ph(\to\gamma\gamma)\Ph(\to\bbbar)$ channel,
we require
\begin{itemize}
\item exactly two tagged $\cPqb$ jets, which together form the $\Ph\to\bbbar$ candidate;
\item the invariant mass \mbb
  of the two tagged $\cPqb$ jets to lie
  in the Higgs boson mass region defined by $95<\mbb<155\GeV$;
\item no identified, isolated electron or muon candidate,
  where the lepton identification criteria are $\pt>15\GeV$ and $\abs{\eta}<2.4$,
  with the isolation requirements $\riso<0.15$
  for electrons and $\riso<0.12$ for muons.
\end{itemize}

The distribution of \mgg for the selected events is shown
in Fig.~\ref{fig:hgg-mgg}.
The principal background arises from
events in which a neutral hadron is misidentified as a photon.

\begin{figure}[tbhp]
  \centering
   \includegraphics[width=0.48\textwidth]{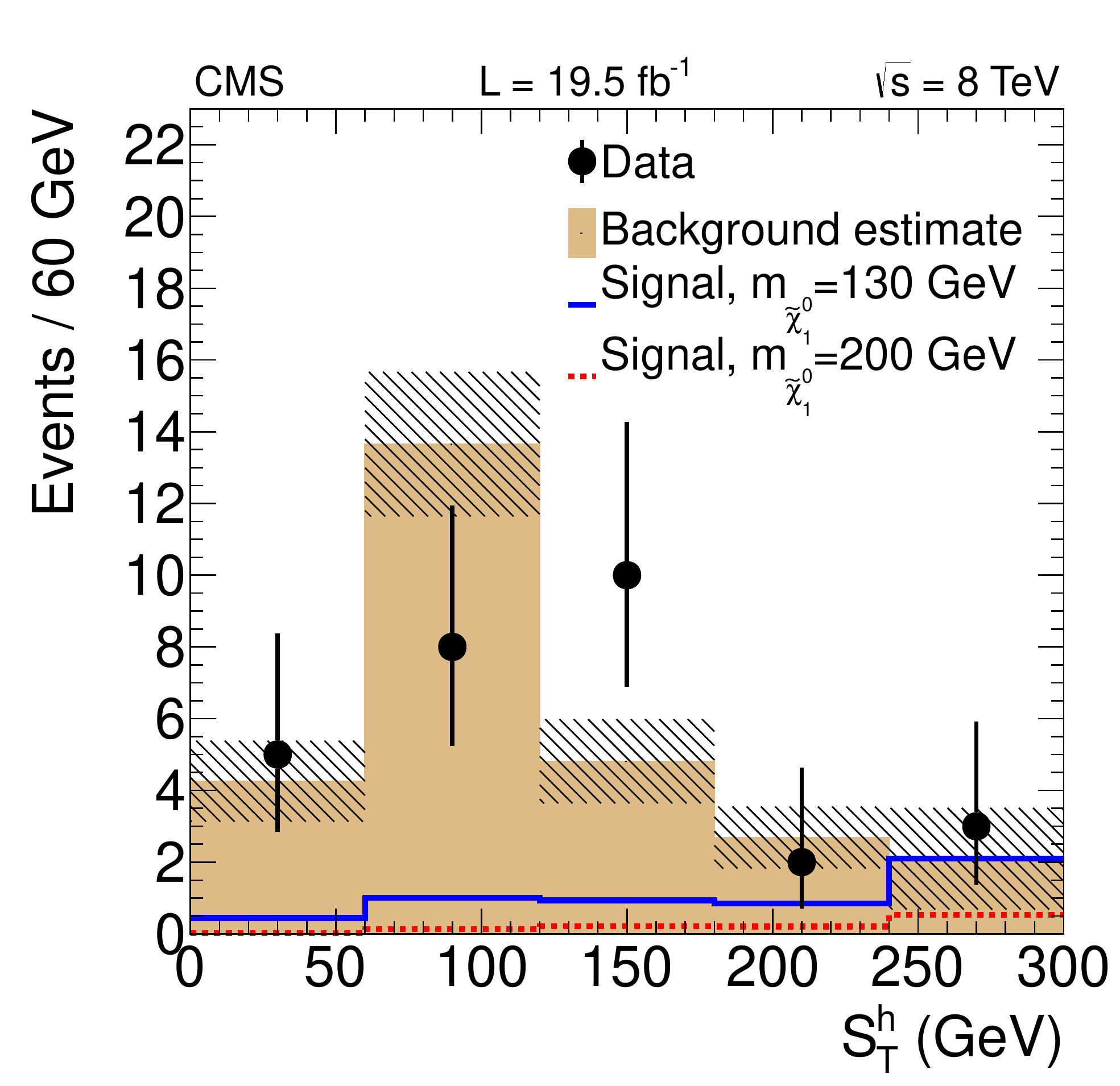}
   \includegraphics[width=0.48\textwidth]{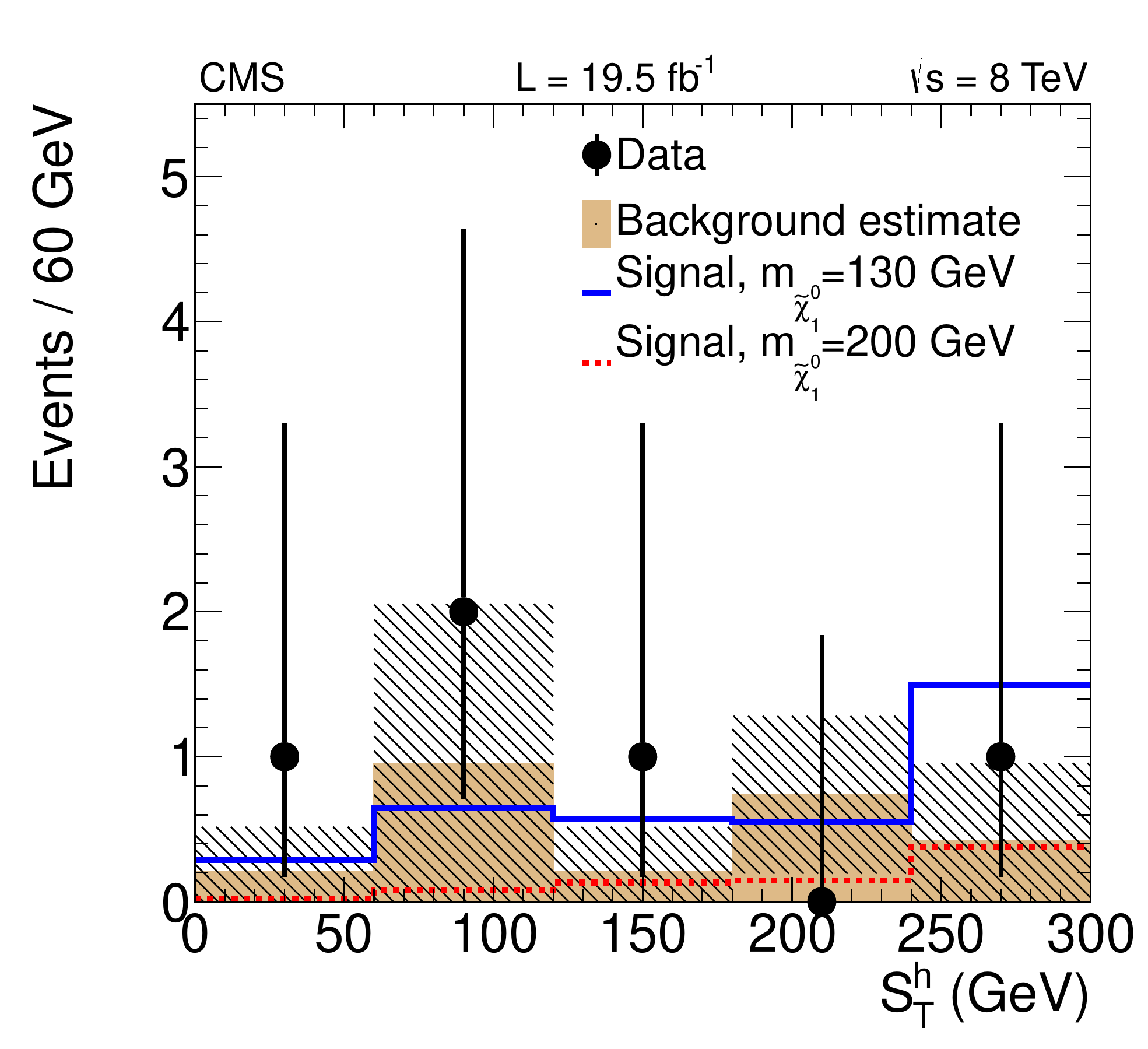}
  \caption{
Observed numbers of events as a function of the scalar sum of
\pt values of the two Higgs boson candidates,
\sthiggs,
for the $\Ph\Ph\to\gamma\gamma\bbbar$ analysis,
in comparison with the SM background estimate,
(\cmsLeft) for a control sample with loose tagging
requirements for $\cPqb$ jets,
and (\cmsRight) for the nominal selection.
The hatched bands show the total uncertainty of the background prediction,
with statistical and systematic terms combined.
The (unstacked) results for signal events,
with higgsino (\PSGczDo) mass values of 130 and 200\GeV
and an LSP (gravitino) mass of 1\GeV,
are also shown.
}
  \label{fig:hgg-kine-01}
\end{figure}

\begin{table*}[tbhp]
\centering
\topcaption{
Observed numbers of events and corresponding SM background estimates,
in bins of Higgs-boson-candidate variable~\sthiggs (see text),
for the $\Ph\Ph\to\gamma\gamma\bbbar$ analysis.
The uncertainties shown for the SM background estimates
are the combined statistical and systematic terms,
while those shown for signal events are statistical.
The expected yields for signal events,
with a higgsino mass value of 130\GeV
and an LSP (gravitino) mass of 1\GeV,
are also shown.
}
\label{tab:hbb-sthiggs}
\begin{scotch}{cccc}
\sthiggs bin (\GeVns{}) & SM background & Data & $\Ph\Ph$ events, \mhiggsino = 130\GeV\\
\hline
& & & \\[-3mm]
0--60    & $0.21^{+0.28}_{-0.21}$ & 1 & $0.28\pm0.03$ \\[1mm]
60--120  & $0.95^{+0.99}_{-0.95}$ & 2 & $0.63\pm0.04$ \\[1mm]
120--180 & $0.21^{+0.29}_{-0.21}$ & 1 & $0.55\pm0.04$ \\[1mm]
180--240 & $0.74 \pm 0.38$        & 0 & $0.53\pm0.04$ \\[1mm]
$>$240  & $0.42^{+0.49}_{-0.42}$ & 1 & $1.46\pm0.06$ \\[1mm]
\end{scotch}
\end{table*}

The SM background,
with the exception of the generally small contribution
from SM Higgs boson production,
is evaluated using \mgg data sidebands defined by
$103\leq\mgg\leq118\GeV$ and $133\leq\mgg\leq163\GeV$.
We construct the quantity \sthiggs,
which is the scalar sum of the \pt values of the two
Higgs boson candidates.
The distribution of \sthiggs is measured separately
in each of the two sidebands.
Each sideband distribution is then
normalized to correspond to the expected number of
background events in the signal region.
To determine the latter,
we perform a likelihood fit of a power-law function
to the \mgg distribution between 103 and 163\GeV,
excluding the $118<\mgg<133\GeV$ region around the Higgs boson mass.
The result of this fit is shown by the solid (blue) curve
in Fig.~\ref{fig:hgg-mgg}.
The scaled distributions of \sthiggs from the two sidebands are
found to be consistent with each other and are averaged.
This average is taken to be the estimate of the SM background
(other than that from SM Higgs boson production),
with half the difference assigned as a systematic uncertainty.

To account for the background from SM Higgs boson production,
which peaks in the \mgg signal region and
is not accounted for with the above procedure,
we use simulated events.
A systematic uncertainty of 30\% is assigned to this result,
which accounts both for the uncertainty of the SM Higgs boson
cross section~\cite{Heinemeyer:2013tqa}
and for potential misrepresentation of the data
by the simulation in the tails of kinematic variables like \sthiggs.

To illustrate the difference in the distribution of \sthiggs
between signal and background events,
Fig.~\ref{fig:hgg-kine-01} (\cmsLeft) shows the distribution of \sthiggs for a sample
of events selected in the same manner as the nominal sample except,
for improved statistical precision,
with loose CSV requirements for the \cPqb-jet tagging.
The distributions for two signal scenarios,
and for the SM background determined as described above,
are also shown.
It is seen that \sthiggs tends to be larger for signal events
than for background events,
providing discrimination between the two.

The corresponding results for the nominal selection
criteria are shown in Fig.~\ref{fig:hgg-kine-01} (\cmsRight),
with numerical values given in Table~\ref{tab:hbb-sthiggs}.

\subsection{\texorpdfstring{$\Ph\Z$ and $\Ph\PW\to\gamma\gamma$+2 jets}{hZ and hW to gamma gamma plus 2 jets}}
\label{sec-gg-2j}

\begin{figure}[tbhp]
  \centering
   \includegraphics[width=0.48\textwidth]{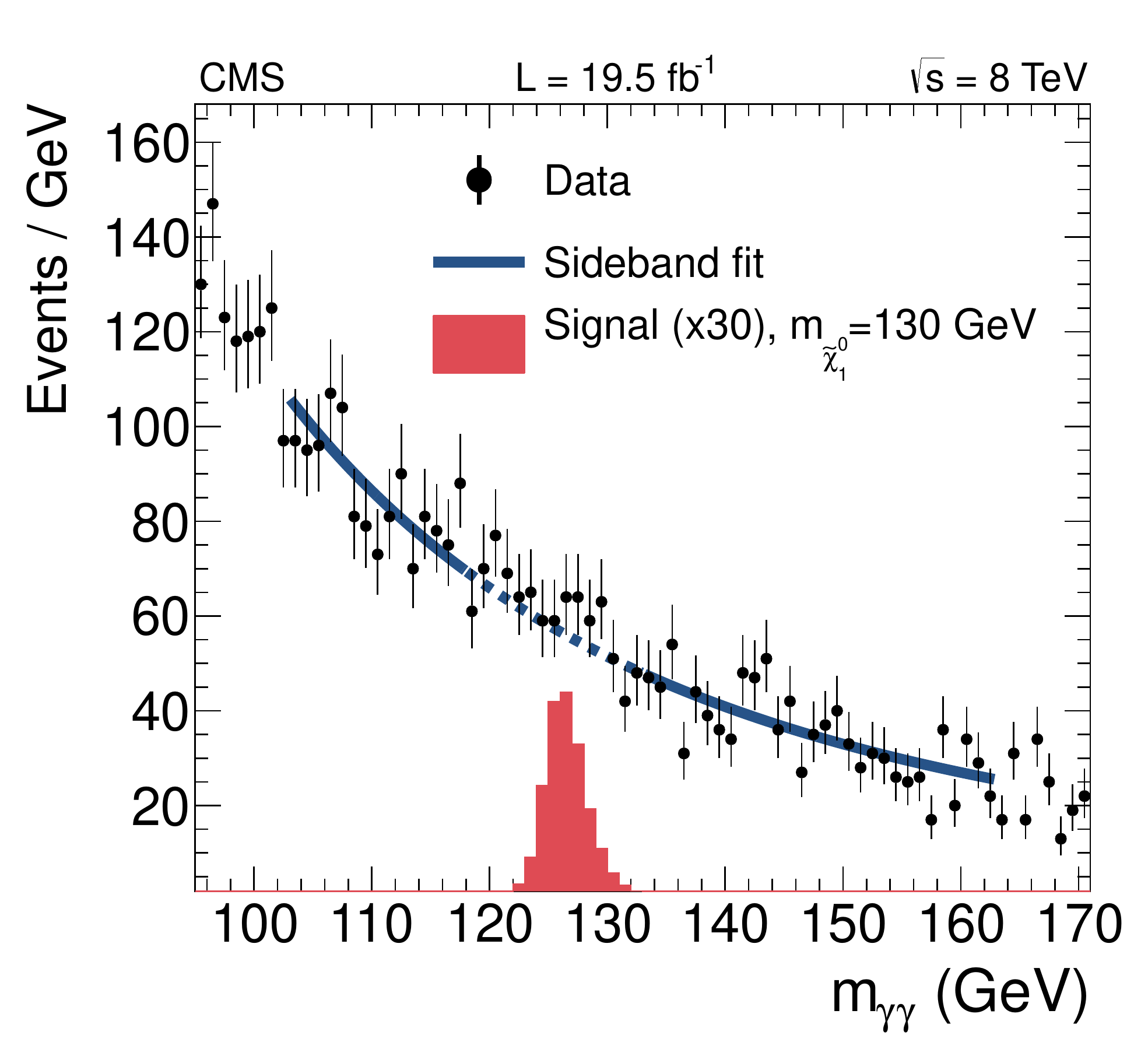}
   \includegraphics[width=0.48\textwidth]{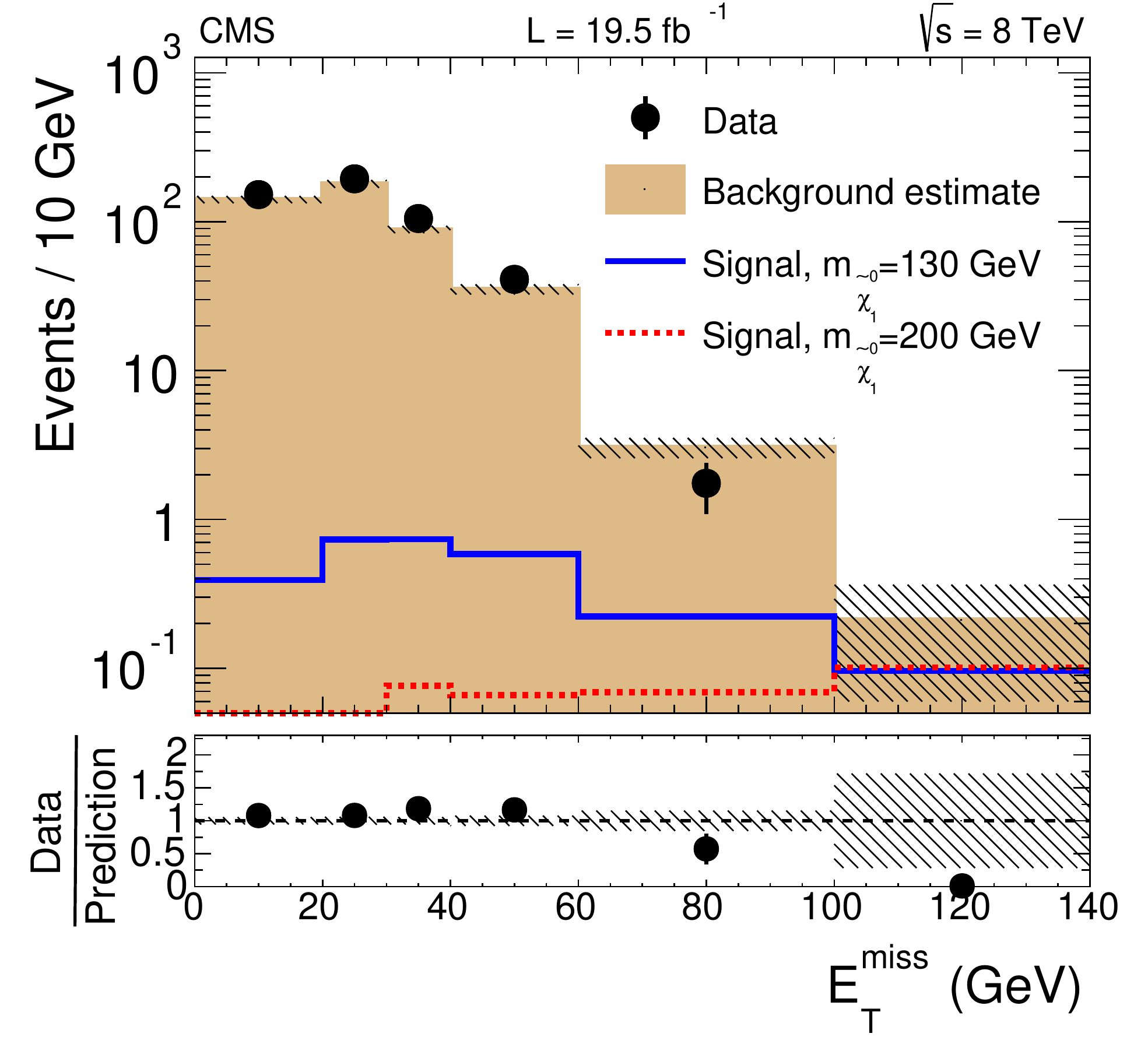}
  \caption{
Results for the
$\Ph\Z$ and $\Ph\PW$ analysis in the
$\gamma\gamma$+2~jets final state
after all selection criteria are applied except for that on
the displayed variable.
(\cmsLLeft) Distribution of diphoton invariant mass \mgg.
The result of a fit to a power-law function using data
in the sideband regions (see text) is indicated by the solid line.
The dotted line shows an interpolation of the fitted function
into the Higgs boson mass region excluded from the fit.
The expected result for $\Ph\Z$ signal events
with a higgsino (\PSGczDo) mass of 130\GeV
and an LSP (gravitino) mass of 1\GeV,
multiplied by a factor of 30 for better visibility,
is also shown.
(\cmsRRight)
Observed numbers of events as a function of \met
in comparison with the SM background estimate.
The hatched bands show the total uncertainty of the background prediction,
with statistical and systematic terms combined.
The expected (unstacked) results for $\Ph\Z$ signal events,
with the indicated values of the higgsino (\PSGczDo) mass
and an LSP (gravitino) mass of 1\GeV,
are also shown.
}
  \label{fig:hgg-kine-02}
\end{figure}

\begin{table*}[tbh]
\centering
\topcaption{
Observed numbers of events and corresponding SM background estimates,
in bins of missing transverse energy~\met,
for the $\Ph\V\to\gamma\gamma$+2 jets analysis,
where \V represents a {\PW} or {\Z} boson.
The uncertainties shown for the SM background estimates
are the combined statistical and systematic terms,
while those shown for signal events are statistical.
The expected yields for $\Ph\Z$ signal events,
with a higgsino mass value of 130\GeV
and an LSP (gravitino) mass of 1\GeV,
are also shown.
}
\label{tab:hz-hw-met}
\begin{scotch}{cccc}
\MET (\GeVns{}) & SM background & Data  & $\Ph\Z$ events, \mhiggsino = 130\GeV \\
\hline
0--20   &   $288 \pm 15$    & 305  &  $0.76\pm0.03$  \\
20--30  &   $183 \pm 10$    & 195  &  $0.71\pm0.03$  \\
30--40  &  $91.1 \pm 4.7$   & 105  &  $0.72\pm0.03$  \\
40--60  &  $72.0 \pm 5.0$   & 82   &  $1.14\pm0.04$  \\
60--100 &  $12.5 \pm 1.9$   & 7    &  $0.87\pm0.03$  \\
$>$100 &  $0.96 \pm 0.61$  & 0    &  $0.37\pm0.02$  \\
\end{scotch}
\end{table*}

For the $\Ph\Z$ and $\Ph\PW$ channels
with $\Ph\to\gamma\gamma$ and
either $\PW\to\,$2~jets or $\Z\to\,$2~jets,
the vector boson candidate is formed from two jets
that yield a dijet mass \mjj consistent with that of
a {\PW} or \Z boson, $70<\mjj<110\GeV$.
Multiple candidates per event are allowed.
The fraction of events with multiple candidates is~16\%.
The average number of candidates per event is~1.2.
Events with isolated electrons and muons are rejected,
using the criteria of Section~\ref{sec-hhggbb}.
To avoid overlap with the sample discussed in Section~\ref{sec-hhggbb},
events are rejected if a loose-tagged $\cPqb$ jet combined
with a medium-tagged $\cPqb$ jet yields an invariant mass
in the range $95<\mbb<155\GeV$.
The distribution of \mgg for the selected events is shown
in Fig.~\ref{fig:hgg-kine-02} (\cmsLeft).

The SM background estimate is obtained
using the procedure described in Section~\ref{sec-hhggbb}
except using the \met variable rather than the \sthiggs variable,
viz.,
from the average of the scaled \met distributions derived from
the two \mgg sidebands,
summed with the prediction from simulated SM Higgs boson events.
The solid (blue) curve in Fig.~\ref{fig:hgg-kine-02} (\cmsLeft)
shows the result of the power-law fit to the \mgg sideband regions.
The scaled \met distributions from the two sidebands are found
to be consistent with each other within their uncertainties.

The measured distribution of \met for the selected events is shown
in Fig.~\ref{fig:hgg-kine-02} (\cmsRight)
in comparison with the SM background estimate
and with the predictions from two signal scenarios.
Numerical values are given in Table~\ref{tab:hz-hw-met}.

\subsection{\texorpdfstring{$\Ph\Ph$, $\Ph\Z$, and $\Ph\PW\to\gamma\gamma$+leptons}{hh, hZ, and hW to gamma gamma + leptons}}
\label{sec-hgglepton}

\begin{figure}[tbh]
  \centering
   \includegraphics[width=0.48\textwidth]{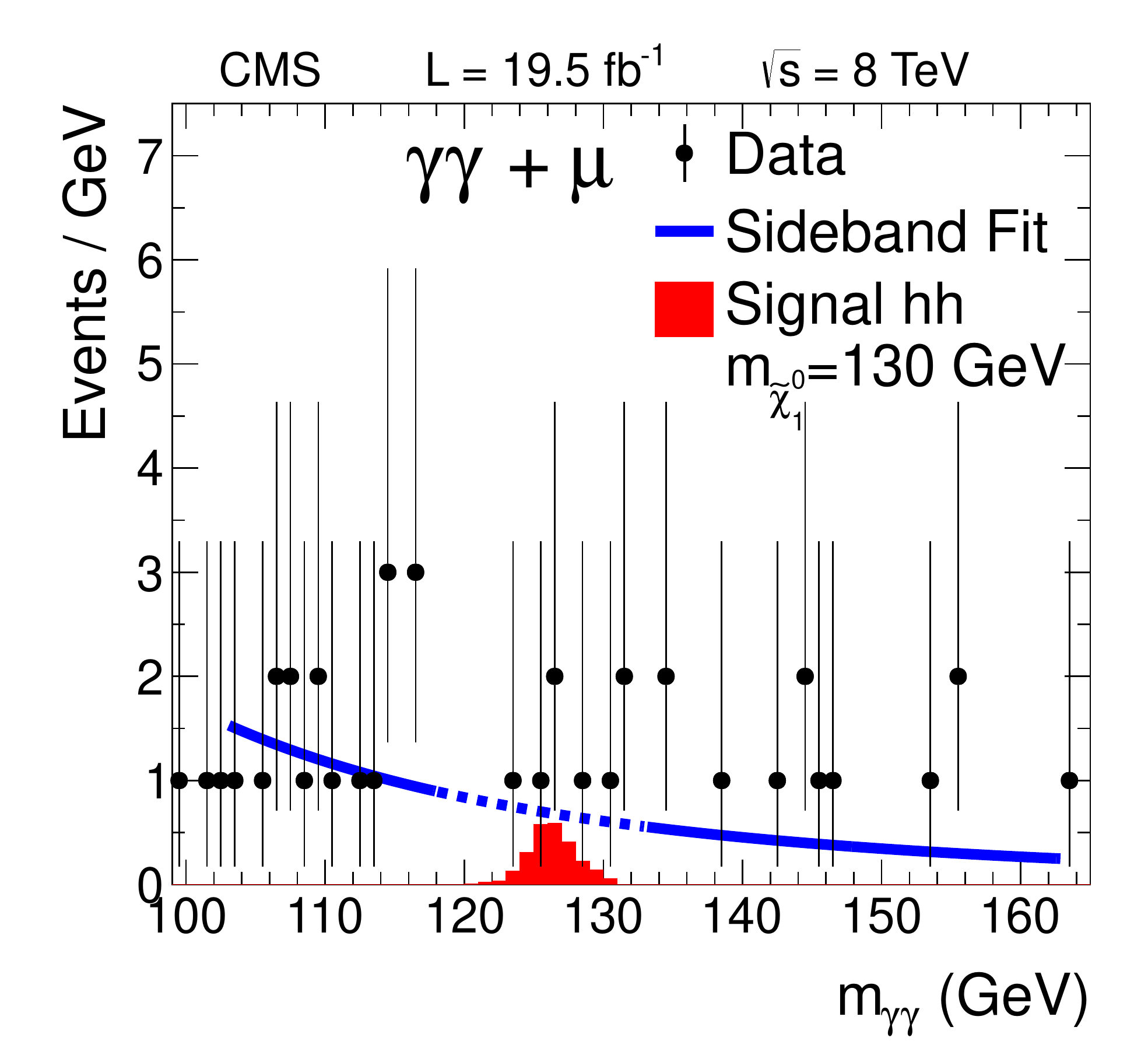}
   \includegraphics[width=0.48\textwidth]{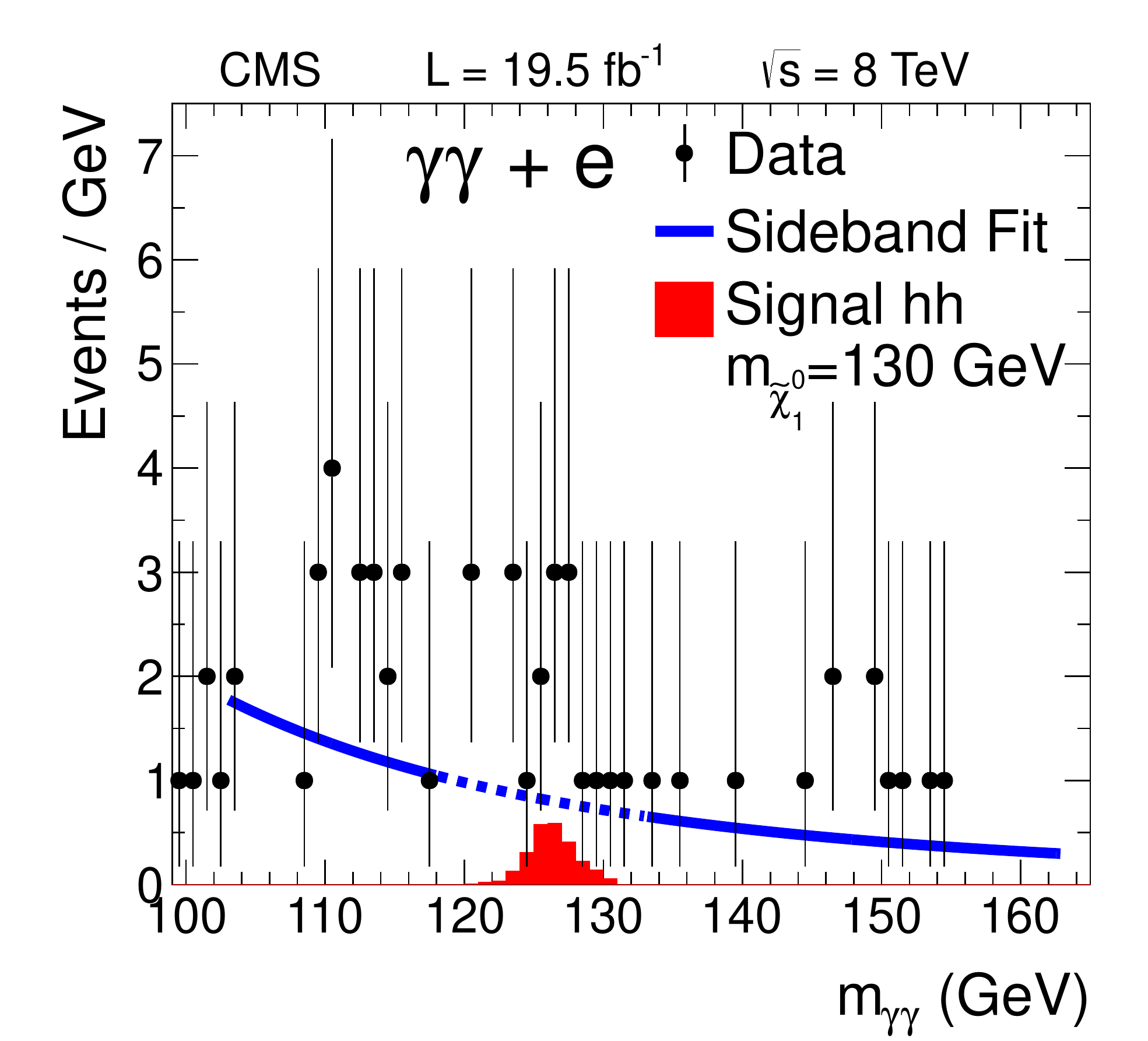}
  \caption{
Distribution of the diphoton invariant mass \mgg
after all selection criteria are applied except for that on~\mgg,
for the
$\Ph\Ph$, $\Ph\Z$, and $\Ph\PW\to\gamma\gamma$+leptons analysis,
for the (\cmsLeft) muon and (\cmsRight) electron samples.
The result of a fit to a power-law function using data
in the sideband regions (see text)
is indicated by the solid line.
The dotted line shows an interpolation of the fitted function
into the Higgs boson mass region excluded from the fit.
The expected results for $\Ph\Ph$ events,
with a higgsino (\PSGczDo) mass value of 130\GeV
and an LSP (gravitino) mass of 1\GeV,
are also shown.
}
  \label{fig:hgg-mgg-2}
\end{figure}

\begin{figure}[tbhp]
\centering
\includegraphics[width=0.49\textwidth]{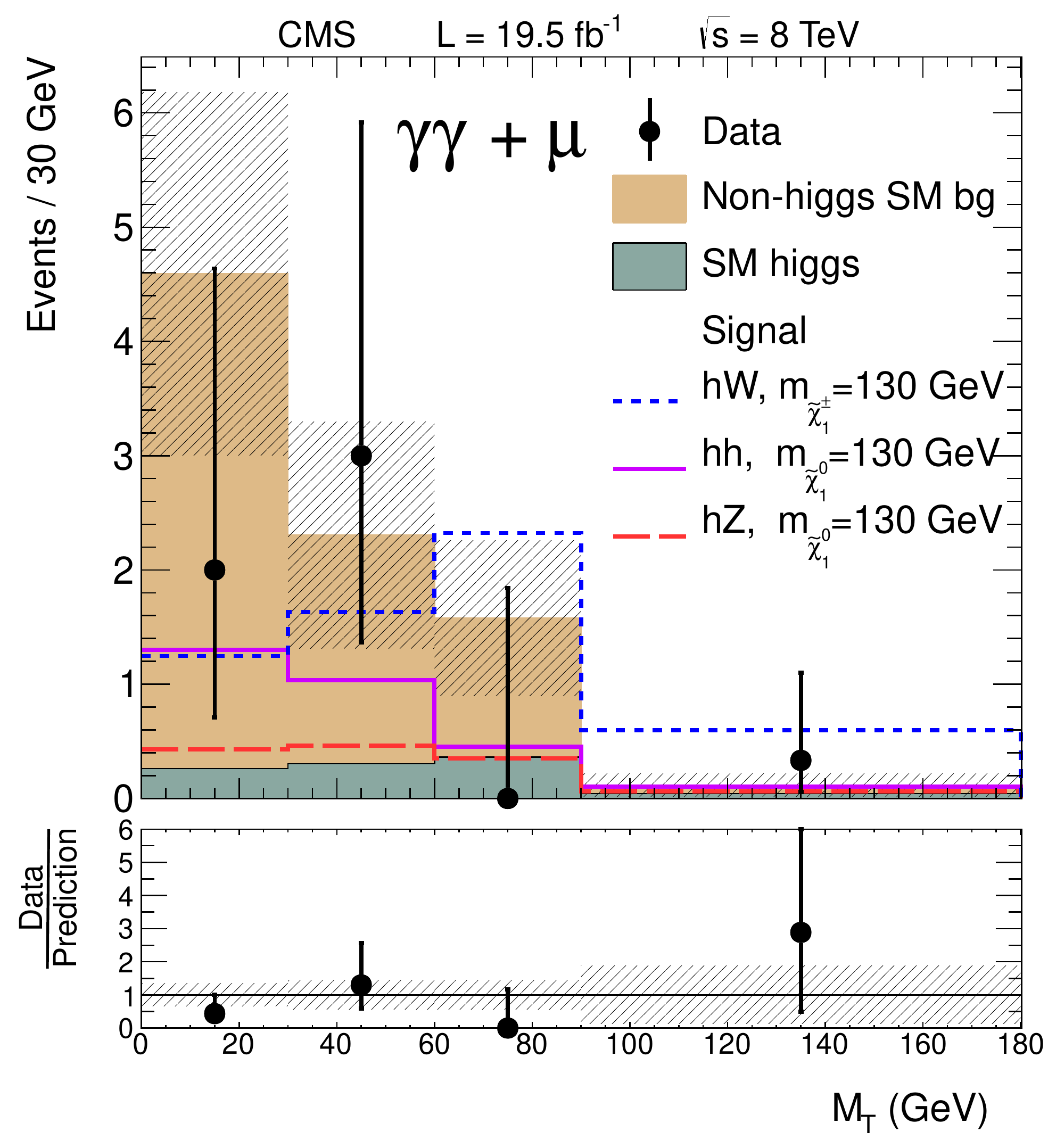}
\includegraphics[width=0.49\textwidth]{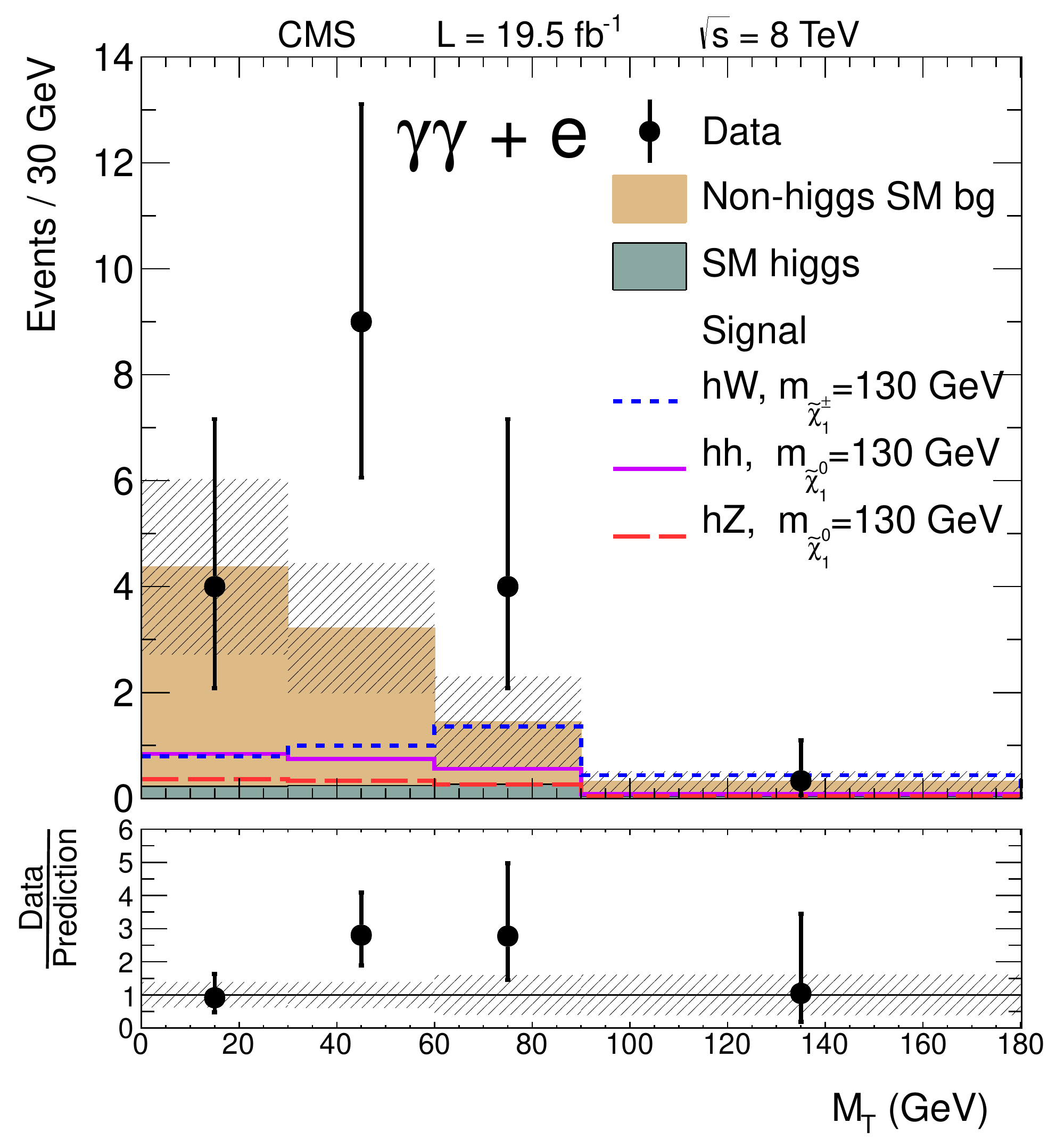}
\caption{
Observed numbers of events as a function of transverse mass \mt for the
$\Ph\Ph$, $\Ph\Z$, and $\Ph\PW\to\gamma\gamma$+leptons analysis,
in comparison with the (stacked) SM background estimates,
for the (\cmsLeft) muon and (\cmsRight) electron samples.
The hatched bands show the total uncertainty of the background prediction,
with statistical and systematic terms combined.
The (unstacked) results for various signal scenarios are also shown.
For the $\Ph\Ph$ and $\Ph\Z$ scenarios,
the higgsino (\PSGczDo) mass is 130\GeV and
the LSP (gravitino) mass is 1\GeV.
For the $\Ph\PW$ scenario,
${m}_{\PSGczDt}={m}_{\PSGcpm_1}=130\GeV$
and $\mhiggsino=1\GeV$
[see Fig.~\ref{fig:event-diagrams} (right)].
}
\label{fig:higgs-mt}
\end{figure}

We next consider $\Ph\Ph$, $\Ph\Z$, and $\Ph\PW$
combinations in which a Higgs boson decays into a pair of photons,
while the other boson
($\Ph$, {\Z}, or {\PW})
decays to a final state with at least one lepton
(electron or muon).
For the $\Ph\Ph$ channel
this signature encompasses events in which the second Higgs boson
decays to
$\Ph\to\Z\Z$, $\PW\PW$, or $\Pgt\Pgt$,
followed by the leptonic decay of at least one
{\Z}, {\PW}, or $\tau$ particle,
including the case where one \Z boson
decays to charged leptons and the other to neutrinos.

The lepton identification criteria are the same
as those presented in Section~\ref{sec-hhggbb}
with the additional requirement that
the $\Delta R$ separation between an electron or muon
candidate and each of the two photon candidates exceed~0.3.
To reduce the background in which an electron is
misidentified as a photon,
events are eliminated if the invariant mass
formed from an electron candidate and one of the
two $\Ph\to\gamma\gamma$ photon candidates lies
in the \Z boson mass region $86<\meg<96\GeV$.
Electron candidates are rejected if they appear within $1.44<\abs{\eta}<1.57$,
which represents a transition region between the barrel and endcap
electromagnetic calorimeters~\cite{Chatrchyan:2008aa},
where the reconstruction efficiency is difficult to model.
To prevent overlap with the other searches,
events are allowed to contain at most one medium-tagged $\cPqb$ jet.

We select a sample with at least one muon
and an orthogonal sample with no muons
but at least one electron.
We refer to these samples as the muon and electron samples,
respectively.
About 93\% of the events in each sample contain only a single electron or muon,
and there are no events for which the sum of electron
and muon candidates exceeds two
(only two events have one electron and one muon).
The \mgg distributions for the two samples are
shown in Fig.~\ref{fig:hgg-mgg-2}.

\begin{table*}[tbh]
\topcaption{
Observed numbers of events and corresponding SM background estimates,
in bins of transverse mass~\mt,
for the
$\Ph\Ph$, $\Ph\Z$, and $\Ph\PW\to\gamma\gamma$+leptons analysis.
The uncertainties shown for the SM background estimates
are the combined statistical and systematic terms,
while those shown for signal events are statistical.
The column labeled ``$\Ph\PW$ events'' shows the expected
number of events from the chargino-neutralino pair-production process of
Fig.~\ref{fig:event-diagrams} (right),
taking $\mchitwoz=\mchionep=130\GeV$ and $\mchionez=1\GeV$.
}
\label{tab:ggMTcounts}
\centering
\begin{scotch}{ccccccc}
         & \multicolumn{3}{c}{Muon sample} & \multicolumn{3}{c}{Electron sample} \\
\mt (\GeVns{}) & SM background & Data & $\Ph\PW$ events
  & SM background & Data & $\Ph\PW$ events \\
\hline
0--30   & $4.6\pm 1.6$    & 2 & $1.2\pm 0.1$   & $4.4\pm 1.7$   & 4 & $0.80\pm0.06$ \\
30--60  & $2.31\pm 0.99$  & 3 & $1.5\pm 0.1$   & $3.2\pm 1.2$   & 9 & $1.0\pm0.1$ \\
60--90  & $1.59\pm 0.68$  & 0 & $2.1\pm 0.1$   & $1.44\pm 0.85$ & 4 & $1.4\pm0.1$ \\
$>$90    & $0.35\pm 0.30$  & 1 & $1.6\pm 0.1$   & $0.96\pm 0.58$ & 1 & $1.3\pm0.1$ \\
\end{scotch}
\end{table*}

\begin{figure}[tbhp]
\centering
\includegraphics[width=0.49\textwidth]{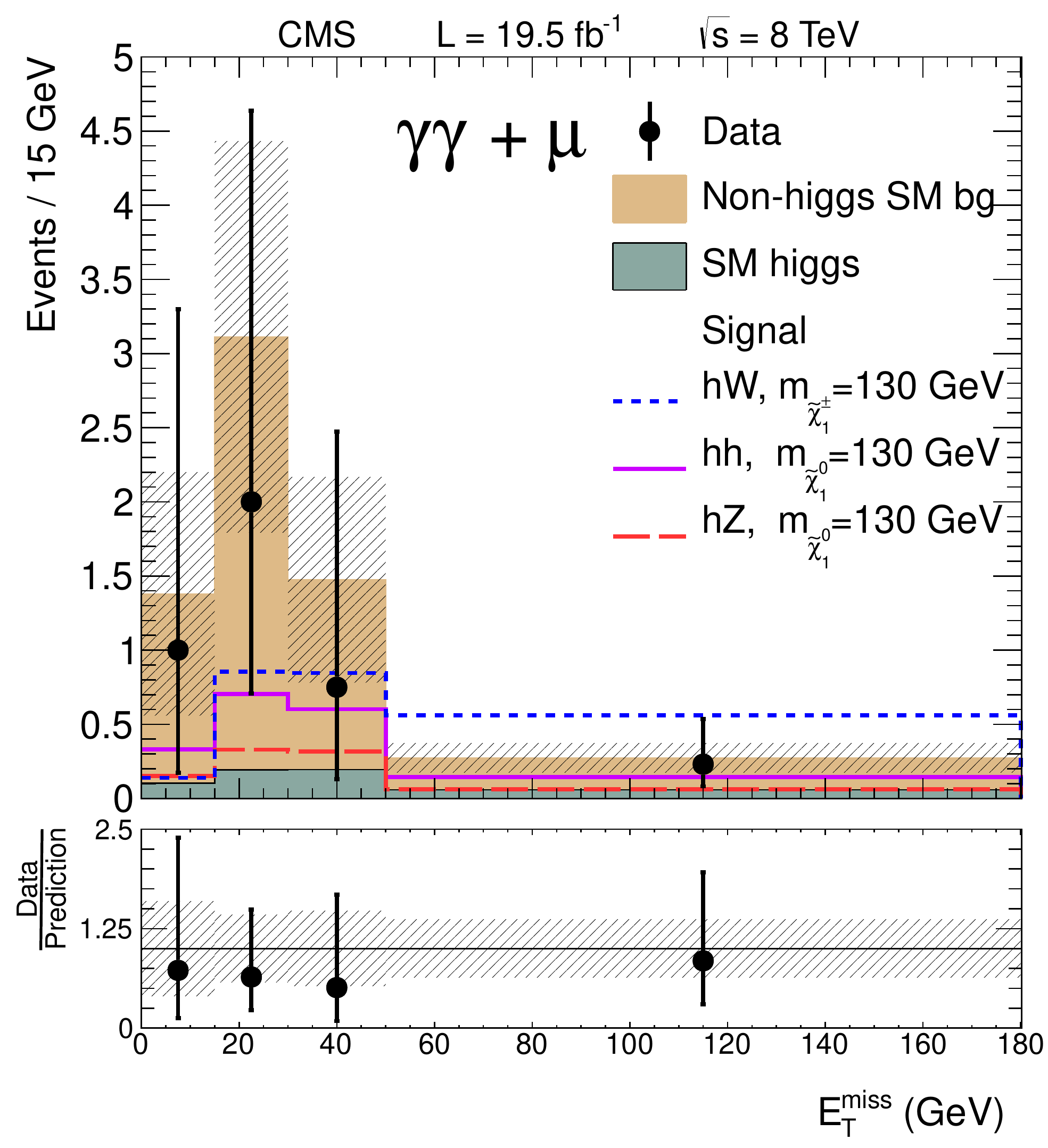}
\includegraphics[width=0.49\textwidth]{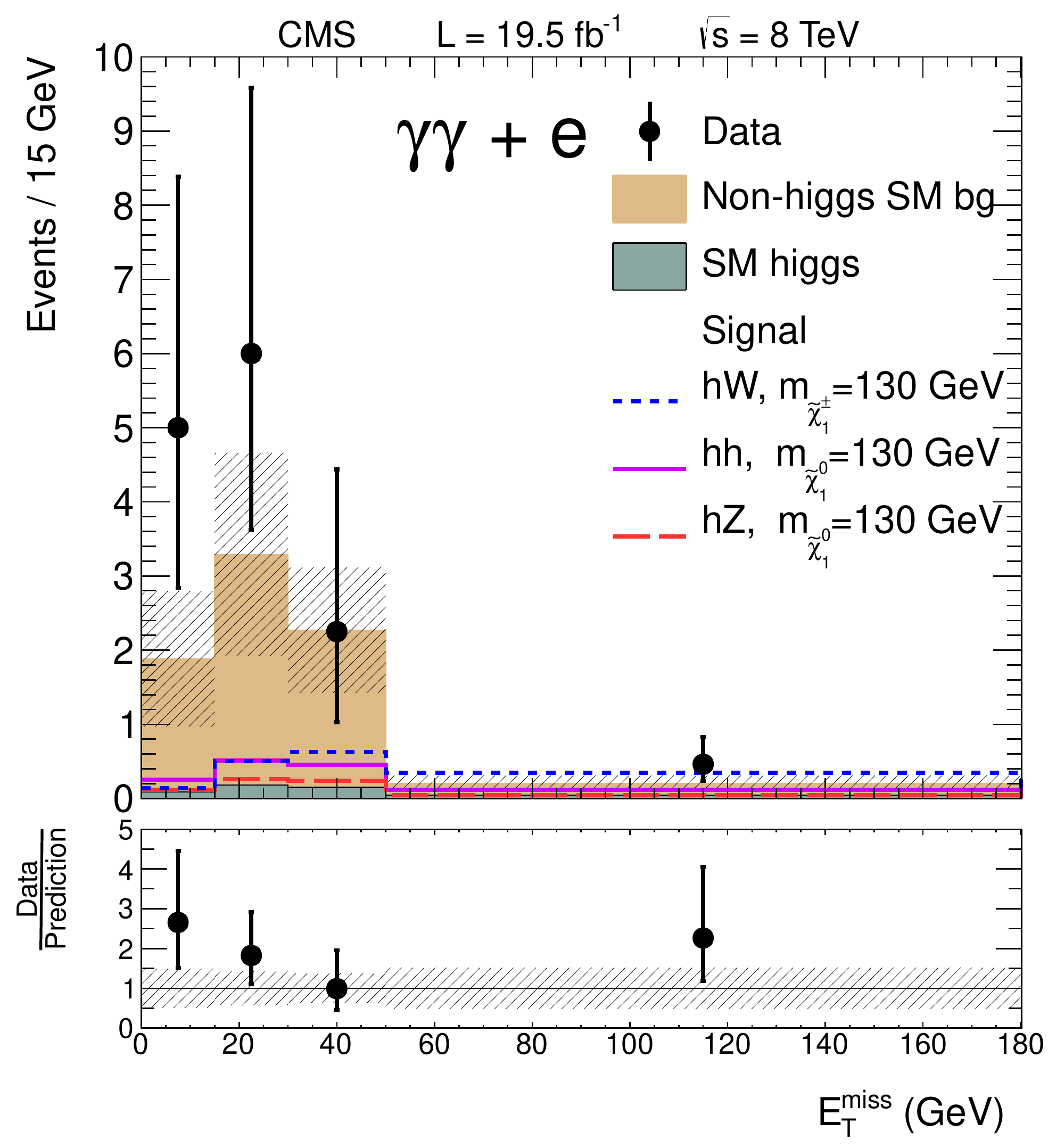}
\caption{
Observed numbers of events as a function of \met for the
$\Ph\Ph$, $\Ph\Z$, and $\Ph\PW\to\gamma\gamma$+leptons analysis
in comparison with the (stacked) SM background estimates,
for the (\cmsLeft) muon and (\cmsRight) electron samples.
The hatched bands show the total uncertainty of the background prediction,
with statistical and systematic terms combined.
The (unstacked) results for various signal scenarios are also shown.
For the $\Ph\Ph$ and $\Ph\Z$ scenarios,
the higgsino (\PSGczDo) mass is 130\GeV and
the LSP (gravitino) mass is 1\GeV.
For the $\Ph\PW$ scenario,
${m}_{\PSGczDt}={m}_{\PSGcpm_1}=130\GeV$
and $\mhiggsino=1\GeV$.
}
\label{fig:higgs-ggl-met}
\end{figure}

The SM background is evaluated in the same manner as described
in Section~\ref{sec-hhggbb}
except using the transverse mass variable
$\mt\equiv\sqrt{\smash[b]{2\met\pt^\ell[1-\cos(\Delta\phi_{\ell,\met})]}}$
in place of the \sthiggs variable,
where $\pt^\ell$ is the transverse momentum of the highest \pt lepton,
with $\Delta\phi_{\ell,\met}$ the difference in
azimuthal angle between the $\pt^\ell$ and \met vectors.
For SM background events with {\PW} bosons,
the \mt distribution exhibits an endpoint near the {\PW} boson mass.
In contrast,
for signal events, the value of \mt can be much larger.
As an alternative,
we tested use of the \met distribution to evaluate the SM background
and found the \mt distribution to be slightly more sensitive.

The SM background estimate is thus
given by the average of the scaled \mt distributions
from the two \mgg sidebands,
summed with the contribution from simulated SM Higgs boson events.
The solid (blue) curves in Fig.~\ref{fig:hgg-mgg-2}
show the results of the power-law fits to the \mgg sideband regions.
For the electron channel
[Fig.~\ref{fig:hgg-mgg-2} (\cmsRight)],
a cluster of events is visible at $\mgg\approx112\GeV$.
We verified that the prediction for the number of background events
is stable within about one standard deviation of the statistical
uncertainty for alternative definitions of the sideband regions,
such as $110<\mgg<118\GeV$ for the lower sideband
rather than $103<\mgg<118\GeV$.

The \mt distributions of the selected events are
presented in Fig.~\ref{fig:higgs-mt}.
Numerical values are given in Table~\ref{tab:ggMTcounts}.
The background estimates and predictions from several
signal scenarios are also shown.
Results for the alternative method to evaluate the SM background,
based on the \met distribution rather than the \mt distribution,
are shown in Fig.~\ref{fig:higgs-ggl-met}.
For the muon channel,
the data exhibit a small deficit with respect to the SM background estimate.
For the electron channel,
there is an excess of 2.1~standard deviations.
Note that this result does not account for the
so-called look-elsewhere effect~\cite{Gross:2010qma}.
The excess of data events in the electron channel above the SM background
prediction clusters at low values $\met\lesssim30\GeV$,
as seen in Fig.~\ref{fig:higgs-ggl-met} (\cmsRight).
Summing the electron and muon channels,
we obtain 24 observed events compared to $18.9\pm3.1$ expected SM events,
corresponding to an excess of 1.3~standard deviations.
To investigate the excess in the electron channel,
we varied the functional form used to fit the sideband data
(an exponential function was used rather than a power-law function),
modified the definitions of the sideband and signal regions,
as mentioned above,
and altered the photon identification criteria.
All variations yielded consistent results,
with an excess in the electron channel of about 2 standard deviations.
An ensemble of MC pseudo-experiments was used to verify that
the background evaluation procedure is unbiased.
Since the excess in the electron channel is neither large nor signal-like,
and since there is not a corresponding excess in the muon channel,
we consider the excess seen in
Fig.~\ref{fig:higgs-mt} (\cmsRight)
to be consistent with a statistical fluctuation.
Note that if we apply looser or tighter photon selection
criteria relative to the nominal criteria,
the significance of the excess decreases in a way that is
consistent with its explanation as a statistical fluctuation.

\section{\texorpdfstring{Search in the $\Ph\Z$ channel with
$\Ph\to\bbbar$ and $\Z\to\ell^+\ell^-$}{Search in the hZ channel with h to b bbar and Z to lepton pairs}}
\label{sec-z-to-ll}

We now describe the search in the SUSY $\Ph\Z$ channel
with $\Ph\to\bbbar$ and $\Z\to\ell^+\ell^-$
($\ell=\Pe$, $\Pgm$).
Electron and muon candidates are required to satisfy
$\pt>20\GeV$, $\abs{\eta}<2.4$, and $\riso<0.15$.
For the \riso variable,
a cone size $\isocone=0.3$ is used for both electrons and muons,
rather than $\isocone=0.4$ for muons as in Sections~\ref{sec-hh-4b}
and~\ref{sec-higgs-to-gg}.
Electron candidates that appear within the transition region $1.44<\abs{\eta}<1.57$
between the barrel and endcap electromagnetic calorimeters are rejected.
Jets must satisfy $\pt>30\GeV$ and $\abs{\eta}<2.5$
and be separated by more than
$\Delta R=0.4$ from an electron or muon candidate.
To be tagged as a $\cPqb$ jet,
the jet must satisfy the CSV-medium criteria.

Events are required to contain:
\begin{itemize}
\item exactly one $\Pep\Pem$ or $\Pgmp\Pgmm$ pair
  with a dilepton invariant mass \mll in the \Z boson
  mass region $81<\mll<101\GeV$;
\item no third electron or muon candidate,
  selected using the above criteria except with $\pt>10\GeV$;
\item no \tauh candidate with $\pt>20\GeV$;
\item  at least two tagged $\cPqb$ jets,
  where the two most b-like jets yield a dijet mass
  in the Higgs boson mass region $100<\mbb<150\GeV$.
\end{itemize}
The reason to reject events with a third lepton is to avoid
overlap with the three-or-more-lepton
sample discussed in Section~\ref{sec-multilepton}.

\begin{figure}[tbh]
  \centering
   \includegraphics[width=0.49\textwidth]{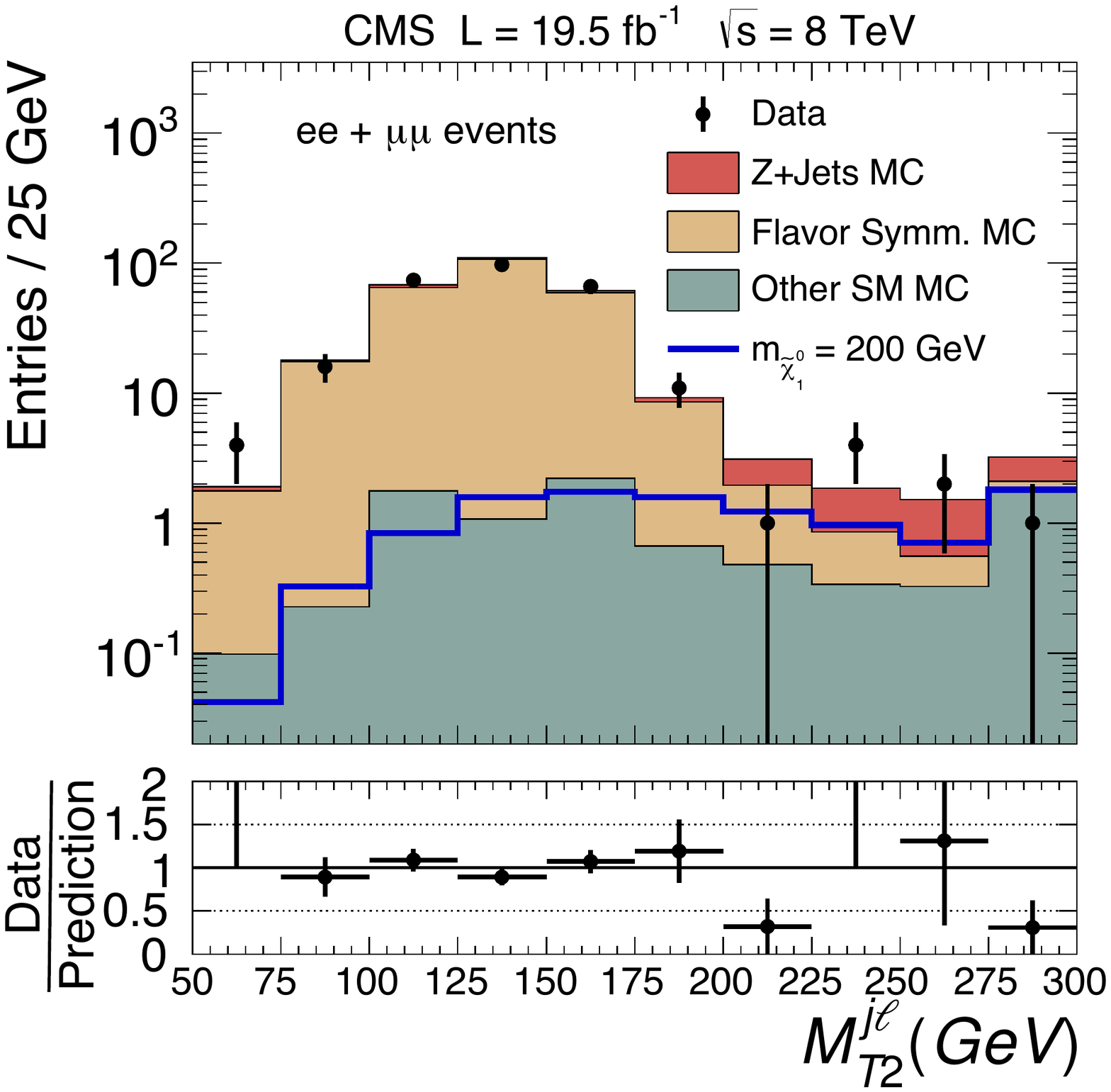}
  \caption{
Distribution of \MTtwoj
for the $\Ph(\to\bbbar)\Z(\to\ell^+\ell^-)$ analysis
after all signal-region requirements are applied except for that on~\MTtwoj,
in comparison with (stacked) SM background estimates taken from simulation.
For this result, $\met>60\GeV$.
The (unstacked) signal results for a higgsino (\PSGczDo) mass of 200\GeV
and an LSP (gravitino) mass of 1\GeV are also shown.
}
  \label{fig:zll-MTtwoj}
\end{figure}

Events with a \ttbar pair represent a large
potential source of background,
especially if both top quarks decay to a state with a lepton.
To reduce this background,
we use the \MTtwoj variable~\cite{Lester:1999tx,Barr:2003rg},
which corresponds to the minimum mass
of a pair-produced parent particle compatible with the
observed four-momenta in the event,
where each parent is assumed to decay to a $\cPqb$ jet,
a charged lepton~$\ell$,
and an undetected particle,
and where the vector sum of the \pt values of the two undetected
particles is assumed to equal the observed result for \met.
For \ttbar events with perfect event reconstruction,
\MTtwoj has an upper bound at the top-quark mass.
For signal events,
\MTtwoj can be much larger.
To account for imperfect reconstruction and
finite detector resolution,
we require ${\MTtwoj}>200\GeV$.
The distribution of \MTtwoj is shown in Fig.~\ref{fig:zll-MTtwoj}.

We further require $\met>60$, 80, or 100\GeV,
where the lower bound on \met depends on which choice
yields the largest expected signal sensitivity
for a given value of the higgsino mass.

The remaining background mostly consists of events from SM {\Z}+jets,
\ttbar, $\PWp\PWm$, \TT,
and $\cPqt\PW$ single-top-quark production.
These backgrounds are evaluated using data,
as described below.
Other remaining SM background processes are
combined into an ``other'' category,
which is evaluated using simulation
and assigned an uncertainty of 50\%.
The ``other'' category includes background from
$\Z\PW$ and $\Z\Z$ boson pair production,
\ttbar processes with an associated {\PW} or \Z boson,
and processes with three vector bosons.

\begin{figure}[tbh]
  \centering
   \includegraphics[width=0.49\textwidth]{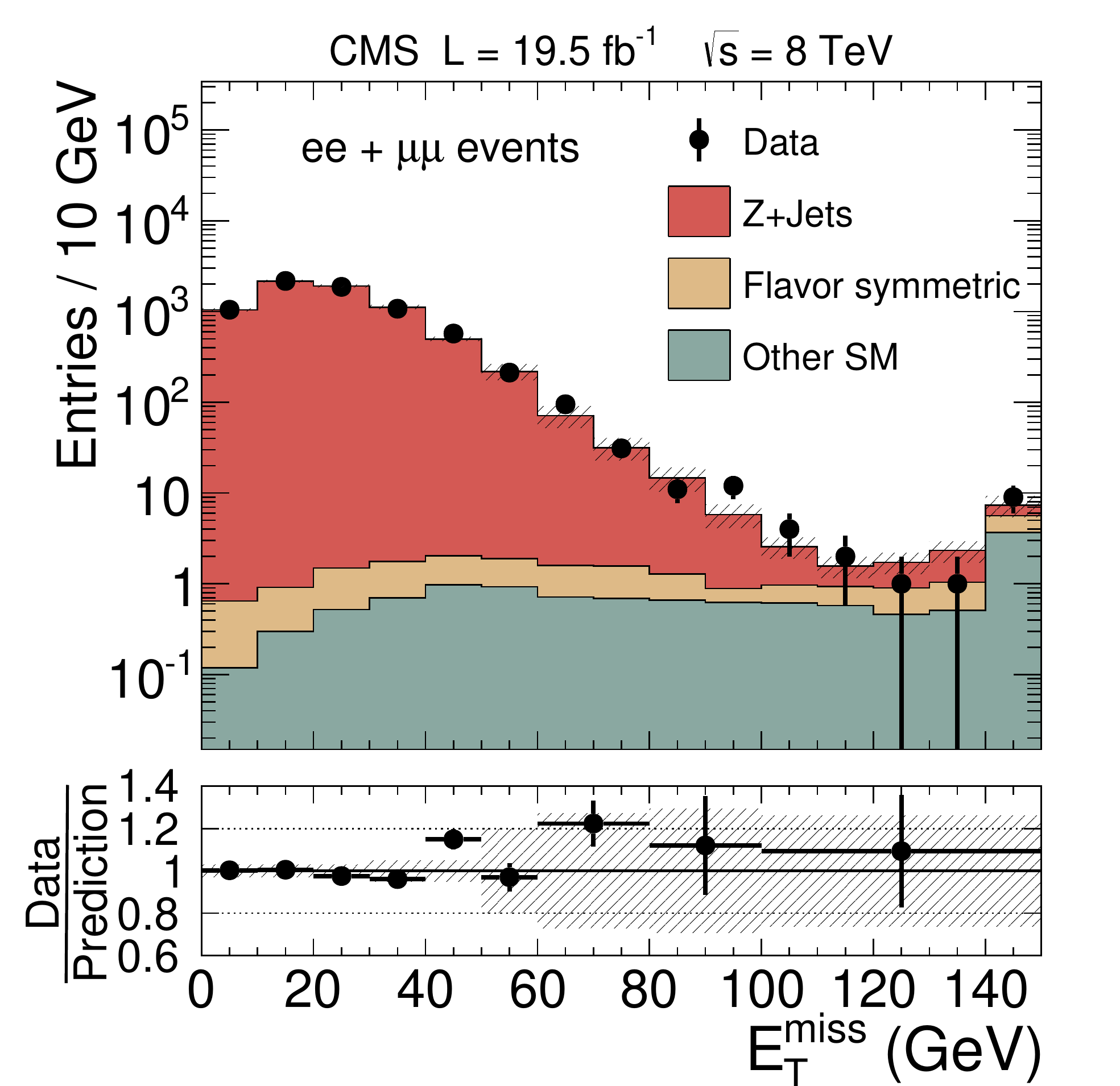}
   \includegraphics[width=0.49\textwidth]{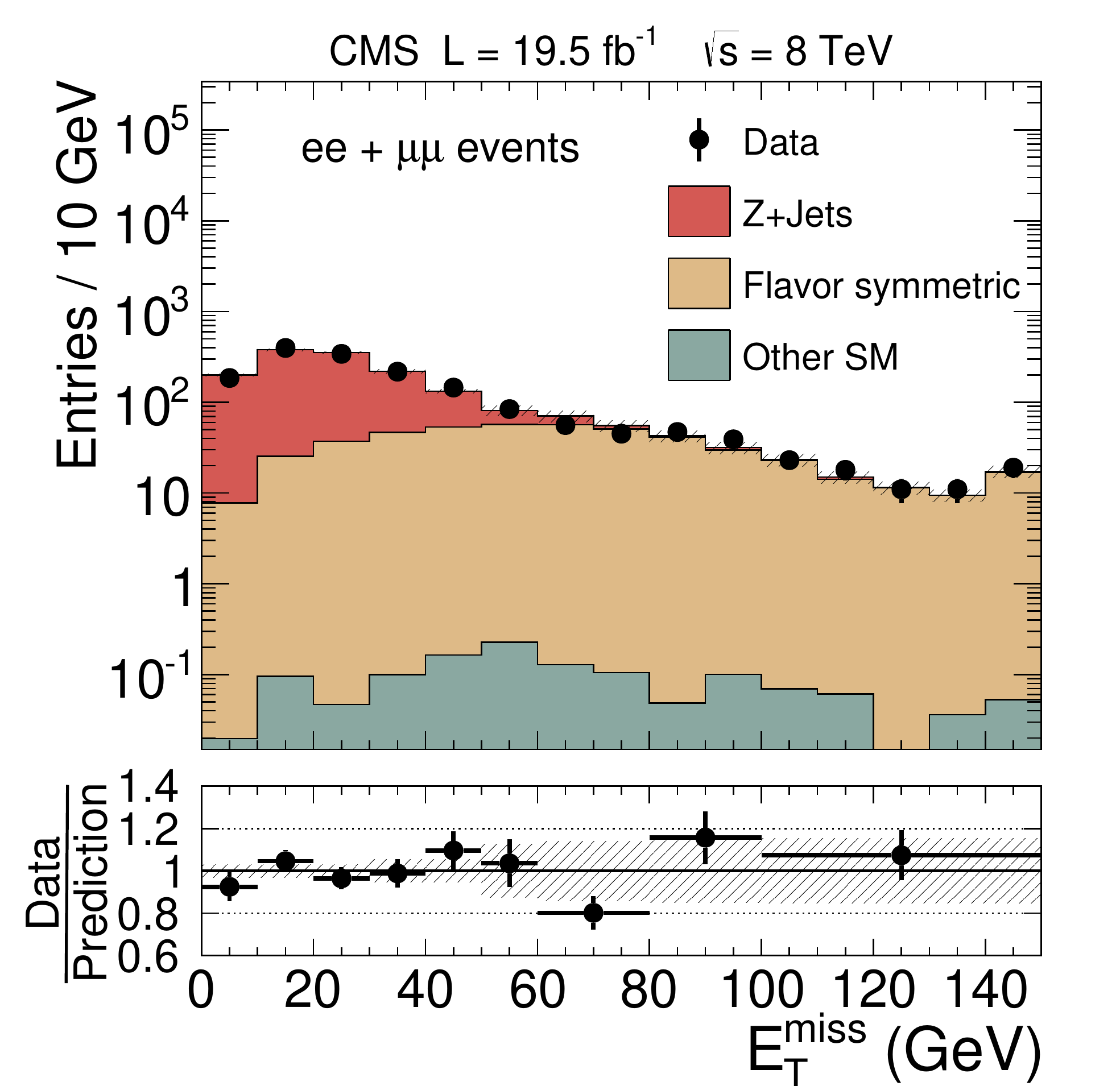}
  \caption{
Distribution of \met in comparison
with the (stacked) SM background estimates
for the $\Ph(\to\bbbar)\Z(\to\ell^+\ell^-)$ analysis,
for data control samples enriched in
(\cmsLeft)~SM {\Z}+jets events,
and (\cmsRight) \ttbar events.
The hatched bands in the ratio plots (lower panels)
indicate the uncertainty of the total background prediction,
with statistical and systematic terms combined.
}
  \label{fig:zll-hbb-control}
\end{figure}

For the SM {\Z}+jets background,
significant values of \met arise primarily because of
the mismeasurement of jet~\pt.
Another source is the
semileptonic decay of charm and bottom quarks.
As in Ref.~\cite{Chatrchyan:2012qka},
we evaluate this background using a sample of $\gamma$+jets events,
which is selected using similar criteria to those used
for the nominal selection,
including the same $\cPqb$-jet tagging requirements and restriction on~\mbb.
We account for kinematic differences
between the $\gamma$+jets and signal samples by reweighting
the \HT and boson-\pt spectra of the former sample
to match those of the latter,
where \HT is the scalar sum of jet \pt values using jets
with $\pt>15\GeV$.
The resulting $\gamma$+jets \met distributions are then normalized
to unit area to define templates.
Two different templates are formed:
one from $\gamma$+jets events with exactly two jets,
and one from the events with three or more jets.
The SM {\Z}+jets background estimate
is given by the sum of the two templates,
each weighted by the number of events in the signal sample
with the respective jet multiplicity.
To account for the small level of background
expected in the signal sample from SM processes other
than SM {\Z}+jets production,
which is mostly due to \ttbar production,
the prediction is normalized to the data yield in the
$0<\met<50\GeV$ region,
where the contribution of SM {\Z}+jets events dominates.
The impact of signal events on the estimate
of the SM {\Z}+jets background is found to be negligible.
The corresponding systematic uncertainty
is evaluated by varying the criteria
used to select $\gamma$+jets events,
by assessing the impact of \ttbar events,
and by determining the difference between the predicted and genuine
SM {\Z}+jets event yields when the simulation is used to describe
the $\gamma$+jets and signal samples.
The three sources of systematic uncertainty are added in
quadrature to define the total systematic uncertainty.

For the \ttbar, $\PWp\PWm$, \TT, and $\cPqt\PW$ background,
the rate of decay to events
with exactly one electron and exactly one muon
is the same as the rate of decay to events with either
exactly one $\Pep\Pem$ or one $\Pgmp\Pgmm$ pair,
once the difference between the electron and muon
reconstruction efficiencies is taken into account.
We therefore refer to this category of events as the
``flavor-symmetric'' background.
The flavor-symmetric background is thus
evaluated by measuring the number of
events in a sample of $\Pe\Pgm$ events,
which is selected in the manner described above for the
$\Pep\Pem$ and $\Pgmp\Pgmm$ samples except
without the requirement on the dilepton mass:
instead of applying an invariant mass restriction $81<\mem<101\GeV$
in analogy with the mass restriction imposed on~\mll,
we apply a factor,
derived from simulation,
that gives the probability
for \mem to fall into this interval,
with a systematic uncertainty defined by the difference between
this factor in data and simulation.
This procedure yields improved statistical precision
compared to the result based on
an \mem requirement~\cite{Chatrchyan:2012qka}.

The background evaluation procedures are validated using
data control samples enriched in the principal background components.
As an example,
Fig.~\ref{fig:zll-hbb-control} (\cmsLeft)
shows the \met distribution
for a control sample
selected in the same
manner as the standard sample
except with the requirement that there be no
tagged $\cPqb$ jet:
this yields a sample dominated by SM {\Z}+jets events.
Figure~\ref{fig:zll-hbb-control} (\cmsRight)
shows the results for a sample selected
with the nominal requirements except
with the \MTtwoj requirement inverted:
this yields a sample dominated by \ttbar events.
For both these control samples,
the SM background estimate is seen to accurately represent the data.

\begin{figure}[tbhp]
  \centering
   \includegraphics[width=0.48\textwidth]{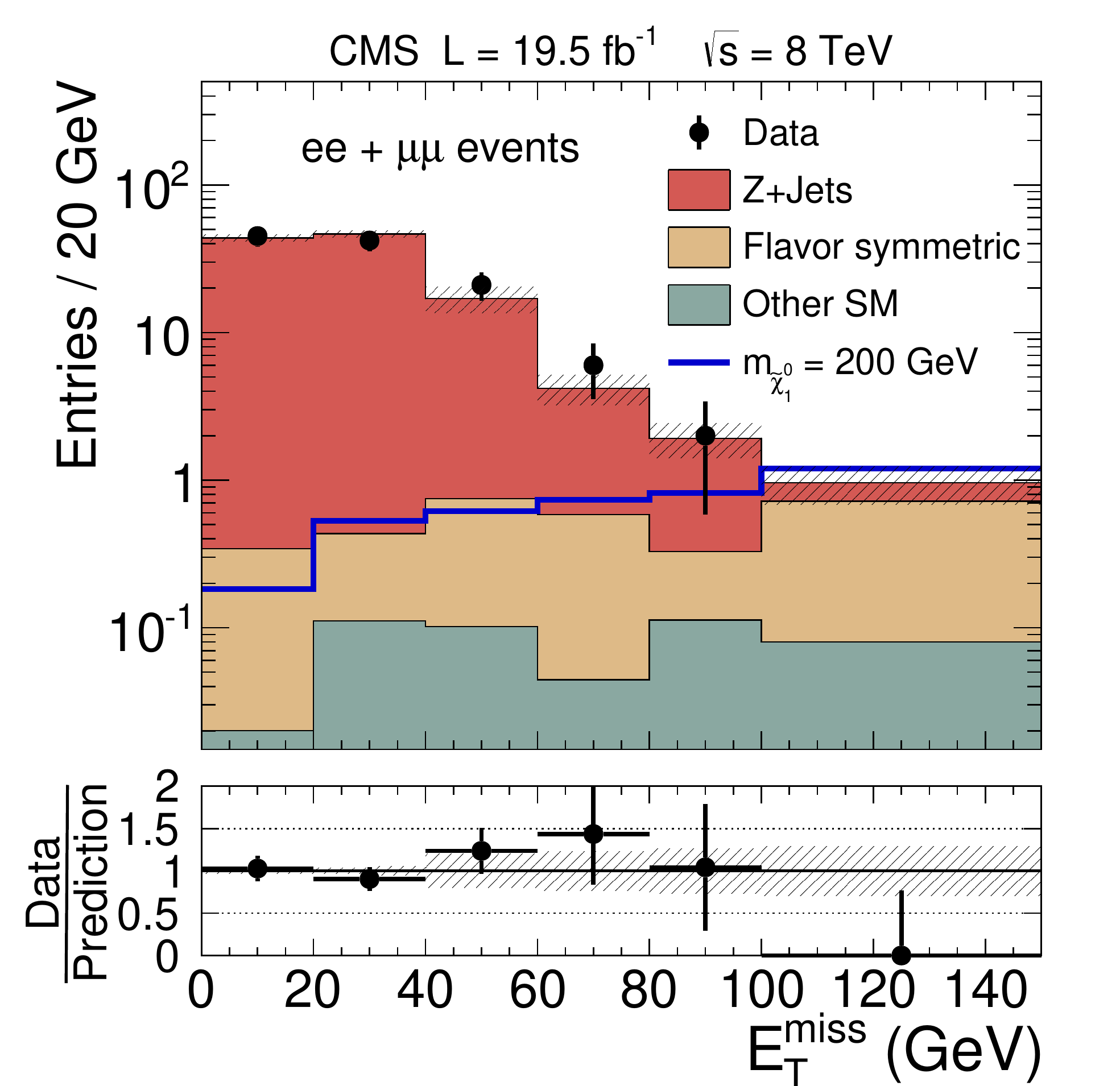}
  \caption{
Observed numbers of events as a function of \met
for the $\Ph(\to\bbbar)\Z(\to\ell^+\ell^-)$ analysis,
in comparison with the (stacked) SM background estimates.
The (unstacked) results for a higgsino (\PSGczDo) mass of 200\GeV
and an LSP (gravitino) mass of 1\GeV
are also shown.
The hatched band in the ratio plot (lower panel)
indicates the uncertainty of the total background prediction,
with statistical and systematic terms combined.
}
  \label{fig:zll-hbb-results}
\end{figure}

The distribution of \met for the selected events
is presented in Fig.~\ref{fig:zll-hbb-results}
in comparison with the corresponding background prediction
and with the prediction from a signal scenario.
Numerical values are given in Table~\ref{tab:zll-hbb-results}.

\begin{table*}[tbhp]
\centering
\topcaption{\label{tab:zll-hbb-results}
Observed numbers of events and corresponding SM background estimates,
in bins of missing transverse energy~\met,
for the $\Ph(\to\bbbar)\Z(\to\ell^+\ell^-)$ analysis.
The uncertainties shown for the SM background estimates are the
combined statistical and systematic terms,
while those shown for signal events are statistical.
For bins with $\met>60\GeV$,
signal event yields are given for four values of the higgsino (\PSGczDo) mass,
with an LSP (gravitino) mass of 1\GeV.
}
\begin{scotch}{lccc}
                              & $\met<25\GeV$  & $25<\met<50\GeV$ & $50<\met<60\GeV$  \\
\hline
  {\Z}+jets background        &  $56.7\pm 1.9$ &    $43.3\pm 2.3$ &    $5.7\pm 1.2$   \\
  Flavor symmetric background &  $0.4\pm 0.3$  &    $0.4\pm 0.3$  &    $0.4\pm 0.3$   \\
  Other SM background         &        $<$0.1  &    $0.1\pm 0.1$  &    $0.1\pm 0.1$   \\
\hline
  Total SM background         &  $57.2\pm 1.9$ &    $43.8\pm 2.3$ &    $6.2\pm 1.2$   \\
  Data                        &   54           &     47           &     7             \\
\hline
\hline
                              & $\met>60\GeV$  & $\met>80\GeV$    & $\met>100\GeV$    \\
\hline
  {\Z}+jets background        &  $5.7\pm 1.8$  &    $2.2\pm 0.9$  &    $0.6\pm 0.3$   \\
  Flavor symmetric background &  $2.4\pm 0.9$  &    $1.8\pm 0.7$  &    $1.6\pm 0.6$   \\
  Other SM background         &  $0.3\pm 0.2$  &    $0.3\pm 0.2$  &    $0.2\pm 0.1$   \\
\hline
  Total SM background         &  $8.5\pm 2.0$  &    $4.3\pm 1.2$  &    $2.4\pm 0.7$   \\
  Data             &   8            &     2            &     0             \\
\hline
  $\Ph\Z$ events    &              &               &                \\
  \mhiggsino = 130\GeV  & $5.4\pm 0.1$ &  $3.1\pm 0.1$ &  $1.7\pm 0.1$  \\
  \mhiggsino = 150\GeV  & $5.3\pm 0.1$ &  $3.3\pm 0.1$ &  $2.0\pm 0.1$  \\
  \mhiggsino = 200\GeV  & $4.7\pm 0.1$ &  $4.2\pm 0.1$ &  $3.3\pm 0.1$  \\
  \mhiggsino = 250\GeV  & $3.5\pm 0.1$ &  $3.2\pm 0.1$ &  $2.8\pm 0.1$  \\
\end{scotch}
\end{table*}

\section{\texorpdfstring{Search in channels with three or more leptons
or with a $\Z\Z\to\ell^+\ell^-$+2 jets combination}{Search in channels with three or more leptons or with a ZZ to lepton pairs plus 2 jets combination}}
\label{sec-multilepton}

The SUSY scenarios of interest to this study
(Fig.~\ref{fig:event-diagrams})
can yield events with three or more leptons if
the \Ph, \Z, or {\PW} bosons decay to final states with leptons.
We therefore combine the results presented here with
our results on final states with three or more leptons~\cite{Chatrchyan:2014aea}
to derive unified conclusions for these scenarios.
The three-or-more-lepton results
provide sensitivity to the SUSY $\Z\Z$ channel,
\ie, to events in which the two Higgs bosons in
Fig.~\ref{fig:event-diagrams} (left)
are each replaced by a \Z boson.
In contrast,
the studies presented in Sections~\ref{sec-hh-4b}--\ref{sec-z-to-ll}
have little sensitivity to $\Z\Z$ production.
In addition,
the three-or-more-lepton results provide
sensitivity to the SUSY $\Ph\Ph$ and $\Ph\Z$ channels,
especially for low values of the higgsino (\PSGczDo) mass.

The analysis of Ref.~\cite{Chatrchyan:2014aea}
requires events to contain at least three charged lepton candidates
including at most one \tauh candidate.
The events are divided into exclusive categories
based on the number and flavor of the leptons,
the presence or absence of an opposite-sign, same-flavor (OSSF) lepton pair,
the invariant mass of the OSSF pair including its consistency with
the \Z boson mass,
the presence or absence of a tagged $\cPqb$ jet,
the \met value,
and the \HT value.
As in Ref.~\cite{Chatrchyan:2014aea},
we order the search channels by their expected sensitivities and,
for the interpretation of results (Section~\ref{sec-interpretation}),
select channels starting with the most sensitive one,
and do not consider additional
channels once the expected number of signal events,
integrated over the retained channels,
equals or exceeds 90\% of the total expected number.

\begin{table*}[htb]
\centering
\topcaption{
The seven most sensitive search channels of the
three-or-more-lepton analysis~\cite{Chatrchyan:2014aea}
for the
$\PSGczDo(\to\Ph\lsp)\PSGczDo(\to\Ph\lsp)$
di-higgsino production scenario
assuming a higgsino mass of 150\GeV
and an LSP (gravitino) mass of 1\GeV.
For all channels,
$\HT<200\GeV$ and the number of tagged $\cPqb$ jets is zero.
The symbols $N_{\ell}$, $N_{\tauh}$, and \nossf indicate the number
of charged leptons,
hadronically decaying $\tau$-lepton candidates,
and  opposite-sign same-flavor (OSSF) lepton pairs,
respectively.
``Below {\Z}'' means that the invariant mass \mll of
the OSSF pair (if present) lies
below the region of the \Z boson ($\mll<75\GeV$),
while ``Off {\Z}'' means that either
$\mll<75\GeV$ or $\mll>105\GeV$.
The uncertainties shown for the SM background estimates
are the combined statistical and systematic terms,
while those shown for signal events are statistical.
The channels are ordered according to the values of
$N_{\ell}$, $N_{\tauh}$, \nossf, and \met.
}
\label{tab:multilepton}
\begin{scotch}{cccccccc}
  $N_{\ell}$ & $N_{\tauh}$ & \nossf & \mll  & \met (\GeVns{}) & SM & Data
  & $\Ph\Ph$ events,\\
  & & & range & & background & & $\mhiggsino=150\GeV$  \\
\hline
3 & 0 & 0 &  \NA     & 0-50   & $51\pm 11$     &  53 & $3.1\pm 0.6$ \\
3 & 0 & 0 &  \NA     & 50-100 & $38\pm 15$     &  35 & $2.7\pm 0.6$ \\
3 & 0 & 1 & Below \Z & 50-100 & $130\pm 27$    & 142 & $7.4\pm 1.6$ \\
3 & 1 & 0 &  \NA     & 50-100 & $400\pm 150$   & 406 & $8.0\pm 1.4$ \\
4 & 0 & 1 & Off \Z   & 50-100 & $0.2\pm 0.1$   &   0 & $0.5\pm 0.2$ \\
4 & 1 & 1 & Off \Z   & 0-50   & $7.5\pm 2.0$   &  15 & $0.8\pm 0.2$ \\
4 & 1 & 1 & Off \Z   & 50-100 & $2.1\pm 0.5$   &   4 & $0.7\pm 0.2$ \\
\end{scotch}
\end{table*}

As an illustration of the information
provided by the three-or-more-lepton analysis,
the seven most sensitive channels for $\Ph\Ph$ signal events,
assuming a higgsino mass of $\mhiggsino=150\GeV$
and a $\PSGczDo\to\Ph\lsp$ branching fraction of unity,
are presented in
Table~\ref{tab:multilepton}.
Similar results are obtained for other values of the higgsino mass.
Table~\ref{tab:multilepton}
includes the observed numbers of events,
the SM background estimates~\cite{Chatrchyan:2014aea},
and the predicted signal yields.
Some excesses in the data relative to the expectations
are seen for the last two channels listed in the table,
for which 15 and 4 events are observed,
compared to $7.5\pm 2.0$ and $2.1\pm 0.5$ events, respectively,
that are expected.
The combined local excess is 2.6~standard deviations.
The excesses in these two search channels
are discussed in Ref.~\cite{Chatrchyan:2014aea},
where it is demonstrated that they are consistent with a
statistical fluctuation once the
large number of search channels in the analysis
is taken into account
(look-elsewhere effect).

We also make use of our results~\cite{Khachatryan:2014qwa}
on final states with two or more jets
and either a $\Z\to\EE$ or $\Z\to\MM$ decay,
which provide yet more sensitivity to the SUSY $\Z\Z$ channel.
In the study of Ref.~\cite{Khachatryan:2014qwa},
events must contain either an \EE or \MM
pair and no other lepton,
at least two jets,
no tagged $\cPqb$ jets,
and large values of \met.
The invariant mass of the lepton pair,
and the dijet mass formed from the two jets with highest~\pt values,
are both required to
be consistent with the \Z boson mass.
Ref.~\cite{Khachatryan:2014qwa} also contains results on the $\Ph\PW$
signal scenario of Fig.~\ref{fig:event-diagrams} (right)
in decay channels that are complementary to those considered here.
We make use of these results in our interpretation
of the $\Ph\PW$ scenario.

\section{Systematic uncertainties}
\label{sec-systematics}

Systematic uncertainties for the various background estimates
are presented in the respective sections above,
or,
in the case of the studies mentioned in Section~\ref{sec-multilepton},
in Refs.~\cite{Chatrchyan:2014aea,Khachatryan:2014qwa}.

Systematic uncertainties associated with the selection efficiency
for signal events arise from various sources.
The uncertainties related to the jet energy scale,
jet energy resolution,
pileup modeling,
trigger efficiencies,
\cPqb-jet tagging efficiency correction factors,
lepton identification and isolation criteria,
and the ISR modeling
are evaluated by varying the respective quantities by their uncertainties,
while those associated with the parton distribution functions
are determined~\cite{Pumplin:2002vw,Martin:2009iq,Ball:2011mu}
using the recommendations of Refs.~\cite{Alekhin:2011sk,PDF4LHC}.
The uncertainty of the luminosity determination
is 2.6\%~\cite{CMS-PAS-LUM-13-001}.
Table~\ref{tab:sigsyst} lists typical values of the uncertainties.
The uncertainty listed for lepton identification and isolation
includes an uncertainty of 1\% per lepton to account for differences
between the fast simulation and GEANT-based modeling of the detector response.
In setting limits (Section~\ref{sec-interpretation}),
correlations between systematic uncertainties
across the different search channels are taken into account,
and the systematic uncertainties are treated as nuisance parameters
as described in Ref.~\cite{cms-note-2011-005}.

\begin{table}[htb]
\topcaption{
Typical values of the systematic uncertainty for signal efficiency,
in percentage.
}
\label{tab:sigsyst}
\centering
\begin{scotch}{lc}
Source                              &        \\
\hline
Jet energy scale                    &  5--10  \\
Jet energy resolution               &  2--4   \\
Pileup modeling                     &  4     \\
Trigger efficiency                  &  1--5   \\
\cPqb--jet tagging efficiency           &  5--10  \\
Lepton identification and isolation &  5     \\
ISR modeling                        &  1     \\
Parton distribution functions       &  1     \\
Integrated luminosity               &  2.6   \\
\end{scotch}
\end{table}

\section{Interpretation}
\label{sec-interpretation}

In this section,
we present the interpretation of our results.
We set 95\% confidence level (\CL) upper limits
on the production cross sections of the considered scenarios
using a modified frequentist CL$_\mathrm{S}$ method
based on the LHC-style test statistic~\cite{Read:2002hq,Junk:1999kv,cms-note-2011-005}.
The input to the procedure is the number of observed events,
the number of expected SM background events (with uncertainties),
and the number of predicted signal events in each bin of the distributions of
Figs.~\ref{fig:result-plots}, \ref{fig:hgg-kine-01} (\cmsRight),
\ref{fig:hgg-kine-02} (\cmsRight), \ref{fig:higgs-mt},
and~\ref{fig:zll-hbb-results},
as well as the relevant results from Refs.~\cite{Chatrchyan:2014aea,Khachatryan:2014qwa}
(see Tables~2--3 of Ref.~\cite{Chatrchyan:2014aea}
and Tables~4--6 of Ref.~\cite{Khachatryan:2014qwa}).
The contributions of signal events 
are incorporated into the likelihood function
for both signal and control regions.
The cross section upper limits are compared to the
predicted cross sections,
which have uncertainties~\cite{PDF4LHC}
of approximately 5\%.

We first present upper limits
for the GMSB higgsino NLSP model~\cite{Matchev:1999ft,Ruderman:2011vv}
discussed in the introduction.
The limits are presented
as a function of the higgsino (\PSGczDo) mass
for the $\Ph\Ph$, $\Z\Z$, and $\Ph\Z$ topologies separately
and then in the two-dimensional plane of the
$\PSGczDo\to\Ph\lsp$ branching fraction
versus~\mhiggsino.
We assume that the higgsino \PSGczDo can decay only to
the $\Ph\lsp$ or $\Z\lsp$ states.
Following our discussion of the GMSB model,
we present limits for the electroweak chargino-neutralino
pair production process of Fig.~\ref{fig:event-diagrams} (right)
as a function of the LSP (\PSGczDo)
and common \PSGczDt, ${\PSGcpm}_1$ masses,
taking the $\PSGczDt\to\Ph\PSGczDo$ and
${\PSGcpm}_1\to\PW^\pm\PSGczDo$
branching fractions each to be unity.

\subsection{Limits on the GMSB di-higgsino NLSP model}

\subsubsection{The \texorpdfstring{$\Ph\Ph$}{hh} topology}
\label{sec-hh-limits}

\begin{figure}[tbh]
  \centering
   \includegraphics[width=\cmsFigWidth]{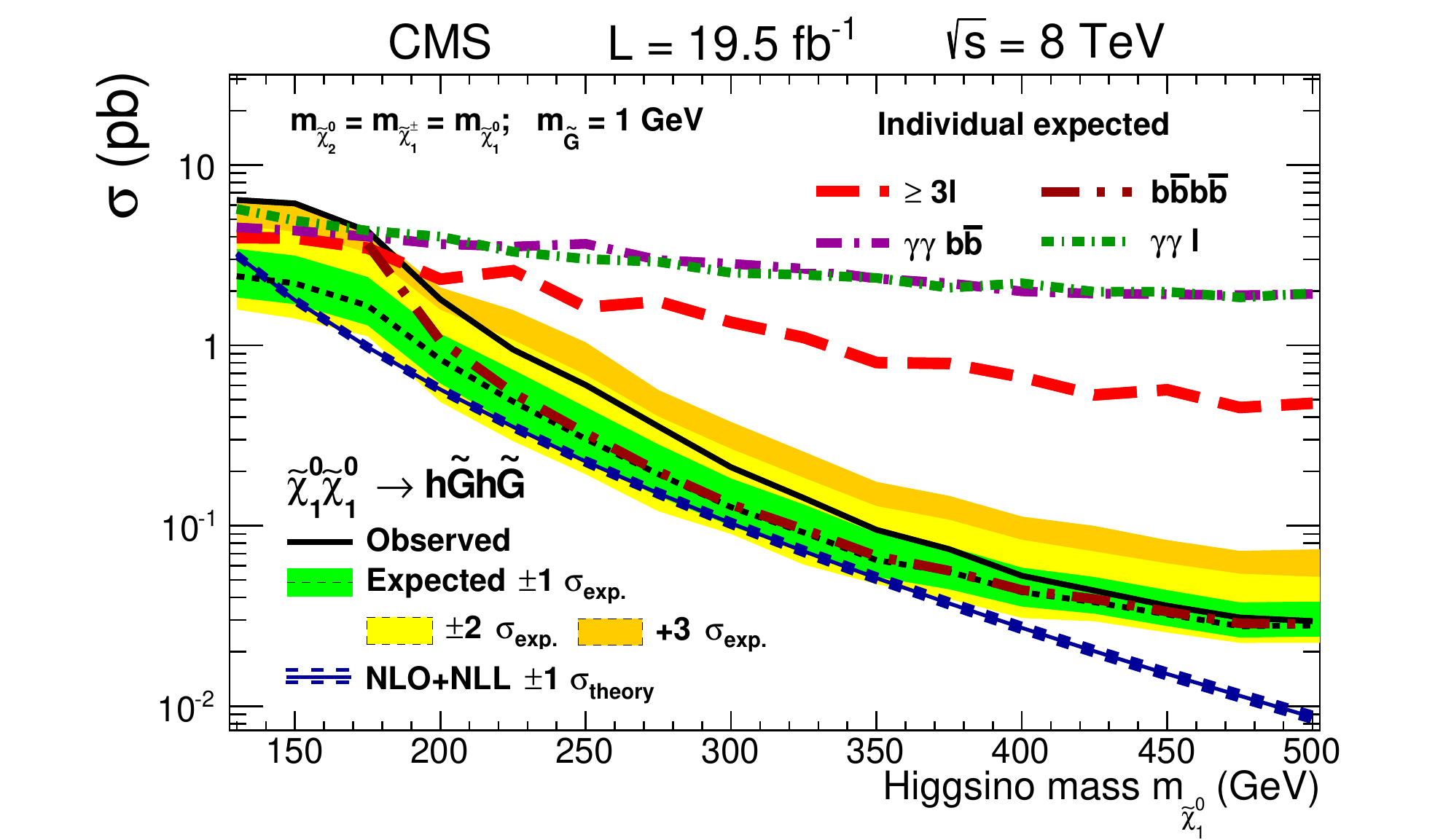}
 \caption{
Observed and expected 95\% confidence level upper limits on
the cross section for higgsino pair production in the
$\Ph\Ph$ topology as a function of the higgsino mass
for the combined
$\bbbar\bbbar$,
$\gamma\gamma\bbbar$,
$\gamma\gamma$+leptons,
and three-or-more-lepton channels.
The dark (green), light (yellow), and medium-dark (orange)
bands indicate the one-, two-,
and three-standard-deviation uncertainty intervals, respectively,
for the expected results.
The theoretical cross section and the
expected curves for the individual search channels are also shown.
}
  \label{fig:hh-limits}
\end{figure}

Figure~\ref{fig:hh-limits} shows the 95\% \CL
cross section upper limits on higgsino pair production
through the $\Ph\Ph$ channel
[Fig.~\ref{fig:event-diagrams} (left)],
\ie, assuming the
$\PSGczDo\to\Ph\lsp$ branching fraction to be unity.
The limits are derived using the combined results from the
$\Ph\Ph\to\bbbar\bbbar$,
$\gamma\gamma\bbbar$,
$\gamma\gamma$+leptons,
and three-or-more-lepton channels,
corresponding to the results presented in
Sections~\ref{sec-hh-4b},
\ref{sec-hhggbb},
\ref{sec-hgglepton},
and~\ref{sec-multilepton}, respectively.
Both the expected and observed limits are shown,
where the expected limits are derived from the SM background estimates.
The expected results are presented with one, two, and
three standard-deviation bands of the experimental uncertainties,
which account for the uncertainties of the background prediction
and for the statistical uncertainties of the signal observables.
The NLO+NLL theoretical cross
section~\cite{Beenakker:1999xh,Fuks:2012qx,Fuks:2013vua}
with its one-standard-deviation uncertainty band is also shown.

The observed exclusion contour in Fig.~\ref{fig:hh-limits} (solid line)
is seen to lie above the theoretical cross section for all examined higgsino mass values.
Therefore, we do not exclude higgsinos for any mass value
in the $\Ph\Ph$ topology scenario.
It is nonetheless seen that
the expected exclusion contour (short-dashed line with uncertainty bands)
lies just above the theoretical higgsino pair production cross section
for higgsino mass values $\mhiggsino\lesssim 360\GeV$.
Most of this sensitivity is provided by the
$\Ph\Ph\to\bbbar\bbbar$ channel,
which dominates the results for $\mhiggsino\gtrsim 200\GeV$.
For lower mass values,
the $\gamma\gamma\bbbar$
and three-or-more-lepton channels
provide the greatest sensitivity.
The $\Ph\Ph\to\bbbar\bbbar$ channel
loses sensitivity for $\mhiggsino\lesssim 200\GeV$
because the \metsig spectrum of signal events becomes
similar to the spectrum from SM events.

The observed limits in Fig.~\ref{fig:hh-limits} are seen to
deviate from the expected ones by
slightly more than three standard deviations for
$\mhiggsino\lesssim 170\GeV$.
The main contribution to this excess
(2.6 standard deviations, discussed in Section~\ref{sec-multilepton})
arises from the three-or-more-lepton channel,
and was also reported in Ref.~\cite{Chatrchyan:2014aea}.
The electron (but not muon) component of the
$\gamma\gamma$+leptons channel
contributes to the excess at the level of 2.1~standard deviations,
as discussed in Section~\ref{sec-hgglepton}
[Fig.~\ref{fig:higgs-mt} (\cmsRight)].
As already mentioned in Sections~\ref{sec-hgglepton} and~\ref{sec-multilepton},
we consider the excesses in the $\gamma\gamma$+electron and
three-or-more-lepton channels to be consistent with
statistical fluctuations.

\subsubsection{The \texorpdfstring{$\Z\Z$ and $\Ph\Z$}{ZZ and hZ} topologies}
\label{sec-hz-limits}

\begin{figure}[tp]
  \centering
   \includegraphics[width=\cmsFigWidth]{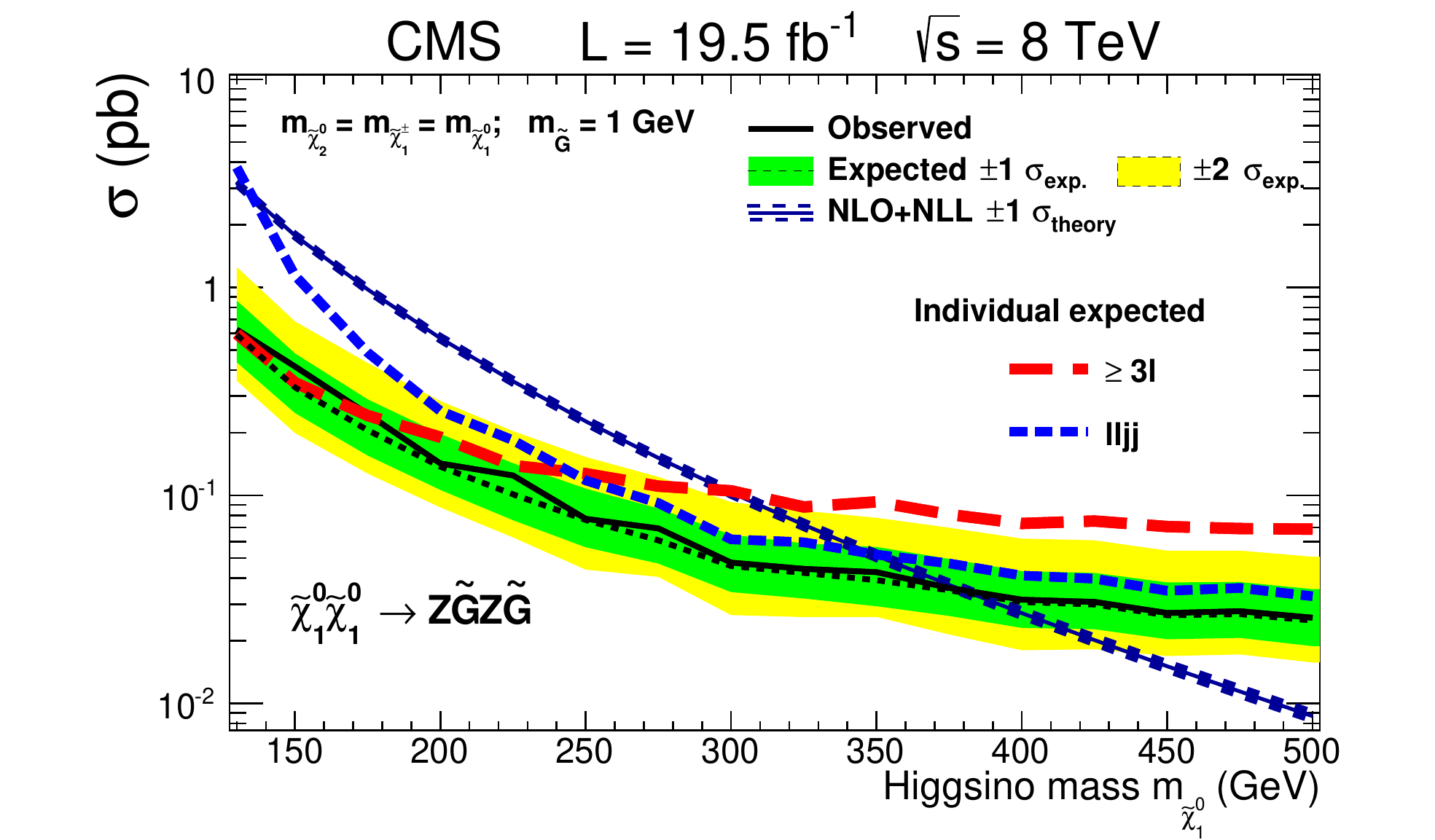} \\
   \includegraphics[width=\cmsFigWidth]{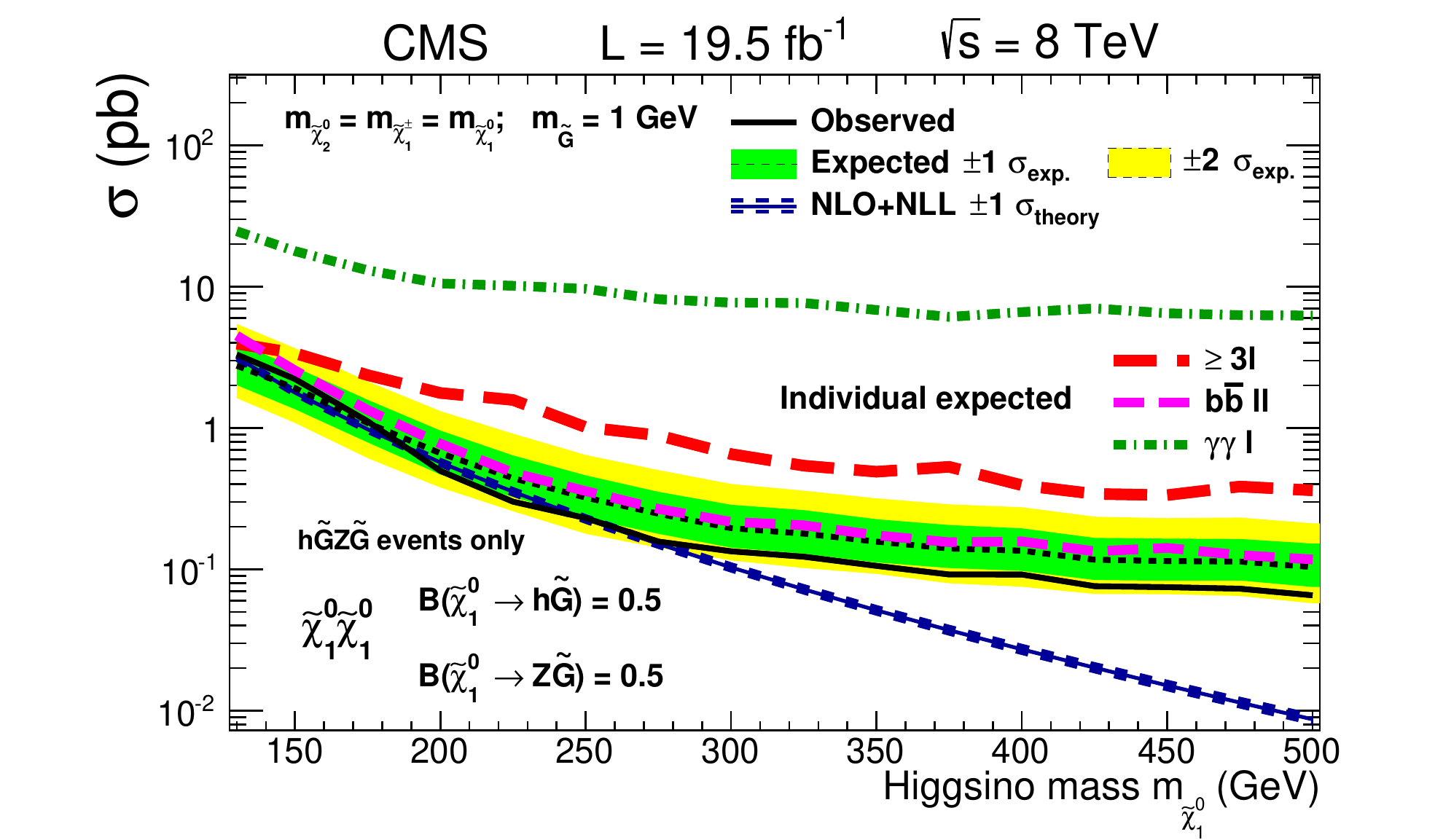}
 \caption{
(Top) Observed and expected 95\% confidence level upper limits on
the cross section for higgsino pair production in the
$\Z\Z$ topology as a function of the higgsino mass
for the combined
three-or-more-lepton and
$\ell^+\ell^-$+2~jets channels.
The dark (green) and light (yellow) bands
indicate the one- and two-standard-deviation
uncertainty intervals, respectively,
for the expected results.
The theoretical cross section and the
expected curves for the individual search channels are also shown.
(Bottom)
Corresponding results for the $\Ph\Z$ topology,
assuming the $\PSGczDo\to\Ph\lsp$
and $\PSGczDo\to\Z\lsp$ branching fractions each to be~0.5,
ignoring contributions from $\Ph\Ph$ and $\Z\Z$ events,
for the individual and combined
$\gamma\gamma$+leptons,
$\bbbar\ell^+\ell^-$,
and three-or-more-lepton channels.
}
  \label{fig:zz-limits}
\end{figure}

The 95\% \CL cross section upper limits on
higgsino pair production through the $\Z\Z$
channel are presented in Fig.~\ref{fig:zz-limits} (top).
For these results,
we assume the
$\PSGczDo\to\Z\lsp$ branching fraction to be unity.
These results are derived using the two search channels
that dominate the sensitivity to the $\Z\Z$ topology:
the three-or-more-lepton and
$\ell^+\ell^-$+2 jets channels
(Section~\ref{sec-multilepton}).
In the context of this scenario,
higgsino masses below 380\GeV are excluded.

To illustrate the sensitivity of our analysis to the $\Ph\Z$ topology
[Fig.~\ref{fig:event-diagrams} (middle)],
we assume the $\PSGczDo\to\Ph\lsp$
and $\PSGczDo\to\Z\lsp$ branching fractions each to be~0.5
and ignore contributions from the $\Ph\Ph$ and $\Z\Z$ channels.
Figure~\ref{fig:zz-limits} (bottom)
shows 95\% \CL cross section upper limits for the $\Ph\Z$ topology
derived from the combined $\gamma\gamma$+leptons, $\bbbar\ell^+\ell^-$,
and three-or-more-lepton samples
(Sections~\ref{sec-hgglepton},
\ref{sec-z-to-ll},
and~\ref{sec-multilepton},
respectively).
The results are dominated by the $\bbbar\ell^+\ell^-$ channel.
The main contribution of the three-or-more-lepton channel arises
for higgsino mass values below around 170\GeV.
The sensitivity of the
$\gamma\gamma$+leptons channel is minimal.
[The $\gamma\gamma$+2~jets channel also contributes minimally
and is not included in the combination of Fig.~\ref{fig:zz-limits} (bottom).]

\subsubsection{Exclusion region as a function of the \texorpdfstring{\PSGczDo mass and
$\PSGczDo\to\Ph\lsp$}{neutralino and neutralino to h LSP} branching fraction}

\begin{figure}[tbh]
  \centering
   \includegraphics[width=\cmsFigWidth]{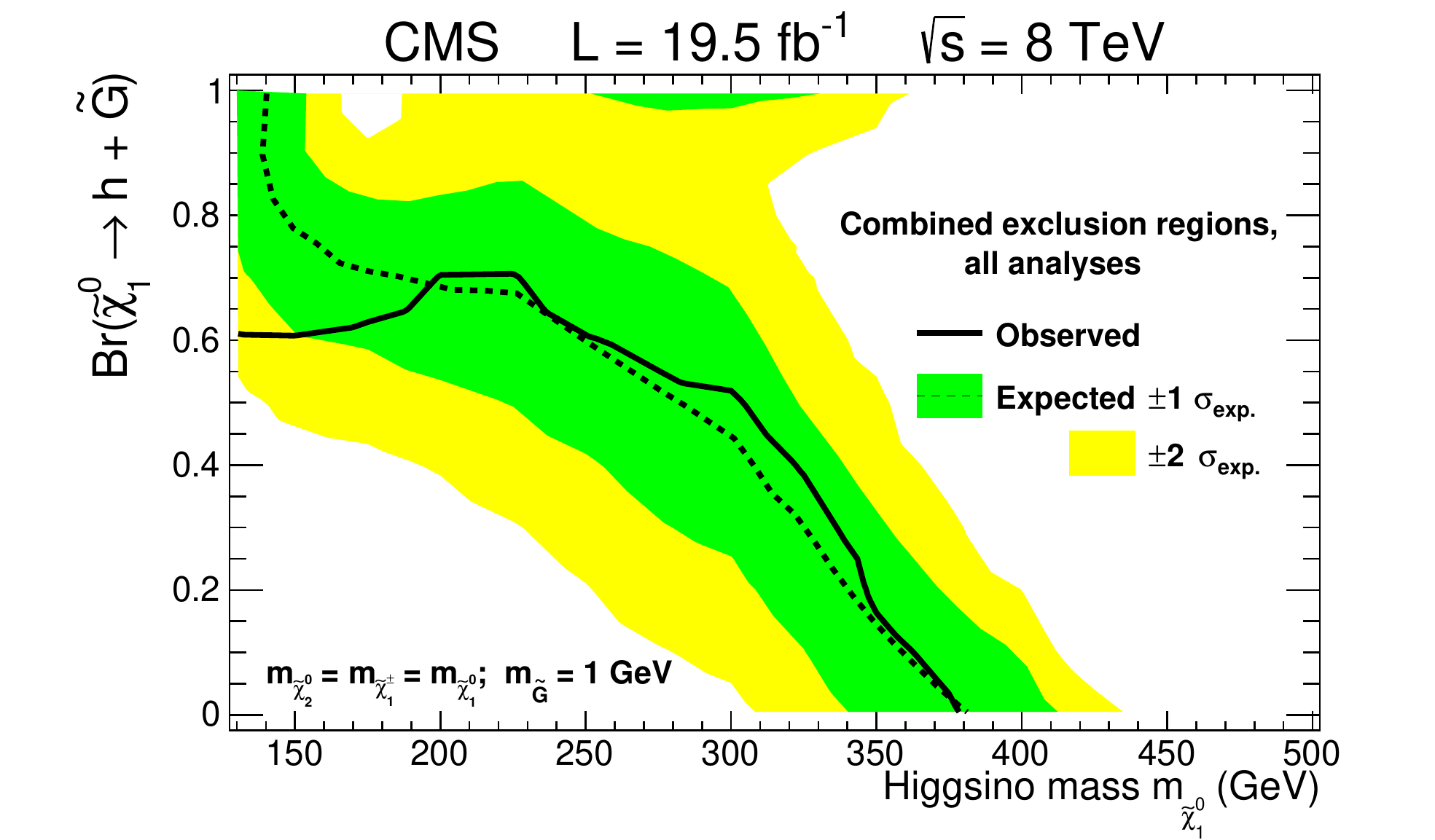}
  \caption{
Observed and expected 95\% confidence level exclusion regions
for higgsino pair production,
with all channels combined,
in the plane of the higgsino branching fraction to a Higgs boson and LSP,
versus the higgsino mass.
The dark (green) and light (yellow) bands
indicate the one- and two-standard-deviation
uncertainty intervals, respectively.
The excluded regions correspond to the area below the contours.
}
  \label{fig:sliding-br}
\end{figure}

\begin{figure}[tbh]
  \centering
   \includegraphics[width=\cmsFigWidth]{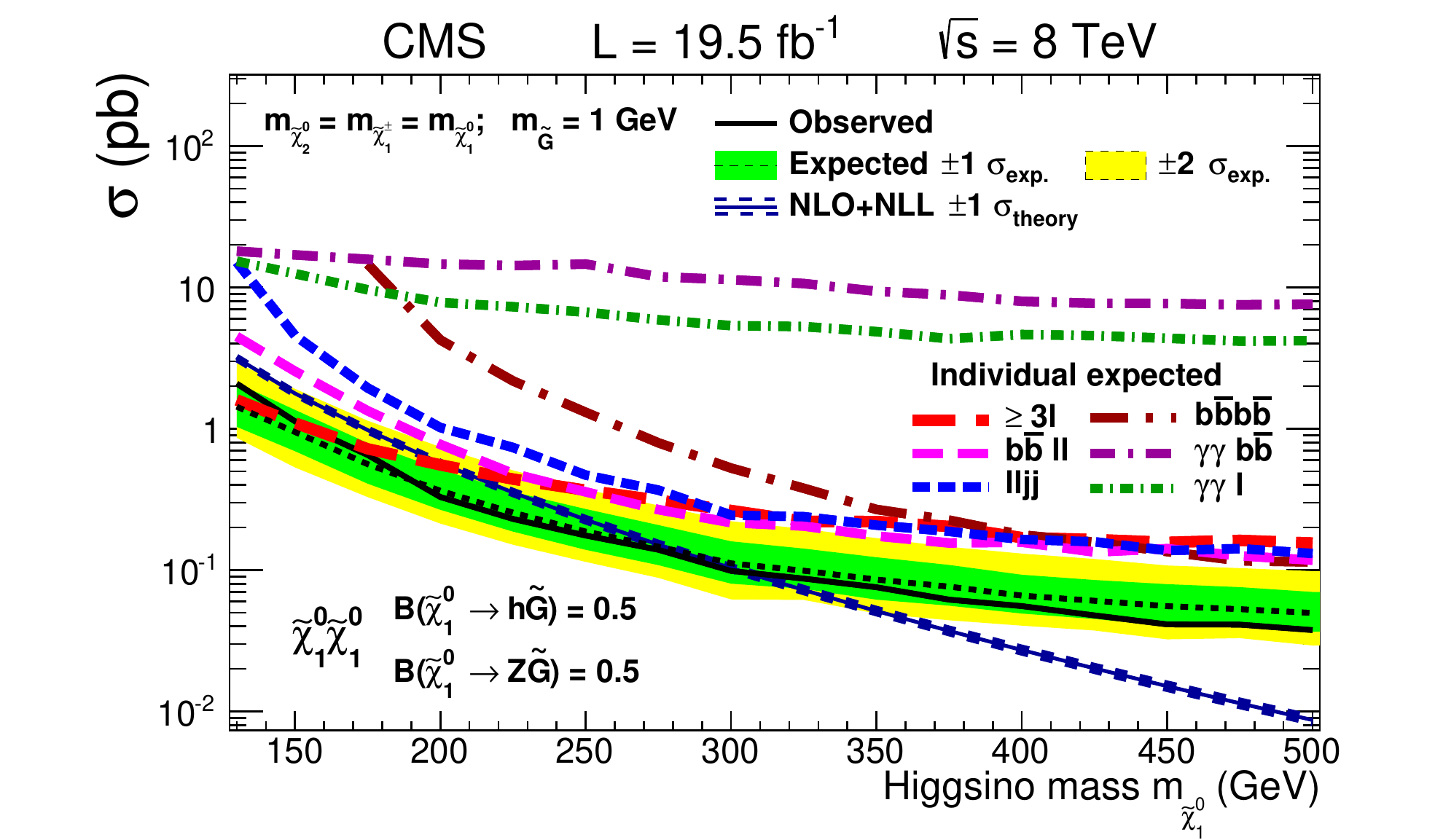}
 \caption{
Observed and expected 95\% confidence level upper limits on
the cross section for higgsino pair production
as a function of the higgsino mass
assuming the $\PSGczDo\to\Ph\lsp$
and $\PSGczDo\to\Z\lsp$ branching fractions each to be~0.5,
including contributions from $\Ph\Ph$ and $\Z\Z$ events,
for the combined
$\bbbar\bbbar$,
$\gamma\gamma\bbbar$,
$\gamma\gamma$+leptons,
$\bbbar\ell^+\ell^-$,
three-or-more-lepton,
and $\ell^+\ell^-$+2~jets channels.
The dark (green) and light (yellow) bands
indicate the one- and two-standard-deviation
uncertainty intervals, respectively,
for the expected results.
The theoretical cross section and the expected curves for the
individual search channels are also shown.
}
  \label{fig:hz-limits-slice}
\end{figure}

Figure~\ref{fig:sliding-br}
presents the 95\% \CL exclusion region
for the GMSB higgsino NLSP scenario
in the two-dimensional plane of the
$\PSGczDo\to\Ph\lsp$
higgsino branching fraction
versus the higgsino mass~\mhiggsino.
The results are based on
all relevant studies discussed in this paper
including those of
Refs.~\cite{Chatrchyan:2014aea,Khachatryan:2014qwa}.
The combined results exclude a significant fraction of
the Fig.~\ref{fig:sliding-br} plane.
For higgsino mass values above around 200\GeV,
the observed results are in agreement with the expected ones
within one standard deviation of the uncertainties.
For smaller higgsino mass values,
the observed exclusion boundary lies below the expected one
because of the excesses in data
discussed in Section~\ref{sec-hh-limits}.
Horizontal slices of Fig.~\ref{fig:sliding-br}
at branching fractions of
one and zero correspond to the results presented
in Figs.~\ref{fig:hh-limits} and~\ref{fig:zz-limits} (top)
for the $\Ph\Ph$ and $\Z\Z$ topologies,
respectively.
The corresponding results for a horizontal slice at
a branching fraction of 0.5 are shown in
Fig.~\ref{fig:hz-limits-slice}.
It is seen that higgsino masses below around 300\GeV are excluded
for this latter scenario.

\begin{figure}[tbp]
  \centering
   \includegraphics[width=\cmsFigWidth]{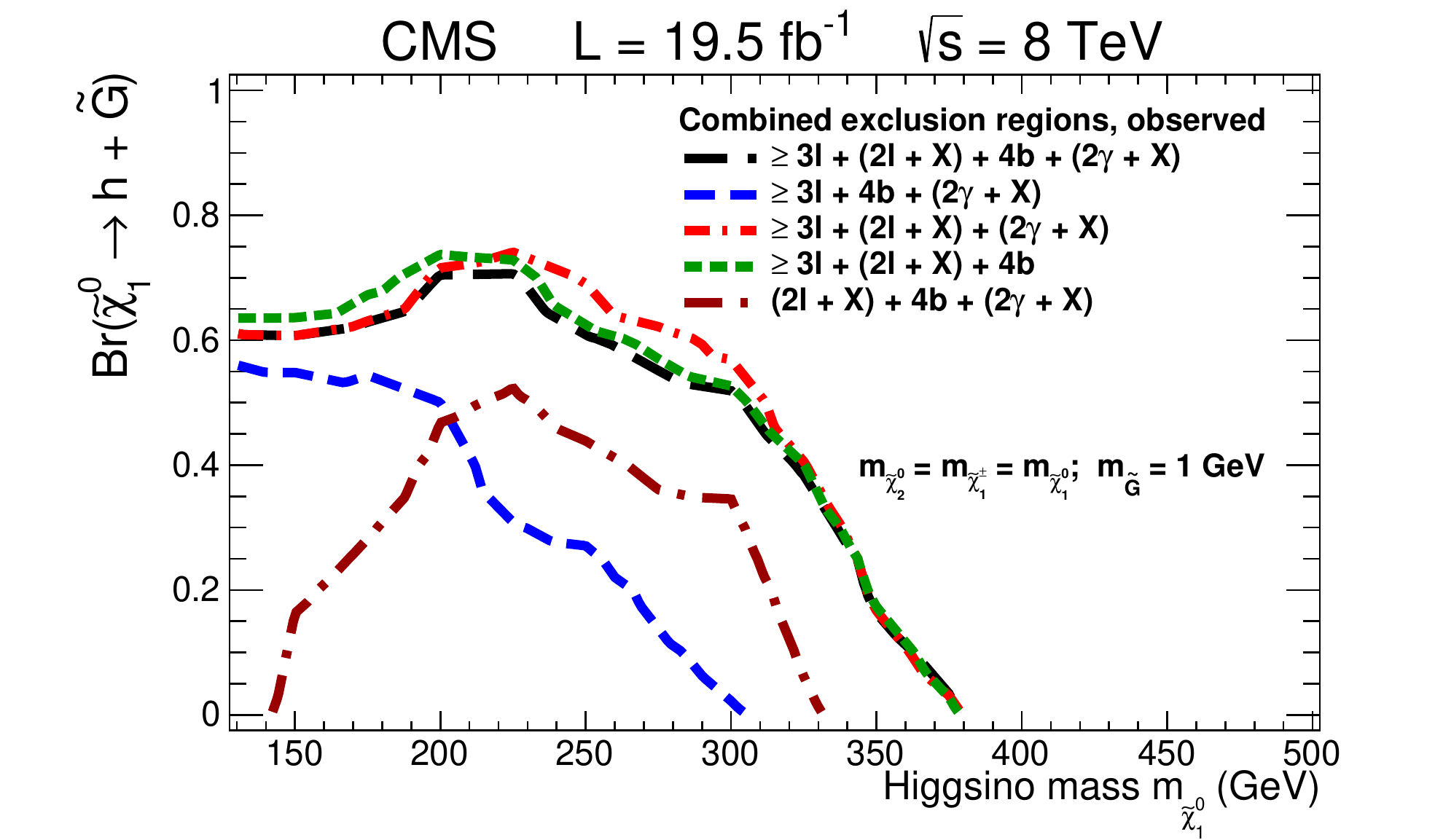}
   \includegraphics[width=\cmsFigWidth]{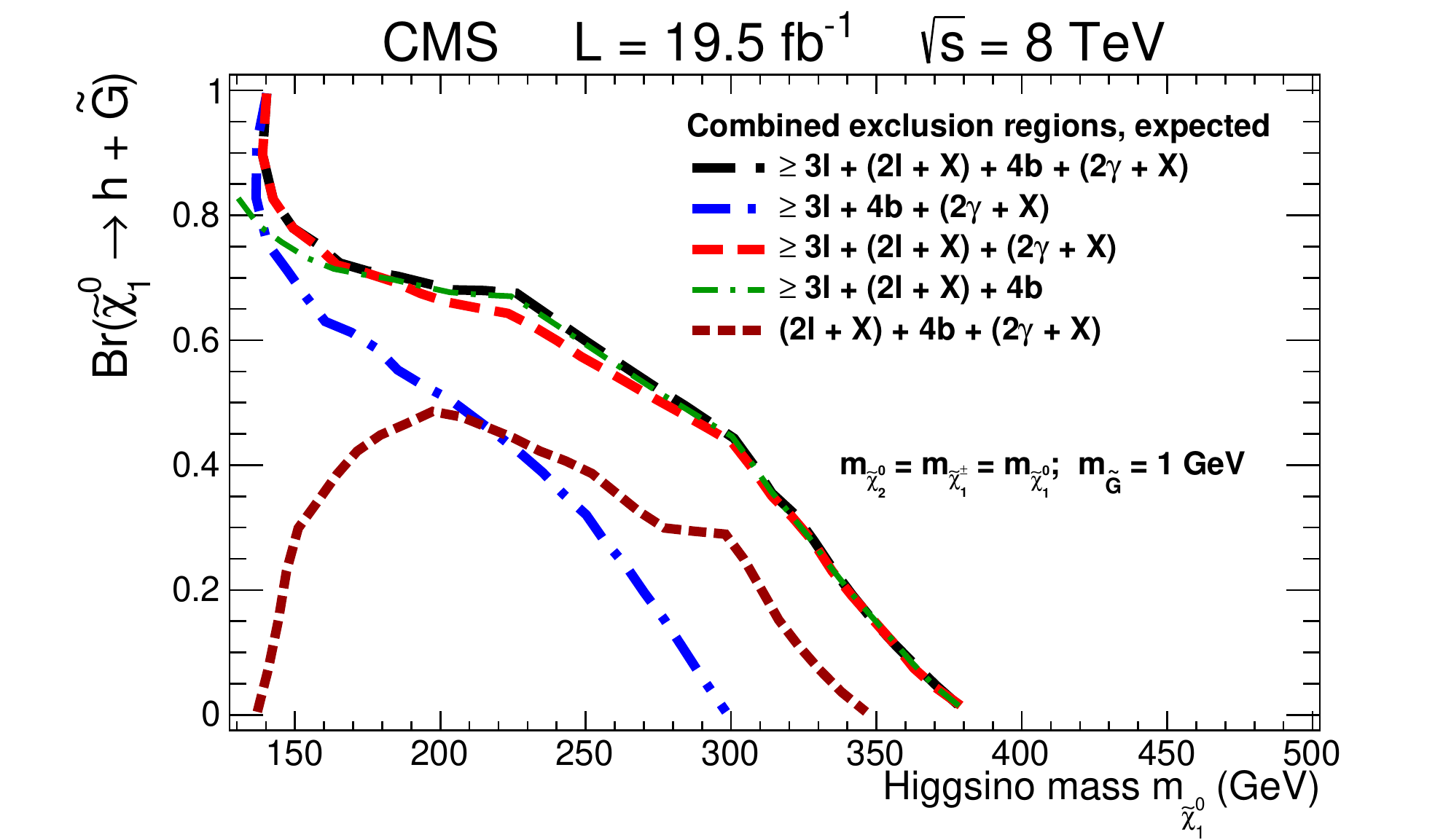}
  \caption{
(Top) Observed, and (bottom) expected
95\% confidence level exclusion regions
for higgsino pair production in the
plane of the higgsino branching fraction to a Higgs boson and the LSP,
versus the higgsino mass,
with each principal search channel group removed in turn
from the combination.
The excluded regions correspond to the area below the contours.
}
  \label{fig:sliding-br-n-minus-one}
\end{figure}

\begin{figure}[tb]p
  \centering
   \includegraphics[width=\cmsFigWidth]{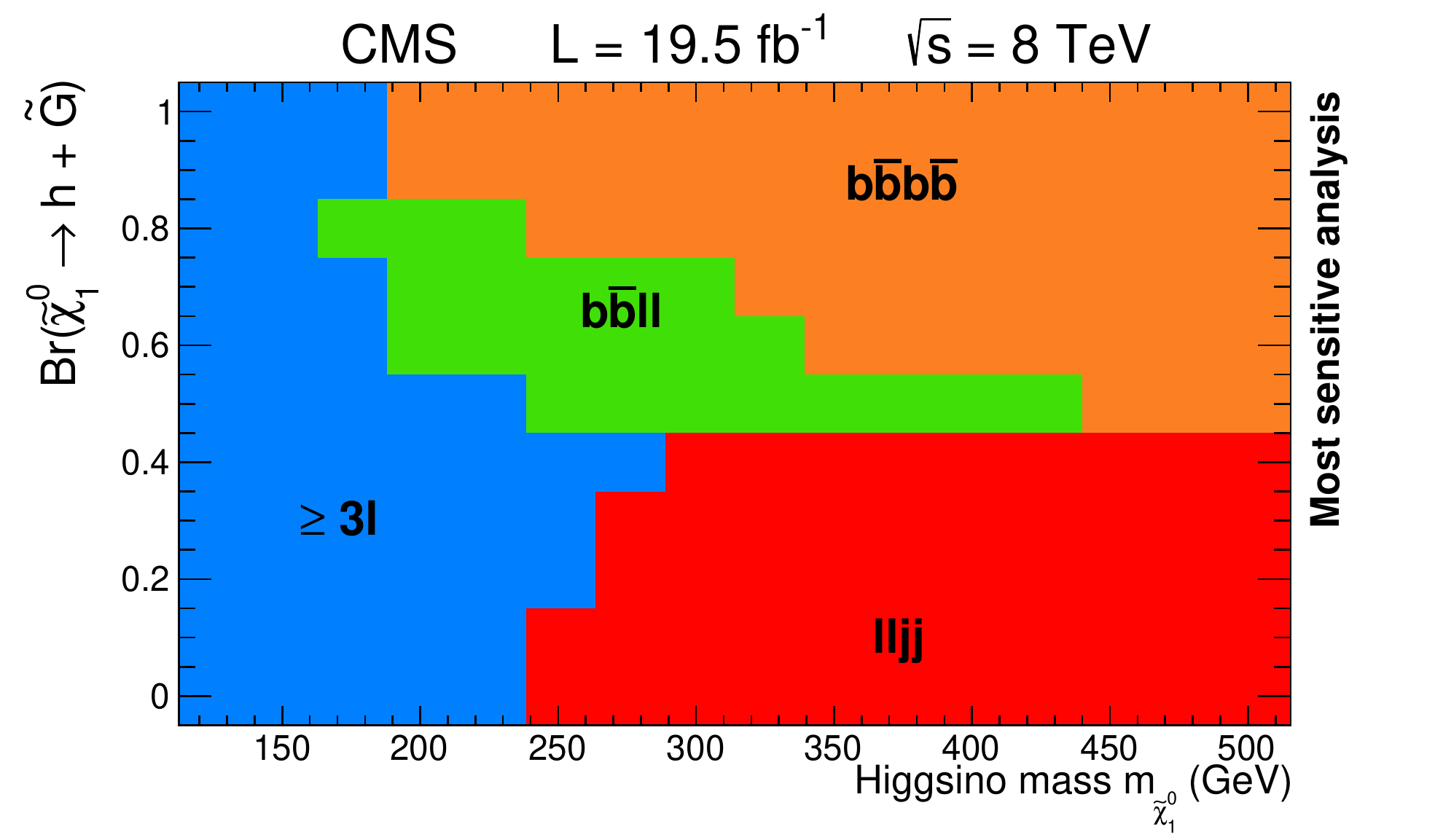}
  \caption{
The search channel that provides the most stringent
95\% confidence level upper limit on \PSGczDo higgsino pair production in the
plane of the higgsino branching fraction to a Higgs boson and the LSP,
versus the higgsino mass.
}
  \label{fig:most-sensitive}
\end{figure}

To illustrate the relative importance of the different
search channels for the results of Fig.~\ref{fig:sliding-br},
we present in Fig.~\ref{fig:sliding-br-n-minus-one}
the observed and expected exclusion regions
when each principal component of the analysis
is in turn removed from the combination.
For this purpose,
the $\Ph\to\gamma\gamma$ studies of
Section~\ref{sec-higgs-to-gg} are grouped together
into a ``$2\gamma$+X'' category,
and the
$\Ph(\to\bbbar)\Z(\to\ell^+\ell^-)$
and $\Z(\to\ell^+\ell^-)\Z(\to\,$2 jets$)$
studies of Sections~\ref{sec-z-to-ll} and~\ref{sec-multilepton}
into a ``$2\ell$+X'' category.
The greatest impact is from
the three-or-more-lepton and combined
$\bbbar\ell^+\ell^-$ and $\ell^+\ell^-$+2~jets channels,
because of the stringent constraints they impose on $\Z\Z$ production
[Fig.~\ref{fig:zz-limits}~(top)].
A distribution showing which search channel provides
the most stringent 95\% \CL cross section upper limit in the plane
of the \PSGczDo branching fraction
versus the \PSGczDo mass
is presented in Fig.~\ref{fig:most-sensitive}.

\subsection{The \texorpdfstring{$\Ph\PW$}{hW} topology}
\label{sec-hw-limits}

In Ref.~\cite{Khachatryan:2014qwa},
we present limits on the chargino-neutralino pair-production
scenario of Fig.~\ref{fig:event-diagrams} (right),
\ie, on a generic new-physics SUSY-like process with
a Higgs boson, a {\PW} boson, and~\met.
The event signatures considered
are those that yield a single electron or muon and a \bbbar pair,
a same-sign $\Pe\Pe$, $\Pgm\Pgm$, or $\Pe\Pgm$ pair and
no third charged lepton,
and three or more charged leptons~\cite{Chatrchyan:2014aea}.
These results target the
$\Ph(\to\bbbar)\PW(\to\ell\nu)$
and $\Ph(\to\Z\Z, \PW\PW, \tau\tau)\PW(\to\ell\nu$)
channels,
with $\ell$ an electron, muon, or leptonically decaying $\tau$ lepton.
With the present work,
we add the search channels with $\Ph\to\gamma\gamma$
and either $\PW\to\,$2~jets or $\PW\to\ell\nu$,
corresponding to the studies of
Sections~\ref{sec-gg-2j} and~\ref{sec-hgglepton}.

\begin{figure}[tp]
  \centering
   \includegraphics[width=\cmsFigWidth]{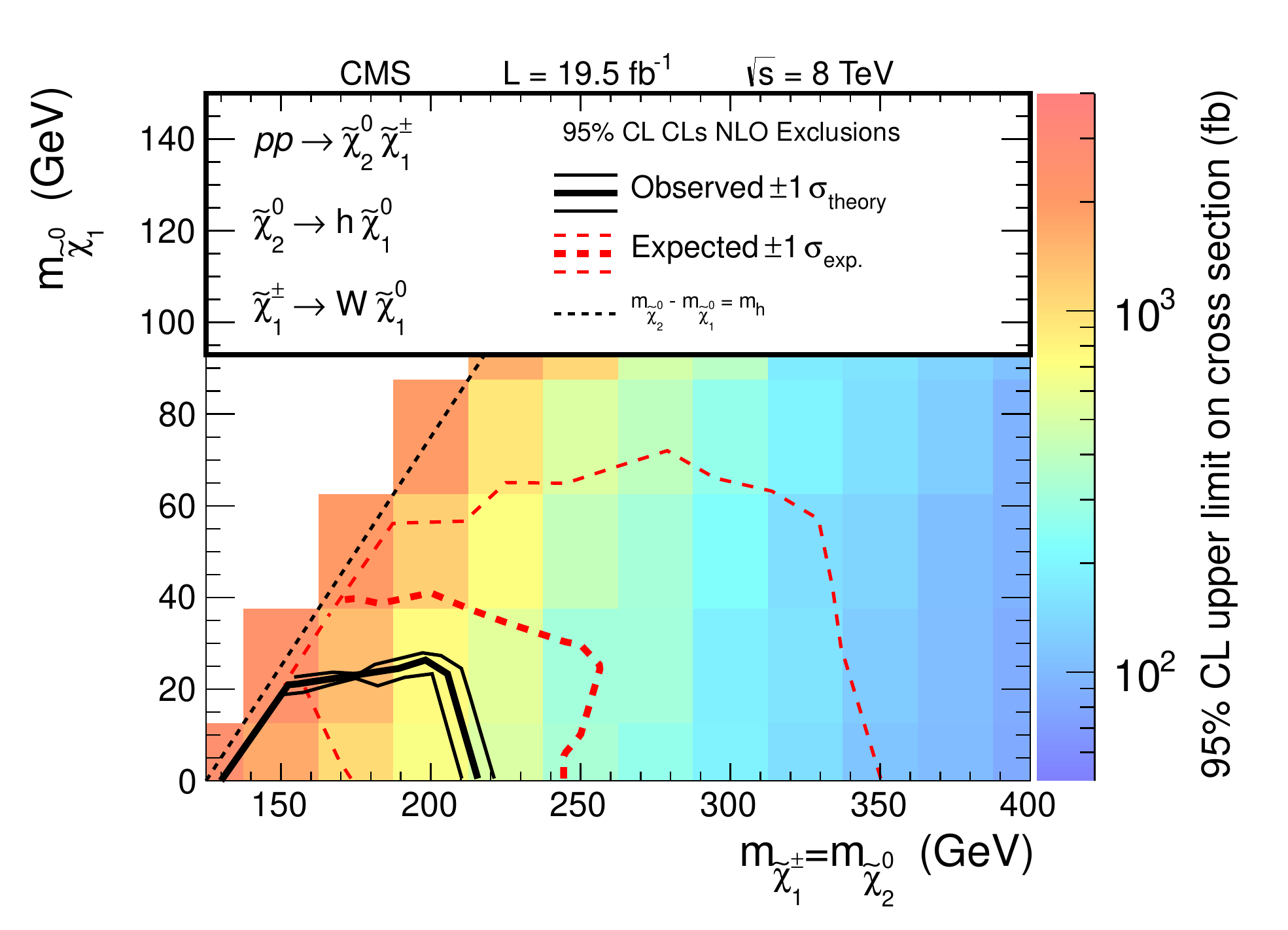}
   \includegraphics[width=\cmsFigWidth]{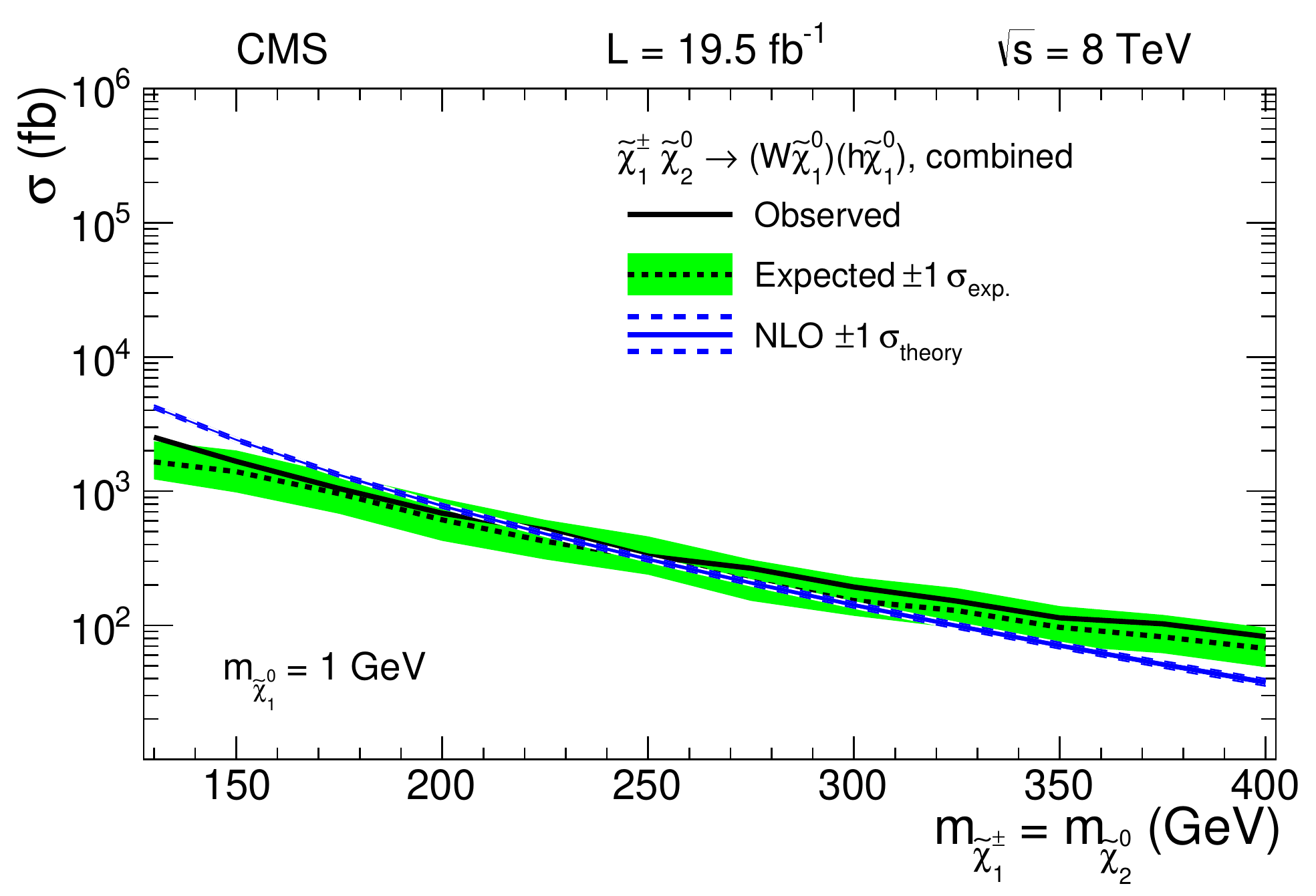}
  \caption{
(Top) Observed and expected 95\% confidence level upper limits on
the cross section for electroweak chargino-neutralino
$\PSGcpm_1\PSGczDt$ pair production
(with ${m}_{\PSGcpm_1}={m}_{\PSGczDt}$)
as a function of the LSP and \PSGczDt masses
for the combined results on single-lepton, same-sign dilepton,
and multilepton data from Ref.~\cite{Khachatryan:2014qwa}
with the diphoton data presented here.
(Bottom)
Corresponding results as a function of the \PSGczDt mass
for an LSP mass of 1\GeV.
The dark (green) band indicates the one-standard-deviation interval.
The theoretical cross section is also shown.
}
  \label{fig:hw-limits-01}
\end{figure}

The 95\% \CL upper bounds on the chargino-neutralino cross section
based on the combination of results from Ref.~\cite{Khachatryan:2014qwa}
with the two $\gamma\gamma$ search channels considered here
are shown in Fig.~\ref{fig:hw-limits-01}.
The top plot shows the cross section limits in the
LSP versus $\PSGczDt=\PSGcpm_1$ mass plane.
The bottom plot shows the limits as a function of the $\PSGczDt=\PSGcpm_1$ mass
assuming an LSP mass of $\mchionez=1\GeV$.
The single most sensitive channel is the single-lepton search
from Ref.~\cite{Khachatryan:2014qwa}.

\begin{figure}[tbp]
  \centering
   \includegraphics[width=0.48\textwidth]{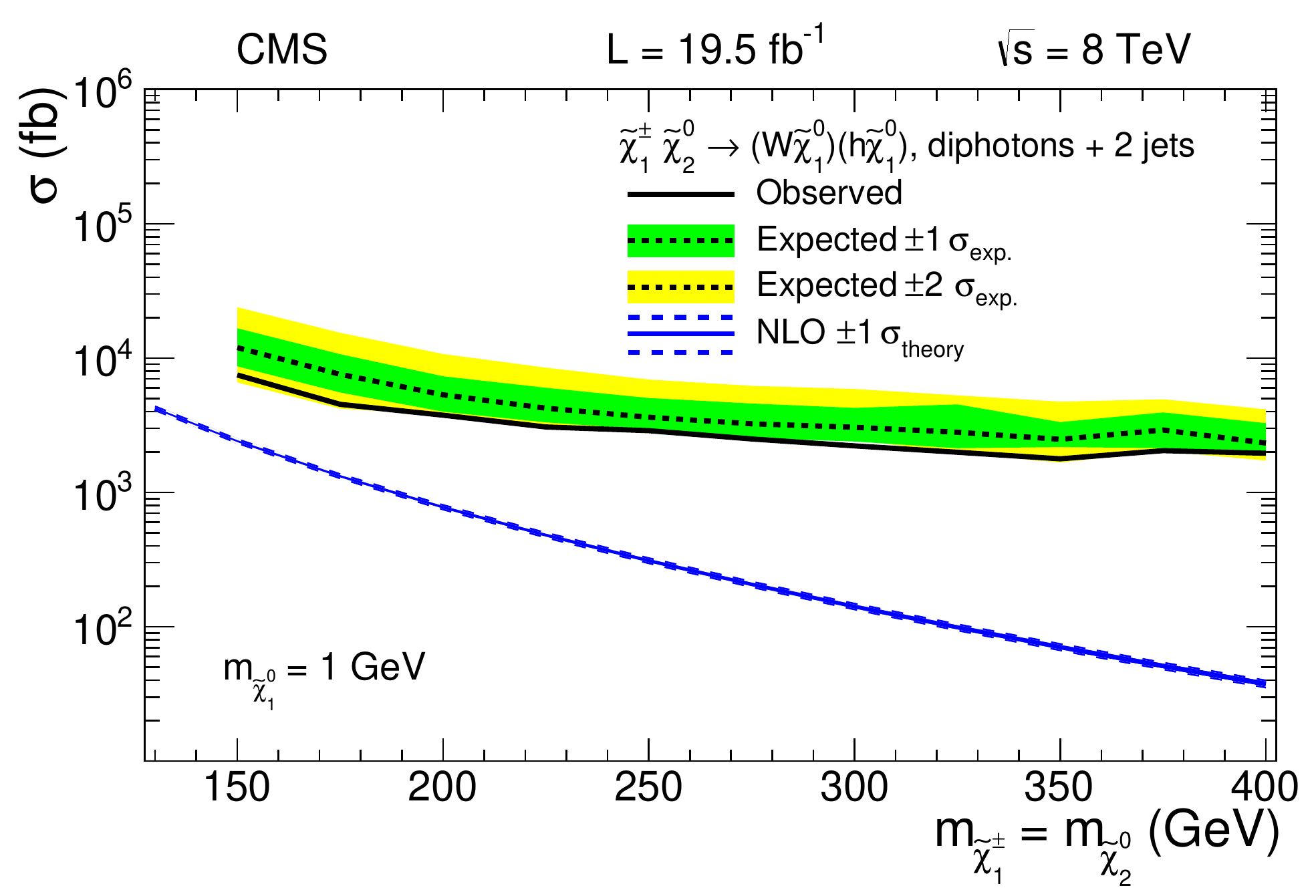}
   \includegraphics[width=0.48\textwidth]{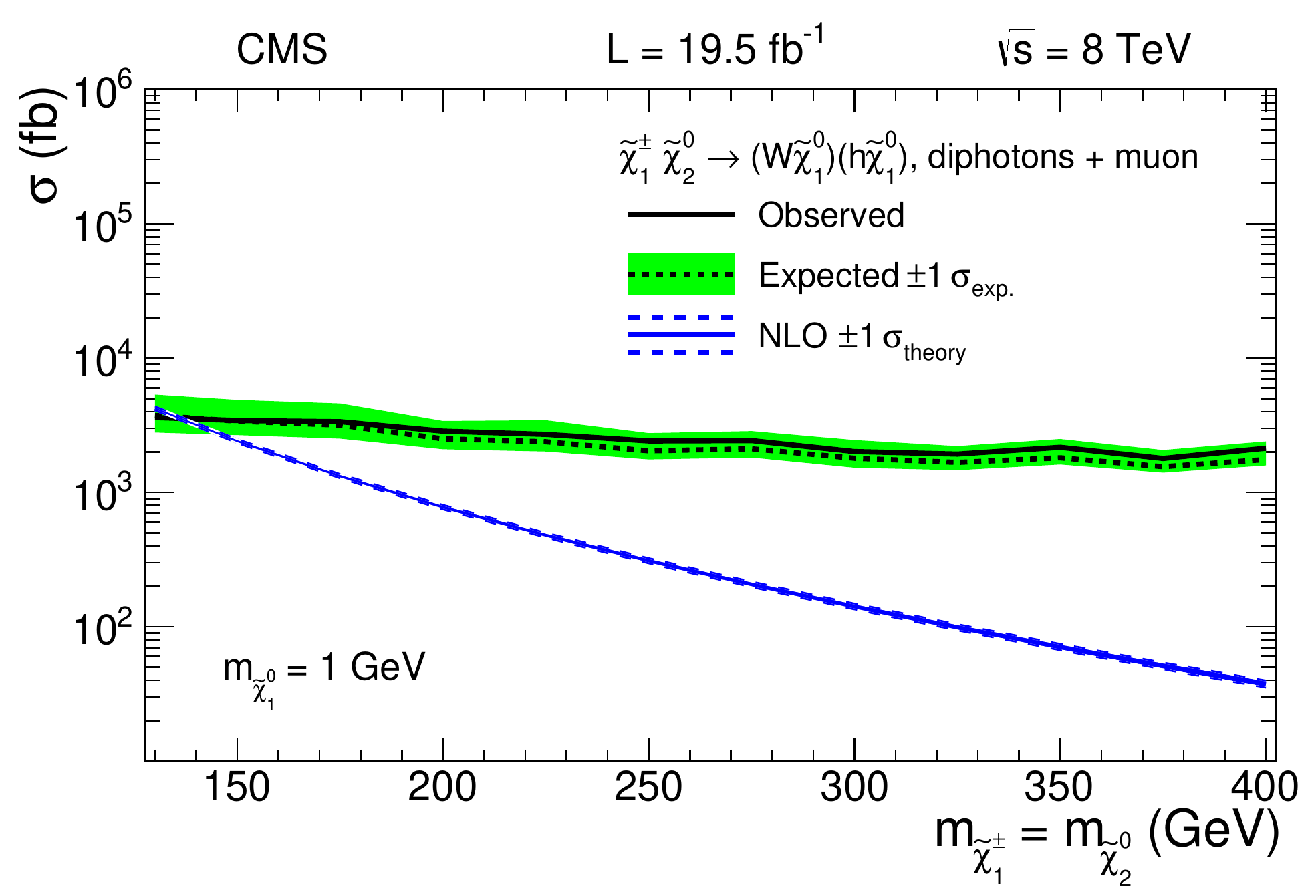}
   \includegraphics[width=0.48\textwidth]{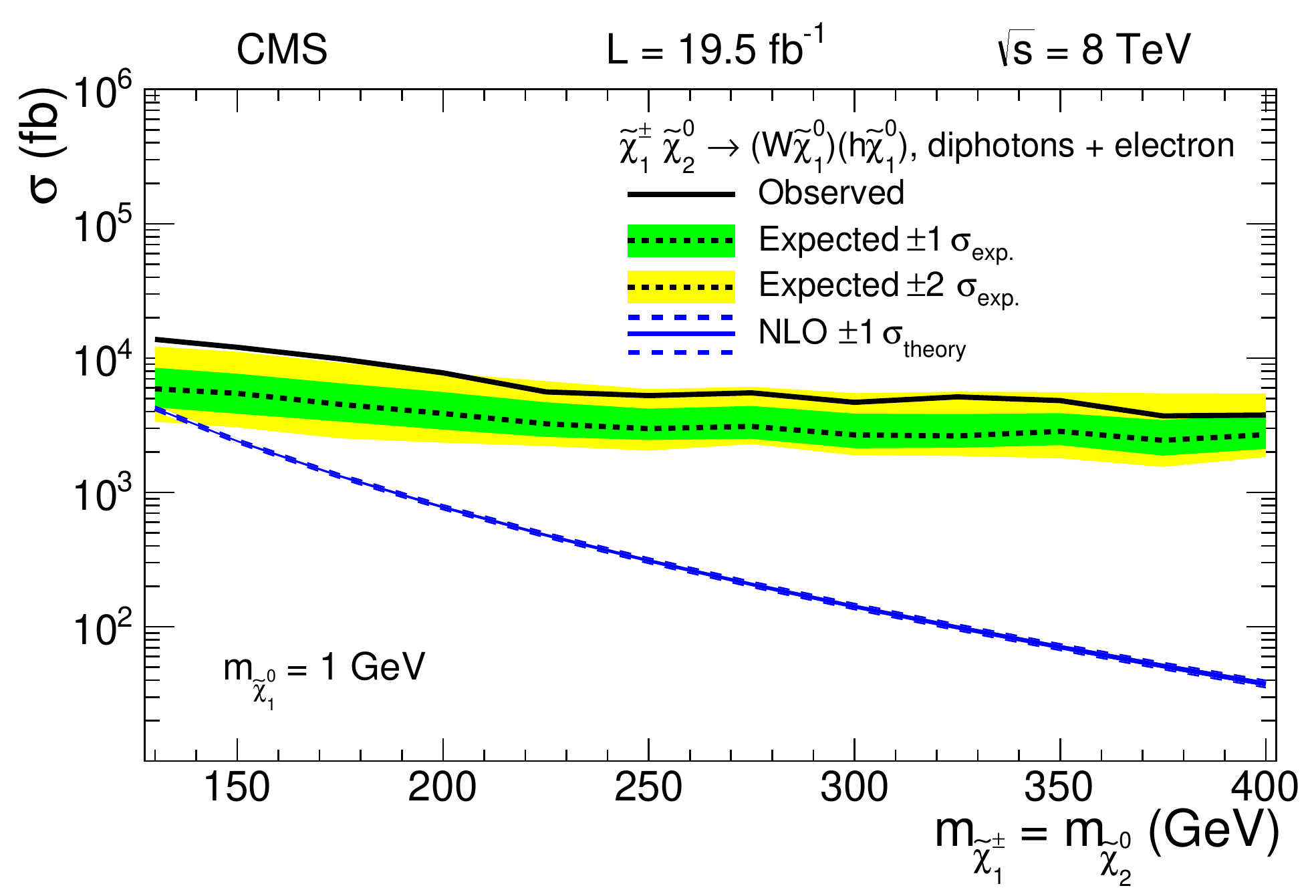}
  \caption{
Observed and expected 95\% confidence level upper limits on
the cross section for chargino-neutralino
$\PSGcpm_1\PSGczDt$ pair production
(with ${m}_{\PSGcpm_1}={m}_{\PSGczDt}$)
as a function of the \PSGczDt mass
assuming an LSP mass of 1\GeV,
for (\cmsLeftTop)
the $\gamma\gamma$+2 jets
study of Section~\ref{sec-gg-2j},
and (\cmsMiddle and bottom),
the $\gamma\gamma$+leptons studies
(for the muon and electron samples, respectively)
of Section~\ref{sec-hgglepton}.
The dark (green) and light (yellow) bands
indicate the one- and two-standard-deviation
uncertainty intervals, respectively.
The theoretical cross section is also shown.
}
  \label{fig:hw-limits-02}
\end{figure}

For small values of the LSP mass,
we exclude chargino-neutralino pair production
for $\PSGczDt=\PSGcpm_1$ mass values up to 210\GeV,
based on the theoretical prediction for the cross section
minus one standard deviation of its uncertainty.
This represents a modest improvement of about 5\%
compared to the corresponding result in Ref.~\cite{Khachatryan:2014qwa}.
The individual diphoton cross section results
assuming $\mchionez=1\GeV$
are presented in Fig.~\ref{fig:hw-limits-02}.

\section{Summary}
\label{sec-summary}

Searches are presented for the electroweak pair production
of higgsinos (\PSGczDo) in proton-proton collisions at 8\TeV,
based on the gauge-mediated-SUSY-breaking
scenario of Ref.~\cite{Matchev:1999ft}.
Each higgsino is presumed to decay to a Higgs boson (\Ph) and
the lightest supersymmetric particle (LSP),
which escapes without detection,
or else to a \Z boson and an LSP,
where the LSP is an almost massless gravitino~\lsp.
We search for an excess,
relative to the expectation from standard model processes,
of events with an $\Ph\Ph$, $\Ph\Z$, or $\Z\Z$ boson pair
produced in association with a large value of either missing transverse energy \met,
transverse mass \mt,
or the scalar sum \sthiggs of the two boson transverse momenta,
depending on the search channel.
In addition,
we perform searches for electroweak chargino-neutralino
($\PSGcpm_1\PSGczDt$) pair production
in channels with an $\Ph\PW$ boson pair and \met.
In the latter case,
the LSP is a massive neutralino,
also denoted \PSGczDo.
The assumed decay modes are $\PSGcpm_1\to\PW\PSGczDo$
and $\PSGczDt\to\Ph\PSGczDo$.
The data sample,
collected with the CMS detector at the LHC in 2012,
corresponds to an integrated luminosity of about 19.5\fbinv.

We select events with four bottom-quark jets ($\cPqb$ jets),
events with two $\cPqb$ jets and two photons,
and events with two $\cPqb$ jets and an $\ell^+\ell^-$ pair
(with $\ell$ an electron or muon),
providing sensitivity to the
$\Ph(\to\bbbar)\Ph(\to\bbbar)$,
$\Ph(\to\gamma\gamma)\Ph(\to\bbbar)$,
and
$\Ph(\to\bbbar)\Z(\to\ell^+\ell^-)$
channels,
respectively.
We also select events with two photons
accompanied by two light-quark jets,
and events with two photons
accompanied by at least one electron or muon,
providing sensitivity to
the $\Ph(\to\gamma\gamma)\Z/\PW(\to\,$2~jets$)$
channels,
and to
the $\Ph(\to\gamma\gamma)\Ph(\to\Z\Z/\PW\PW/\tau\tau)$
and $\Ph(\to\gamma\gamma)\Z/\PW$
channels where the \Z and {\PW} bosons decay leptonically.
As an aid for studies of signal scenarios
other than those considered in this paper,
Tables~\ref{tab:sigeff250}-\ref{table:m-200-cutflow}
of the Appendix provide results for the signal yields at
different stages of the event selection process
for the studies presented herein.
We incorporate results from
Refs.~\cite{Chatrchyan:2014aea} and~\cite{Khachatryan:2014qwa}
to gain sensitivity to higgsino pair production in the $\Z\Z$ channel
and to access complementary $\PSGcpm_1\PSGczDt$ decay modes.

The results are combined in a likelihood fit to derive 95\%
confidence level upper limits on the higgsino pair production cross section
in the two-dimensional plane of the higgsino branching fraction
to the $\Ph\lsp$ state versus the higgsino mass \mhiggsino,
where $\PSGczDo\to\Ph\lsp$ and $\PSGczDo\to\Z\lsp$
are taken as the only possible higgsino decay modes.
With the $\PSGczDo\to\Z\lsp$ branching fraction set to unity,
higgsinos with a mass value below 380\GeV are excluded.
With the $\PSGczDo\to\Ph\lsp$ branching fraction set to unity,
higgsinos are not excluded for any mass value,
but we obtain an expected exclusion region
that lies just above the theoretical higgsino pair production cross section
for higgsino mass values $\mhiggsino\lesssim 360\GeV$.

We also determine 95\% confidence level upper limits on
the cross section for electroweak chargino-neutralino
$\PSGcpm_1\PSGczDt$ pair production,
adding the search channels
with $\Ph\to\gamma\gamma$
and either $\PW\to\,$2~jets or $\PW\to\ell\nu$
to the results presented in Ref.~\cite{Khachatryan:2014qwa}.
For small values of the LSP mass,
we exclude this process for chargino mass values up to 210\GeV,
where the $\PSGcpm_1$ and \PSGczDt masses are taken to be equal.

\begin{acknowledgments}
\hyphenation{Bundes-ministerium Forschungs-gemeinschaft Forschungs-zentren} We congratulate our colleagues in the CERN accelerator departments for the excellent performance of the LHC and thank the technical and administrative staffs at CERN and at other CMS institutes for their contributions to the success of the CMS effort. In addition, we gratefully acknowledge the computing centers and personnel of the Worldwide LHC Computing Grid for delivering so effectively the computing infrastructure essential to our analyses. Finally, we acknowledge the enduring support for the construction and operation of the LHC and the CMS detector provided by the following funding agencies: the Austrian Federal Ministry of Science, Research and Economy and the Austrian Science Fund; the Belgian Fonds de la Recherche Scientifique, and Fonds voor Wetenschappelijk Onderzoek; the Brazilian Funding Agencies (CNPq, CAPES, FAPERJ, and FAPESP); the Bulgarian Ministry of Education and Science; CERN; the Chinese Academy of Sciences, Ministry of Science and Technology, and National Natural Science Foundation of China; the Colombian Funding Agency (COLCIENCIAS); the Croatian Ministry of Science, Education and Sport, and the Croatian Science Foundation; the Research Promotion Foundation, Cyprus; the Ministry of Education and Research, Estonian Research Council via IUT23-4 and IUT23-6 and European Regional Development Fund, Estonia; the Academy of Finland, Finnish Ministry of Education and Culture, and Helsinki Institute of Physics; the Institut National de Physique Nucl\'eaire et de Physique des Particules~/~CNRS, and Commissariat \`a l'\'Energie Atomique et aux \'Energies Alternatives~/~CEA, France; the Bundesministerium f\"ur Bildung und Forschung, Deutsche Forschungsgemeinschaft, and Helmholtz-Gemeinschaft Deutscher Forschungszentren, Germany; the General Secretariat for Research and Technology, Greece; the National Scientific Research Foundation, and National Innovation Office, Hungary; the Department of Atomic Energy and the Department of Science and Technology, India; the Institute for Studies in Theoretical Physics and Mathematics, Iran; the Science Foundation, Ireland; the Istituto Nazionale di Fisica Nucleare, Italy; the Korean Ministry of Education, Science and Technology and the World Class University program of NRF, Republic of Korea; the Lithuanian Academy of Sciences; the Ministry of Education, and University of Malaya (Malaysia); the Mexican Funding Agencies (CINVESTAV, CONACYT, SEP, and UASLP-FAI); the Ministry of Business, Innovation and Employment, New Zealand; the Pakistan Atomic Energy Commission; the Ministry of Science and Higher Education and the National Science Centre, Poland; the Funda\c{c}\~ao para a Ci\^encia e a Tecnologia, Portugal; JINR, Dubna; the Ministry of Education and Science of the Russian Federation, the Federal Agency of Atomic Energy of the Russian Federation, Russian Academy of Sciences, and the Russian Foundation for Basic Research; the Ministry of Education, Science and Technological Development of Serbia; the Secretar\'{\i}a de Estado de Investigaci\'on, Desarrollo e Innovaci\'on and Programa Consolider-Ingenio 2010, Spain; the Swiss Funding Agencies (ETH Board, ETH Zurich, PSI, SNF, UniZH, Canton Zurich, and SER); the Ministry of Science and Technology, Taipei; the Thailand Center of Excellence in Physics, the Institute for the Promotion of Teaching Science and Technology of Thailand, Special Task Force for Activating Research and the National Science and Technology Development Agency of Thailand; the Scientific and Technical Research Council of Turkey, and Turkish Atomic Energy Authority; the National Academy of Sciences of Ukraine, and State Fund for Fundamental Researches, Ukraine; the Science and Technology Facilities Council, UK; the US Department of Energy, and the US National Science Foundation.

Individuals have received support from the Marie-Curie programme and the European Research Council and EPLANET (European Union); the Leventis Foundation; the A. P. Sloan Foundation; the Alexander von Humboldt Foundation; the Belgian Federal Science Policy Office; the Fonds pour la Formation \`a la Recherche dans l'Industrie et dans l'Agriculture (FRIA-Belgium); the Agentschap voor Innovatie door Wetenschap en Technologie (IWT-Belgium); the Ministry of Education, Youth and Sports (MEYS) of the Czech Republic; the Council of Science and Industrial Research, India; the HOMING PLUS programme of Foundation for Polish Science, cofinanced from European Union, Regional Development Fund; the Compagnia di San Paolo (Torino); the Consorzio per la Fisica (Trieste); MIUR project 20108T4XTM (Italy); the Thalis and Aristeia programmes cofinanced by EU-ESF and the Greek NSRF; and the National Priorities Research Program by Qatar National Research Fund.
\end{acknowledgments}

\bibliography{auto_generated}   % will be created by the tdr script.

\appendix

\clearpage

\section{Event selection flow tables}
\label{app:event-flow}

In this appendix,
we present tables that illustrate the
event selection process, or ``flow'',
for the analyses presented in
Sections~\ref{sec-hh-4b}--\ref{sec-z-to-ll}.
For each analysis,
the selection flow is illustrated for two or more signal points.
These tables are intended as an aid for
those wishing to replicate
these analyses
using signal scenarios other than those considered
in the present work.

\begin{table*}[htb]
\topcaption{
Number of signal events remaining after each stage of the event
selection for the $\Ph\Ph\to\bbbar\bbbar$ search,
with a higgsino mass of 250\GeV
and an LSP (gravitino) mass of 1\GeV.
The results are normalized to an integrated luminosity of 19.3\fbinv
using NLO+NLL calculations.
The uncertainties are statistical.
``\metsig bin 0'' corresponds to $0<\metsig<30$.
The baseline selection accounts for the primary vertex criteria
and for quality requirements applied to the \met distribution.
This search is described in Section~\ref{sec-hh-4b}.
}
\label{tab:sigeff250}
\centering
\begin{scotch}{l|ccccc}
$\Ph\Ph$ events, $m_{\PSGczDo} = 250\GeV$
         &  $\metsig$ bin 0 &  $\metsig$ bin 1 &  $\metsig$ bin 2 &  $\metsig$ bin 3 &  $\metsig$ bin 4\\
\hline
All events                    & $590\pm 2$ & $264\pm 2$ & $376\pm 2$ & $107\pm 1$ & $22.7\pm 0.5$\\
Baseline selection            & $548\pm 2$ & $257\pm 2$ & $369\pm 2$ & $106\pm 1$ & $22.1\pm 0.5$\\
$\pt>50\GeV$, leading 2 jets  & $470\pm 2$ & $220\pm 1$ & $321\pm 2$ & $95\pm 1$ & $20.7\pm 0.5$\\
Number of jets = 4 or 5       & $288\pm 2$ & $132\pm 1$ & $196\pm 1$ & $58.3\pm 0.8$ & $12.2\pm 0.4$\\
Lepton vetoes                 & $280\pm 2$ & $128\pm 1$ & $190\pm 1$ & $56.7\pm 0.8$ & $11.7\pm 0.4$\\
Isolated track veto           & $253\pm 2$ & $116\pm 1$ & $173\pm 1$ & $51.9\pm 0.7$ & $10.8\pm 0.3$\\
$\mdp$ requirement            & $111\pm 1$ & $64.3\pm 0.8$ & $133\pm 1$ & $42.6\pm 0.7$ & $9.1\pm 0.3$\\
\hline
3$\cPqb$  selection                 & $15.3\pm 0.4$ & $8.6\pm 0.3$ & $19.0\pm 0.4$ & $6.3\pm 0.3$ & $1.3\pm 0.1$\\
$\dRmax < 2.2$                & $6.6\pm 0.3$ & $3.4\pm 0.2$ & $7.6\pm 0.3$ & $2.5\pm 0.2$ & $0.53\pm 0.08$\\
Higgs boson SIG region        & $2.7\pm 0.2$ & $1.3\pm 0.1$ & $2.7\pm 0.2$ & $0.87\pm 0.10$ & $0.14\pm 0.04$\\
Trigger emulation             & $0.41\pm 0.06$ & $0.83\pm 0.08$ & $2.3\pm 0.1$ & $0.82\pm 0.09$ & $0.13\pm 0.04$\\
\hline
4$\cPqb$  selection                 & $20.3\pm 0.5$ & $12.3\pm 0.4$ & $26.3\pm 0.5$ & $8.4\pm 0.3$ & $1.7\pm 0.1$\\
$\dRmax < 2.2$                & $9.8\pm 0.3$ & $5.9\pm 0.2$ & $11.6\pm 0.3$ & $3.6\pm 0.2$ & $0.79\pm 0.09$\\
Higgs boson SIG region        & $4.7\pm 0.2$ & $3.0\pm 0.2$ & $5.1\pm 0.2$ & $1.5\pm 0.1$ & $0.30\pm 0.06$\\
Trigger emulation             & $0.55\pm 0.07$ & $1.8\pm 0.1$ & $4.4\pm 0.2$ & $1.4\pm 0.1$ & $0.28\pm 0.05$\\
\end{scotch}
\end{table*}

\begin{table*}[htb]
\topcaption{
Number of signal events remaining after each stage of the event
selection for the $\Ph\Ph\to\bbbar\bbbar$ search,
with a higgsino mass of 400\GeV
and an LSP (gravitino) mass of 1\GeV.
The results are normalized to an integrated luminosity of 19.3\fbinv
using NLO+NLL calculations.
The uncertainties are statistical.
``\metsig bin 0'' corresponds to $0<\metsig<30$.
The baseline selection accounts for the primary vertex criteria
and for quality requirements applied to the \met distribution.
This search is described in Section~\ref{sec-hh-4b}.
}
\label{tab:sigeff400}
\centering
\begin{scotch}{l|ccccc}
$\Ph\Ph$ events, $m_{\PSGczDo} = 400\GeV$
     &  $\metsig$ bin 0 &  $\metsig$ bin 1 &  $\metsig$ bin 2 &  $\metsig$ bin 3 &  $\metsig$ bin 4\\
\hline
All events                   & $28.8\pm 0.3$ & $15.9\pm 0.2$ & $35.3\pm 0.3$ & $31.1\pm 0.3$ & $51.9\pm 0.4$\\
Baseline selection           & $26.9\pm 0.3$ & $15.6\pm 0.2$ & $34.6\pm 0.3$ & $30.5\pm 0.3$ & $50.9\pm 0.4$\\
$\pt>50\GeV$, leading 2 jets & $25.3\pm 0.2$ & $14.6\pm 0.2$ & $32.4\pm 0.3$ & $28.8\pm 0.3$ & $49.3\pm 0.3$\\
Number of jets = 4 or 5      & $15.7\pm 0.2$ & $9.1\pm 0.1$ & $19.8\pm 0.2$ & $17.6\pm 0.2$ & $30.4\pm 0.3$\\
Lepton vetoes                & $15.3\pm 0.2$ & $8.8\pm 0.1$ & $19.3\pm 0.2$ & $17.1\pm 0.2$ & $29.8\pm 0.3$\\
Isolated track veto          & $13.9\pm 0.2$ & $8.0\pm 0.1$ & $17.6\pm 0.2$ & $15.6\pm 0.2$ & $27.3\pm 0.3$\\
$\mdp$ requirement           & $5.9\pm 0.1$ & $4.25\pm 0.10$ & $13.3\pm 0.2$ & $12.9\pm 0.2$ & $24.4\pm 0.2$\\
\hline
3$\cPqb$  selection                & $0.85\pm 0.04$ & $0.56\pm 0.04$ & $1.90\pm 0.07$ & $1.70\pm 0.06$ & $3.64\pm 0.09$\\
$\dRmax < 2.2$               & $0.44\pm 0.03$ & $0.31\pm 0.03$ & $1.03\pm 0.05$ & $0.91\pm 0.05$ & $2.12\pm 0.07$\\
Higgs boson SIG region       & $0.22\pm 0.02$ & $0.13\pm 0.02$ & $0.45\pm 0.03$ & $0.30\pm 0.03$ & $0.88\pm 0.05$\\
Trigger emulation            & $0.029\pm 0.007$ & $0.09\pm 0.01$ & $0.39\pm 0.03$ & $0.29\pm 0.03$ & $0.83\pm 0.04$\\
\hline
4$\cPqb$  selection                & $1.18\pm 0.05$ & $0.85\pm 0.04$ & $2.44\pm 0.08$ & $2.57\pm 0.08$ & $4.6\pm 0.1$\\
$\dRmax < 2.2$               & $0.77\pm 0.04$ & $0.52\pm 0.04$ & $1.40\pm 0.06$ & $1.59\pm 0.06$ & $3.02\pm 0.09$\\
Higgs boson SIG region       & $0.45\pm 0.03$ & $0.29\pm 0.03$ & $0.77\pm 0.04$ & $0.83\pm 0.04$ & $1.56\pm 0.06$\\
Trigger emulation            & $0.07\pm 0.01$ & $0.20\pm 0.02$ & $0.68\pm 0.04$ & $0.78\pm 0.04$ & $1.47\pm 0.06$\\
\end{scotch}
\end{table*}

\begin{table*}[htb]
\topcaption{
Number of signal events remaining after each stage of the event
selection for the
$\Ph\Ph\to\gamma\gamma\bbbar$ search,
described in Section~\ref{sec-hhggbb},
and for the
$\Ph\Z$ and $\Ph\PW\to\gamma\gamma$+2 jets search,
described in Section~\ref{sec-gg-2j}.
The $\Ph\Ph$ and $\Ph\Z$ scenarios assume a
higgsino mass value of 130\GeV
and an LSP (gravitino) mass of 1\GeV.
For the $\Ph\PW$ scenario,
${m}_{\PSGcpm_1}={m}_{\PSGczDt}=130\GeV$
and the LSP (\PSGczDo) mass is 1\GeV.
The results are normalized to an integrated luminosity of 19.7\fbinv
using NLO+NLL calculations for the
$\Ph\Ph$ and $\Ph\Z$ results
and NLO calculations for the $\Ph\PW$ results.
The uncertainties are statistical.
}
\centering
\begin{scotch}{lccc}
  & $\Ph\Ph$ events & $\Ph\Z$ events & $\Ph\PW$ events \\
\hline
All events                                         & $71.5\pm 0.4$ &  $63.3\pm 0.3$ & $118\pm 1$ \\
Trigger emulation                                  & $53.6\pm 0.4$ &  $48.3\pm 0.2$ & $89.9\pm 0.4$  \\
Photon selection (except for $\eta$ requirement)   & $34.0\pm 0.3$ &  $30.9\pm 0.2$ & $57.2\pm 0.4$  \\
$120<\mgg<131\GeV$                                 & $31.1\pm 0.3$ &  $28.0\pm 0.2$ & $51.9\pm 0.3$  \\
$\abs{\eta}<1.4442$ for photons                        & $20.0\pm 0.2$ &  $17.9\pm 0.1$ & $32.9\pm 0.3$  \\
Lepton vetoes                                      &  $4.1\pm 0.1$ &  $16.7\pm 0.1$ & $27.5\pm 0.2$  \\
Reject events with $95<\mbb<155\GeV$               &   \NA        &   $7.7\pm 0.1$ & $13.0\pm 0.2$  \\
$70<\mjj<110$\GeV                                  &   \NA         &   $4.6\pm 0.1$ &  $7.9\pm 0.1$  \\
Exactly two $\cPqb$ jets                                 &  $4.1\pm 0.1$ &    \NA  &   \NA   \\
$95<\mbb<155\GeV$                                  &  $3.5\pm 0.1$ &    \NA  &   \NA   \\
\end{scotch}
\end{table*}

\begin{table*}[htb]
\topcaption{\label{Table:ggLeptonCutflow}
Number of signal events remaining after each stage of the event
selection for the
$\Ph\Ph$ and $\Ph\PW\to\gamma\gamma$+leptons searches.
The $\Ph\Ph$ scenario assumes a
higgsino mass value of 130\GeV
and an LSP (gravitino) mass of 1\GeV.
For the $\Ph\PW$ scenario,
${m}_{\PSGcpm_1}={m}_{\PSGczDt}=130\GeV$
and the LSP (\PSGczDo) mass is 1\GeV.
The results are normalized to an integrated luminosity of 19.5\fbinv
using NLO+NLL calculations for the $\Ph\Ph$ results
and NLO calculations for the $\Ph\PW$ results.
The uncertainties are statistical.
The baseline selection accounts for the primary vertex criteria
and for quality requirements applied to the \met distribution.
This search is described in Section~\ref{sec-hgglepton}.
}
\centering
\begin{scotch}{lcccc}
  & \multicolumn{2}{c}{$\Ph\Ph$ events} & \multicolumn{2}{c}{$\Ph\PW$ events} \\
  & $\gamma\gamma+$muon & $\gamma\gamma+$electron & $\gamma\gamma+$muon & $\gamma\gamma+$electron \\
\hline
All events                              & $90.3\pm 0.6$ & $90.3\pm 0.6$ & $261\pm 1$     & $261\pm 1$    \\
Baseline selection                      & $90.3\pm 0.6$ & $90.3\pm 0.6$ & $261\pm 1$     & $261\pm 1$    \\
Trigger emulation                       & $70.7\pm 0.5$ & $70.7\pm 0.5$ & $200\pm 1$     & $200\pm 1$    \\
Photon selection                        & $27.4\pm 0.3$ & $27.4\pm 0.3$ & $77.8\pm 0.6$  & $77.8\pm 0.6$ \\
Lepton selection                        &  $3.3\pm 0.1$ &  $3.5\pm 0.1$ & $6.8\pm 0.2$   & $7.2\pm 0.2$  \\
At most one $\cPqb$ jet                       &  $3.3\pm 0.1$ &  $3.5\pm 0.1$ & $6.8\pm 0.2$   & $7.2\pm 0.2$  \\
$\Delta R(\gamma,\text{lepton})>0.3$  &  $3.3\pm 0.1$ &  $3.5\pm 0.1$ & $6.8\pm 0.2$   & $7.1\pm 0.2$  \\
Reject events with $86<\meg<96\GeV$     &  $3.3\pm 0.1$ &  $2.5\pm 0.1$ & $6.8\pm 0.2$   & $4.8\pm 0.1$  \\
$120<\mgg<131\GeV$                      &  $3.1\pm 0.1$ &  $2.3\pm 0.1$ & $6.4\pm 0.2$   & $4.5\pm 0.1$  \\
\end{scotch}
\end{table*}

\begin{table*}[htb]
\centering
\topcaption{\label{table:m-200-cutflow}
Number of signal events remaining after each stage of the event
selection for the $\Ph\Z$ search with
$\Ph\to\bbbar$ and $\Z\to\ell^+\ell^-$,
with higgsino mass values of~130 and 200\GeV
and an LSP (gravitino) mass of 1\GeV.
The results are normalized to an integrated luminosity of 19.5\fbinv
using NLO+NLL calculations.
The uncertainties are statistical.
The baseline selection accounts for the primary vertex criteria
and for quality requirements applied to the \met distribution.
This search is described in Section~\ref{sec-z-to-ll}.
}
\begin{scotch}{lcccccc}
& \multicolumn{3}{c}{$m_{\PSGczDo} = 130\GeV$} & \multicolumn{3}{c}{$m_{\PSGczDo} = 200\GeV$} \\
$\Ph\Z$ events & $\Pe\Pe$ & $\Pgm\Pgm$ & $\Pe\Pe$ + $\Pgm\Pgm$ & $\Pe\Pe$ & $\Pgm\Pgm$ & $\Pe\Pe$ + $\Pgm\Pgm$ \\
\hline
Baseline selection           & $579\pm 2$    & $576\pm 2$    & $1154\pm 2$   & $100\pm 1$    & $102\pm 1$    & $202\pm 1$ \\
Trigger emulation            & $548\pm 1$    & $494\pm 1$    & $1042\pm 2$   & $95.5\pm 0.6$ & $87.2\pm 0.5$ & $183\pm 1$ \\
Lepton ID \& isolation       & $262\pm 1$    & $315\pm 1$    & $577\pm 1$    & $50.0\pm 0.4$ & $60.9\pm 0.5$ & $111\pm 1$ \\
2 leptons ($\pt>20\GeV$)     & $238\pm 1$    & $287\pm 1$    & $525\pm 1$    & $47.2\pm 0.4$ & $57.3\pm 0.4$ & $105\pm 1$ \\
$81<\mll<101\GeV$            & $231\pm 1$    & $277\pm 1$    & $507\pm 1$    & $45.7\pm 0.4$ & $55.3\pm 0.4$ & $101\pm 1$ \\
Third-lepton veto              & $230\pm 1$    & $276\pm 1$    & $505\pm 1$    & $45.5\pm 0.4$ & $55.1\pm 0.4$ & $101\pm 1$ \\
Hadronic $\tau$-lepton veto  & $226\pm 1$    & $271\pm 1$    & $496\pm 1$    & $44.8\pm 0.4$ & $54.3\pm 0.4$ & $99.1\pm 0.5$ \\
$\geq$2 jets                & $148\pm 1$    & $176\pm 1$    & $323\pm 1$    & $31.0\pm 0.3$ & $37.5\pm 0.3$ & $68.5\pm 0.4$ \\
$\geq$2 $\cPqb$ jets        & $44.1\pm 0.4$ & $51.1\pm 0.4$ & $95.2\pm 0.6$ & $9.2\pm 0.2$ & $11.1\pm 0.2$ & $20.3\pm 0.3$ \\
$100<\mbb<150\GeV$           & $34.6\pm 0.3$ & $40.0\pm 0.3$ & $74.6\pm 0.5$ & $7.2\pm 0.2$ & $8.7\pm 0.2$ & $15.9\pm 0.3$ \\
$\MTtwoj>200\GeV$            & $7.6\pm 0.1$  & $8.4\pm 0.1$  & $16.0\pm 0.1$ & $3.0\pm 0.1$ & $3.3\pm 0.1$ & $6.3\pm 0.1$ \\
\hline
$\met>60\GeV$                & $2.6\pm 0.1$  & $2.8\pm 0.1$  & $5.4\pm 0.1$  & $2.2\pm 0.1$ & $2.5\pm 0.1$ & $4.7\pm 0.1$ \\
$\met>80\GeV$                & $1.5\pm 0.1$  & $1.6\pm 0.1$  & $3.1\pm 0.1$  & $2.0\pm 0.1$ & $2.2\pm 0.1$ & $4.2\pm 0.1$ \\
$\met>100\GeV$               & $0.8\pm 0.1$  & $0.9\pm 0.1$  & $1.7\pm 0.1$  & $1.6\pm 0.1$ & $1.7\pm 0.1$ & $3.3\pm 0.1$ \\
\end{scotch}
\end{table*}

\cleardoublepage \section{The CMS Collaboration \label{app:collab}}\begin{sloppypar}\hyphenpenalty=5000\widowpenalty=500\clubpenalty=5000\textbf{Yerevan Physics Institute,  Yerevan,  Armenia}\\*[0pt]
V.~Khachatryan, A.M.~Sirunyan, A.~Tumasyan
\vskip\cmsinstskip
\textbf{Institut f\"{u}r Hochenergiephysik der OeAW,  Wien,  Austria}\\*[0pt]
W.~Adam, T.~Bergauer, M.~Dragicevic, J.~Er\"{o}, C.~Fabjan\cmsAuthorMark{1}, M.~Friedl, R.~Fr\"{u}hwirth\cmsAuthorMark{1}, V.M.~Ghete, C.~Hartl, N.~H\"{o}rmann, J.~Hrubec, M.~Jeitler\cmsAuthorMark{1}, W.~Kiesenhofer, V.~Kn\"{u}nz, M.~Krammer\cmsAuthorMark{1}, I.~Kr\"{a}tschmer, D.~Liko, I.~Mikulec, D.~Rabady\cmsAuthorMark{2}, B.~Rahbaran, H.~Rohringer, R.~Sch\"{o}fbeck, J.~Strauss, A.~Taurok, W.~Treberer-Treberspurg, W.~Waltenberger, C.-E.~Wulz\cmsAuthorMark{1}
% ...
\vskip\cmsinstskip
% ...
1:~~Also at Vienna University of Technology, Vienna, Austria\\
2:~~Also at CERN, European Organization for Nuclear Research, Geneva, Switzerland\\
% ...

\end{sloppypar}
\end{document}